\font\twelvrm=cmr12
\font\ninerm=cmr9
\font\twelvi=cmmi12
\font\ninei=cmmi9
\font\twelvex=cmex10 scaled\magstep1
\font\twelvbf=cmbx12
\font\ninebf=cmbx9
\font\twelvit=cmti12
\font\twelvsy=cmsy10 scaled\magstep1
\font\ninesy=cmsy9
\font\twelvtt=cmtt12

\font\twelvsl=cmsl12

\font\abstractfont=cmr10
\font\abstractitalfont=cmti10

\def\twelvepoint{\def\rm{\fam0\twelvrm}
  \textfont0=\twelvrm \scriptfont0=\ninerm \scriptscriptfont0=\sevenrm
  \textfont1=\twelvi \scriptfont1=\ninei \scriptscriptfont1=\seveni
  \textfont2=\twelvsy \scriptfont2=\ninesy \scriptscriptfont2=\sevensy
  \textfont3=\twelvex \scriptfont3=\tenex \scriptscriptfont3=\tenex
	\textfont\itfam=\twelvit \def\it{\fam\itfam\twelvit}
	\textfont\slfam=\twelvsl \def\sl{\fam\slfam\twelvsl}
	\textfont\ttfam=\twelvtt \def\tt{\fam\ttfam\twelvtt}
	\textfont\bffam=\twelvbf \def\bf{\fam\bffam\twelvbf}
	\scriptfont\bffam=\ninebf  \scriptscriptfont\bffam=\sevenbf
        \skewchar\ninei='177
        \skewchar\twelvi='177
        \skewchar\seveni='177
}

\newdimen\normalwidth
\newdimen\double
\newdimen\single
\newdimen\indentlength		\indentlength=.5in

\newif\ifdrafton

\def\galley{
 \draftonfalse
 \twelvepoint
 \rm
 \font\chapterfont=cmbx10 scaled\magstep1
 \font\sectionfont=cmbx12
 \font\subsectionfont=cmbx12
 \font\headingfont=cmr10 scaled\magstep2
 \font\titlefont=cmbx10 scaled\magstep2
 \normalwidth=5.7in
 \double=.34in
 \single=.17in
 \hsize=\normalwidth
 \vsize=8.7in
 \hoffset=0.48in
 \voffset=0.1in
 \hfuzz=0.5pt
 \baselineskip=\double plus 2pt minus 2pt }

\parindent=\indentlength
\clubpenalty=10000
\widowpenalty=10000
\displaywidowpenalty=500
\overfullrule=2pt
\tolerance=100

\newcount\chapterno	\chapterno=0
\newcount\sectionno	\sectionno=0
\newcount\appno		\appno=0
\newcount\subsectionno	\subsectionno=0
\newcount\eqnum	\eqnum=0
\newcount\refno \refno=0
\newcount\chap
\newcount\figno \figno=0
\newcount\tableno \tableno=0
\newcount\lettno \lettno=0

\def\bodypaging{
 \headline={\ifodd\chap \hfil \else \tenrm\hfil\twelvrm\folio \fi}
 \footline={\rm \ifodd\chap \global\chap=0 \tenrm\hfil\twelvrm\folio\hfil
 \else \hfil \fi}}

%
\def\preq{\mainid.\the\eqnum}
\def\Elabel#1{\xdef#1{\mainid.\the\eqnum}}
\def\eq{\global\advance\eqnum by1 \eqno(\mainid.\the\eqnum)}
\def\eqlabel#1{\eq {\xdef#1{\mainid.\the\eqnum}}}
\def\quieteqlabel#1{\advance\eqnum by1 {\xdef#1{\mainid.\the\eqnum}}}
\def\newlett{\global\lettno=1\global\advance\eqnum by1
     \eqno(\mainid.\the\eqnum a)}
\def\lett{\global\advance\lettno by 1
  \eqno(\mainid.\the\eqnum{\ifcase\lettno\or a\or b\or c\or d\or e\or
    f\or g\or h\or \fi})}
\def\newlettlabel#1{\newlett {\xdef#1{\mainid.\the\eqnum}}}
\def\lettlabel#1{\lett {\xdef#1{\mainid.\the\eqnum}}}
%
\newwrite\refs
\def\startrefs#1{\immediate\openout\refs=#1
\immediate\write\refs{\global\chap=1}
\immediate\write\refs{\vfill\noexpand\eject\noexpand\vglue.2in}
\immediate\write\refs{\noexpand\centerline{\headingfont REFERENCES}}
\immediate\write\refs{\noexpand\vglue.5in\baselineskip=\single}
\immediate\write\refs{\parindent=16pt \parskip=\single}}

\def\ref#1{\advance\refno by1 \the\refno \immediate\write\refs
{\noexpand\item{\the\refno.}#1 \hfill\par}}
\def\andref#1{\advance\refno by1\kern-.4em[\the\refno]\immediate\write\refs
{\noexpand\item{\the\refno.}#1 \hfill\par}}
\def\cite#1{[#1]}

\def\refname#1{ \xdef#1{\the\refno}}

\newif\iftoc \tocfalse
\newif\ifpart \partfalse
\newwrite\conts
\def\startcontents{\iftoc\immediate\openout\conts=contents
\immediate\write\conts{\noexpand\centerline{\headingfont CONTENTS}}
\immediate\write\conts{\noexpand\vglue.5in\baselineskip=\single}
\immediate\write\conts{\parskip=0pt \parindent=0pt}\else\relax\fi}

\def\chaptercont#1{\iftoc\immediate\write\conts{
  \vskip\single\the\chapterno.\ #1\hfill\folio\vskip0cm}\else\relax\fi}
\def\andchaptercont#1{\iftoc\immediate\write\conts{
  \hskip .5in \ #1\hfill}\else\relax\fi}
\def\sectioncont#1{\iftoc\immediate\write\conts{\hskip.5cm
  \the\chapterno.\the\sectionno\ #1\hfill\folio\vskip0cm}\else\relax\fi}
\def\subsectioncont#1{\iftoc\immediate\write\conts{\hskip1cm
  \the\chapterno.\the\sectionno.\the\subsectionno\
  #1\hfill\folio\vskip0cm}\else\relax\fi}
\def\appendixcont#1{\iftoc\immediate\write\conts{\vskip\single
  Appendix \mainid.\ #1\hfill\folio\vskip0cm}\else\relax\fi}

\def\partcont#1{\iftoc
 \ifpart\immediate\write\conts{\noexpand\vfill\noexpand\eject}\fi
 \immediate\write\conts{\vskip\single
 \noexpand\centerline{- PART #1 -}}
 \parttrue
 \else\relax\fi}


\newif\ifpskip

\def\chapter#1{
  \ifpskip\vfill\eject \else \bigskip\bigskip\bigskip \fi
 \global\advance\chapterno by1 \sectionno=0 \subsectionno=0 \eqnum=0
 \def\mainid{\the\chapterno}
 \vglue.8cm\centerline{\chapterfont
 \uppercase\expandafter{\romannumeral\the\chapterno.\ #1}}
 \nobreak\vskip.2cm\nobreak
 \ifpskip\global\chap=1\fi
 \chaptercont{#1}}

\def\andchapter#1{\centerline{\chapterfont\raise.1cm
 \hbox{\uppercase{#1}}}\vskip.2cm\nobreak\andchaptercont{#1}}

\def\section#1{
 \global\advance\sectionno by1 \subsectionno=0
 \vskip.8cm\centerline{\sectionfont \the\chapterno.\the\sectionno\ #1}
 \nobreak\vskip.2cm\nobreak
 \sectioncont{#1}}

\def\andsection#1{\centerline{\sectionfont\raise.1cm\hbox{#1}}
 \nobreak\vskip.2cm\nobreak}
\def\subsection#1{
 \global\advance\subsectionno by1
 \vskip.2cm\leftline{\subsectionfont
 \the\chapterno.\the\sectionno.\the\subsectionno\ #1}
 \subsectioncont{#1}}
\def\appendix#1{
 \vfil\eject
 \global\advance\appno by1 \subsectionno=0 \eqnum=0
 \def\mainid{\ifcase\appno\or A\or B\or C\or D\or E\or F\or G\or H\or \fi}
 \vglue.8cm\centerline{\sectionfont
 \uppercase{\mainid .\ \ #1}}
 \nobreak\vskip.2cm\nobreak
 \global\chap=1
 \appendixcont{#1}}

\def\loneappendix#1{
 \vfil\eject
 \global\advance\appno by1 \subsectionno=0 \eqnum=0
 \def\mainid{\ifcase\appno\or A\or B\or C\or D\or E\or F\or G\or H\or \fi}
 \vglue.8cm\centerline{\sectionfont
 \uppercase{APPENDIX: #1}}
 \nobreak\vskip.2cm\nobreak
 \global\chap=1
 \appendixcont{#1}}
\def\Slabel#1{\xdef#1{\mainid
              \ifnum\sectionno>0
                    { {.\the\sectionno}
                        \ifnum\subsectionno>0
                             { .\the\subsectionno}
                         \fi}
                \fi}}
%
\newif\iftable \tablefalse
\newwrite\tablelist
\def\starttablelist{\iftable
 \immediate\openout\tablelist=tablelist
 \immediate\write\tablelist{\noexpand\centerline{\headingfont LIST
     OF TABLES}}
 \immediate\write\tablelist{\vskip\double\vskip\single\baselineskip=\single}
 \immediate\write\tablelist{\parskip=0pt \parindent=0pt}\else\relax\fi}
\def\inserttable#1#2{\global\advance\tableno by 1 \vfill\vskip\double
 \centerline{Table \the\tableno:\ #2}\nobreak\vskip\double\nobreak
 \begingroup #1 \endgroup \vskip\double
 \iftable\immediate\write\tablelist{\hskip1cm
 \the\tableno. #2 \hfill\folio\vskip0cm}\else\relax\fi}
\def\andinserttable#1#2{\vfill\vskip\double
 \centerline{Table \the\tableno\ (continued):\ #2}
 \nobreak\vskip\double\nobreak
 \begingroup #1 \endgroup \vskip\double}

\def\noadvancetable#1#2{\vfill\vskip\double
 \centerline{Table \the\tableno:\ #2}\nobreak\vskip\double\nobreak
 \begingroup #1 \endgroup \vskip\double
 \iftable\immediate\write\tablelist{\hskip1cm
 \the\tableno. #2 \hfill\folio\vskip0cm}\else\relax\fi}
\def\continuetable#1#2{\vfill\vskip\double
 \centerline{Table \the\tableno{ }(continued):\ #2}\nobreak
 \vskip\double\nobreak
 \begingroup #1 \endgroup \vskip\double}
\def\Tlabel#1{\global\advance\tableno by 1
     \xdef#1{\the\tableno} \the\tableno}
%
\newif\iffig \figfalse
\newwrite\figlist
\newwrite\figs
\def\startfiglist{\iffig\immediate\openout\figlist=figlist
 \immediate\openout\figs=figs
 \immediate\write\figlist{\noexpand\centerline{\headingfont LIST
    OF FIGURES}}
 \immediate\write\figlist{\vskip\double\vskip\double\baselineskip=\single}
 \immediate\write\figlist{\parskip=0pt \parindent=0pt}\else\relax\fi}
\def\insertpagefig#1{\iffig\global\advance\figno by 1
 \immediate\write\figs{\vfil\noexpand\eject\bodypaging\pageno=\the\pageno
 \noexpand\vglue 20cm}
 \immediate\write\figs{\noexpand\centerline{Figure \the\figno. #1}}
 \immediate\write\figlist{\hskip1cm
 \the\figno. #1 \hfill\folio\vskip0cm}\advance\pageno by1 \else\relax\fi}
\def\insertfig#1#2{\iffig\global\advance\figno by 1
 \vglue #2
 \centerline{Figure \the\figno. #1}
 \immediate\write\figlist{\hskip1cm
 \the\figno. #1 \hfill\folio\vskip0cm} \else\relax\fi}
\def\insertfig#1#2{\iffig\global\advance\figno by 1
 \vglue #2
 \centerline{Figure \the\figno. #1}
 \immediate\write\figlist{\hskip1cm
 \the\figno. #1 \hfill\folio\vskip0cm} \else\relax\fi}

\def\Flabel#1{\global\advance\figno by 1 \xdef#1{\the\figno} \the\figno}
\def\noadvancefig#1#2{\iffig
 \vglue #2
 \centerline{Figure \the\figno. #1}
 \immediate\write\figlist{\hskip1cm
 \the\figno. #1 \hfill\folio\vskip0cm} \else\relax\fi}

%
%

\input tables

\galley

%
%
\def \arglt#1#2#3#4{(#1;#2,#3,\ldots,#4)}
\def \argrt#1#2#3#4{(#1,#2,\ldots,#3;#4)}
\def \argltpos#1#2#3#4{(#1^{+};#2,#3,\ldots,#4)}
\def \argltneg#1#2#3#4{(#1^{-};#2,#3,\ldots,#4)}
\def \argrtpos#1#2#3#4{(#1,#2,\ldots,#3;#4^{+})}
\def \argrtneg#1#2#3#4{(#1,#2,\ldots,#3;#4^{-})}

%
\def \positronon{\bar\psi}
\def \electronon{\psi}

\def \Wnorm{W}
\def \Wms{{\cal{W}}}
\def \Wbar{{\overline{\Wms}}}

\def \PHI{{\mit\Phi}}

\def \pole{{\mit\Pi}}

\def \amp{{\cal M}}

%
%
\def \permsum#1#2{\sum_{{\cal{P}}(#1\ldots #2)}}

\def \braket#1#2{ \langle #1 \thinspace\thinspace #2 \rangle }
\def \bra#1{ \langle #1 | }
\def \ket#1{ | #1 \rangle }

\def \eps{\epsilon}
\def \vareps{\varepsilon}
\def \down{{}_\downarrow}
\def \up{{}_\uparrow}

\def \Poff{{\cal{P}}}
\def \Qoff{{\cal{Q}}}

\def \link#1#2#3{ {{\braket{#1}{#3}}\over{\bra{#1}#2\ket{#3}}} }
\def \invlink#1#2#3{ {{\bra{#1}#2\ket{#3}}\over{\braket{#1}{#3}}} }

\def \linkstar#1#2#3{ { {{\braket{#1}{#3}}^{*}}
\over{{\bra{#1}#2\ket{#3}}^{*}}} }


\def \dyadic#1{\vbox{\ialign{##\crcr
     $\hfil
{\thinspace\scriptstyle\leftrightarrow}
\hfil$\crcr\noalign{\kern-.01pt\nointerlineskip}
     $\hfil\displaystyle{#1}\hfil$\crcr}}}

\def\centeronto#1#2{{\setbox0=\hbox{#1}\setbox1=\hbox{#2}\ifdim
\wd1>\wd0\kern.5\wd1\kern-.5\wd0\fi
\copy0\kern-.5\wd0\kern-.5\wd1\copy1\ifdim\wd0>\wd1
\kern.5\wd0\kern-.5\wd1\fi}}
\def\slash#1{\centeronto{$#1$}{$/$}}

\def \varsp{\thinspace}


\hyphenation{ap-pen-dix  spin-or  spin-ors}
\def \PHImod{{\it{\Theta}}}

\def \Wmod{{\cal V}}
\def \Wmodbar{{\overline{\Wmod}}}

\def \curlyZ{{\cal Z}}

\def \curlyK{{\cal K}}

\def \zhe{{\cal Z}}

\def\upsilon{\theta}


\def \msreplacefermion{A.23}
\def \msreplacedotprod{A.22}

\def \msconvent{A}


\def \slashsqr{A.6}

\def \dopermsum{A.19}

\def \fierz{A.14}
\def \Shouten{A.8}

\def \linkidnosum{A.15}
\def \linkidsummed{A.16}

\def \splitid{A.27}



\def \antisym{A.13a}

\def \reverseid{A.28}

\def \prodtosum{A.25}

\def \Xid{B}

\def \XidsimpleA{B.3}
\def \XidsimpleB{B.16}

\def \XidmessyA{B.17}
\def \XidmessyB{B.41}

{\nopagenumbers
\hfill\hbox{
CLNS 92/1154}

\hfill\hbox{
September 1992}
\vfill
\baselineskip=\double
{\titlefont
        \centerline{MULTIPHOTON PRODUCTION AT HIGH}
        \centerline{ENERGIES IN THE  STANDARD MODEL II}
}
\bigskip
\bigskip
\bigskip
\centerline{Gregory MAHLON$^{1}$}
\medskip

{\baselineskip=.20 in plus 2pt minus 2 pt
\centerline{
{\it Newman Laboratory of Nuclear Studies,}
}
\centerline{
{\it Cornell University,
Ithaca, NY 14853, USA}
}
}

\bigskip
\bigskip
\bigskip
\centerline{ABSTRACT}
{\narrower\medskip\baselineskip=.20in plus 2pt minus 2pt
{\abstractfont
We examine multiphoton production in the electroweak
sector of the Standard Model  in the high energy
limit using the equivalence theorem in combination with spinor
helicity techniques.  We utilize currents consisting of a
charged scalar, spinor, or vector line that radiates
{\abstractitalfont n}\ photons.  Only one end of the charged line
is off shell in these currents, which are known for the cases of
like-helicity and one unlike-helicity photons.
We obtain a wide variety of helicity amplitudes
for processes involving two pairs of charged particles by
considering combinations of four currents.
We examine the situation with respect to currents which
have both ends of the charged line off-shell, and present
solutions for the case of like-helicity photons.
These new currents may be combined with two of the
original currents to produce additional amplitudes
involving Higgs, longitudinal
{\abstractitalfont Z}\ or neutrino pairs.
}\smallskip}

\vfill

\noindent
\hrule
\smallskip

\noindent
$^1$ e-mail:  gdm@beauty.tn.cornell.edu

\eject}

\pageno=2  \bodypaging
\startrefs{refs1}

\chapter{INTRODUCTION}

In this paper we will conclude a study of the high energy scatterings
involving many vector bosons and Higgs bosons in a spontaneously
symmetry-broken  gauge theory begun in references
[\ref{C. Dunn and T.--M. Yan, Nucl. Phys. {\bf B352}, 402 (1991).}]
\refname\DY
and
[\ref{G. Mahlon and T.--M. Yan, Cornell preprint CLNS 91/1119 (1992).}].
\refname\firstpaper
This program is a generalization of the work on QCD by
Berends and Giele
[\ref{F. A. Berends and W. T. Giele, Nucl. Phys. {\bf B306},
759 (1988).}].
\refname\BG
In particular, we will consider the Weinberg-Salam-Glashow
model
[\ref{S. L. Glashow, Nucl. Phys. {\bf 22}, 579 (1961);
S. Weinberg, Phys. Rev. Lett. {\bf 19}, 1264 (1967);
A. Salam in {\it Proc. 8th Nobel Symposium,
Aspen\"asgarden,} edited by  N. Svartholm, (Almqvist and
Wiksell, Stockholm, 1968), p. 367.}],
with a focus upon processes involving
an arbitrary number of photons, plus one or two pairs of  charged
particles ($W^{+}W^{-}$, $\ell^{\pm}W^{\mp}$, or $\ell^{+}\ell^{-}$),
and possibly one or two $Z$'s, Higgs bosons, or neutrinos.
\refname\WSGmodel{\kern-.5em}

Our work is based upon three main ideas:  the equivalence theorem
[\ref{J. M. Cornwall, D. N. Levin, and G. Tiktopoulous,
Phys. Rev. {\bf D10}, 1145 (1974);
B. W. Lee, C. Quigg,
and H. Thacker, Phys. Rev. {\bf D16}, 1519 (1977);
M. S. Chanowitz and M. K. Gaillard, Nucl. Phys. {\bf B261},
379 (1985);
G. J. Gounaris, R. Kogerler, and H. Neufeld,
Phys. Rev. {\bf D34}, 3257 (1986).}],
\refname\ET
the multispinor
[\ref{J. Schwinger, {\it Particles, Sources and Fields,}
(Addison-Wesley, Redwood City, 1970), Vol. I;
Ann. Phys. {\bf 119}, 192 (1979).},
\refname\SCHWINGER  {\kern-.75em}
\ref{
The spinor technique was first introduced by the
CALCUL collaboration, in the context
of massless Abelian gauge theory:
P. De Causmaecker, R. Gastmans, W. Troost, and T.T. Wu,
Phys. Lett. {\bf 105B}, 215 (1981);
P. De Causmaecker, R. Gastmans, W. Troost, and T.T. Wu,
Nucl. Phys. {\bf B206}, 53 (1982);
F. A. Berends, R. Kleiss, P. De Causmaecker, R. Gastmans,
W. Troost, and T.T. Wu, Nucl. Phys. {\bf B206}, 61 (1982);
F.A. Berends, P. De Causmaecker, R. Gastmans, R. Kleiss,
W. Troost, and T.T. Wu, Nucl. Phys. {\bf B239}, 382 (1984);
{\bf B239}, 395 (1984); {\bf B264}, 243 (1986); {\bf B264}, 265 (1986).
\smallskip
\noexpand\item{ }
By now, many papers have been published on the subject.  A
partial list of references follows.
\smallskip
\noexpand\item{ }
P. De Causmaecker, thesis, Leuven University, 1983;
R. Farrar and F. Neri, Phys. Lett. {\bf 130B}, 109 (1983);
R. Kleiss, Nucl. Phys. {\bf B241}, 61 (1984);
Z. Xu, D.H. Zhang and Z. Chang, Tsingua University
preprint TUTP-84/3, 84/4, and 84/5a (1984);
Nucl. Phys. {\bf B291}, 392 (1987);
J.F. Gunion and Z. Kunszt, Phys. Lett. {\bf 161B}, 333 (1985);
F.A. Berends, P.H. Davereldt, and R. Kleiss, Nucl. Phys. {\bf B253},
441 (1985);
R. Kleiss and W.J. Stirling, Nucl. Phys. {\bf B262} 235 (1985);
J.F. Gunion and Z. Kunszt, Phys. Lett. {\bf 159B}, 167 (1985);
{\bf 161B}, 333 (1985);
S.J. Parke and T.R. Taylor, Phys. Rev. Lett. {\bf 56}, 2459 (1986);
Z. Kunszt, Nucl. Phys. {\bf B271}, 333 (1986);
J.F. Gunion and J. Kalinowski, Phys. Rev. {\bf D34}, 2119 (1986);
R. Kleiss and W.J. Stirling, Phys. Lett. {\bf 179B}, 159 (1986);
M. Mangano and S.J. Parke, Nucl. Phys. {\bf B299}, 673 (1988);
M. Mangano, S.J. Parke, and Z. Xu, Nucl. Phys. {\bf B298},
653 (1988);
D.A. Kosower, B.--H. Lee, and V.P. Nair, Phys. Lett. {\bf 201B},
85 (1988);
M. Mangano and S.J. Parke, Nucl. Phys. {\bf B299}, 673 (1988);
F.A. Berends and W.T. Giele, Nucl. Phys. {\bf B313}, 595 (1989);
M. Mangano, Nucl. Phys. {\bf B315}, 391 (1989);
D.A. Kosower, Nucl. Phys. {\bf B335}, 23 (1990);
Phys. Lett. {\bf B254}, 439 (1991);
Z. Bern and D.A. Kosower, Nucl. Phys. {\bf B379}, 451 (1992);
C.S. Lam, McGill preprint McGill/92-32, 1992.},
\refname\REVIEWy {\kern-.75em}
\ref{Many of the results for processes containing six or fewer
particles are collected in R. Gastmans and T.T. Wu,
{\it The Ubiquitous Photon:  Helicity Method for
QED and QCD} (Oxford University Press, New York, 1990).},
\ref{
The excellent review by Mangano and Parke provides a guide to the
various approaches to  and extensive literature on the subject:
M. Mangano and S.J. Parke,  Phys. Reports {\bf 200}, 301 (1991).}]
\refname\REVIEW
representation of a vector field, and the use of recursion
relations \cite{\DY,\firstpaper,\BG}.
The equivalence theorem allows us to identify
the longitudinal degrees of freedom of the $W^{\pm}$ and $Z$ bosons
with the corresponding would-be Goldstone bosons  $\phi^{\pm}$ and
$\phi_2$, up to corrections of the order of the vector boson mass
divided by the center-of-mass energy.
The multispinor representation of a vector field allows us to treat
fermions and vector bosons on an equal footing by replacing the
conventional Lorentz 4-vector with  a second rank spinor which may be
thought of as a combination of two spin-$1\over2$ objects.
We will use Weyl-van der Waerden spinors in this work
(see Appendix \msconvent{ }for a summary of our conventions).

Reference \cite\firstpaper\ develops recursion relations
for currents consisting of a single charged line, with $n$
on-shell photons attached.  This line could be scalar,
spinor, or vector in nature, and has one end off shell.
It is possible to solve the recursion relations in the
cases where either all, or all but one, of the photons have
the same helicity.  From the solutions for the currents,
we may obtain amplitudes for various processes.  Those amplitudes
which are calculable from the currents taken either singly
or in pairs are discussed in reference \cite\firstpaper.  In this
paper, we will discuss processes involving combinations
of four currents, such as
$$
e^{+} e^{-}\longrightarrow W^{+} W^{-} \gamma\ldots\gamma.
\eqlabel\sampleprocess
$$
By choosing various combinations of the four currents, a wide
range of processes is covered.  All of these amplitudes
involve diagrams which contain some type of neutral particle propagating
between the charged lines.  The next logical question to
ask concerns processes for which the propagating particle
itself carries a charge.  In that case, charge conservation
dictates that we combine three currents, one of which has two
off-shell particles.  The complications involved in this case
is the topic of the remainder of the paper.

We have organized our presentation as follows.  Section 2 is
a brief review of the electroweak recursion relations and
their solutions.  All of the ingredients necessary to
compute quadruple current amplitudes  like
(\sampleprocess) are
presented there.  Section 3 describes, using an explicit example,
how to perform calculations involving four currents.
In addition, the results for the helicity combinations we are
able to solve for are collected there.  Next, we proceed
to discuss the prospects for obtaining double-off-shell
currents.
As we see in Section 4, although we cannot always obtain expressions
for the currents themselves, we are able to obtain enough
information to compute the desired amplitudes.  Section 5
presents an example of such a computation, along with the
rest of the results involving three currents.  The
final section contains a few concluding remarks.
\vfill\eject
\chapter{THE ELECTROWEAK RECURSION RELATIONS}
\andchapter{AND THEIR SOLUTIONS}

In this section we will review the recursion relations
for currents containing a charged line plus $n$ photons
presented in reference \cite\firstpaper.
All of the photons will be on shell, and one end of the
charged line will be off shell.  Within the framework
of the Weinberg-Salam-Glashow model, the charged line can have
spin 0, spin $1\over2$, or spin 1.  We will consider each possibility
in turn.


\section{The longitudinal $W$ current}

We define the quantity $\PHI(P;1,2,\ldots,n)$ to represent
the sum of all tree-level diagrams consisting of an
unbroken scalar line (the would-be Goldstone bosons $\phi^{\pm}$)
with $n$ photons attached in all possible ways.  By the equivalence
theorem, this represents a longitudinally polarized $W$ ($\equiv W_L$)
that radiates
$n$ photons.  Choose the convention that all momenta flow into
the diagram.  Then, the on-shell $\phi^{+}$ has momentum $P$,
the $n$ on-shell photons have momenta $k_1,k_2,\ldots,k_n$, and
the off-shell $\phi^{-}$ has momentum $Q=-[P+k_1+\ldots+k_n]\equiv
-[P+\kappa(1,n)]$.  In reference \cite\firstpaper, we show that
$\PHI(P;1,2,\ldots,n)$ satisfies the recursion relation
$$
\eqalign{
&\PHI\arglt{P}{1}{2}{n}=
\cr& \enspace =
{{-e\sqrt2}\over{[P+\kappa(1,n)]^2}}
\Biggl[
\permsum{1}{n} \negthinspace
{1\over{(n-1)!}}
\bar\eps^{\dot\alpha\alpha}(n) [P{+}\kappa(1,n)]_{\alpha\dot\alpha}
\thinspace
\PHI\arglt{P}{1}{2}{n{-}1}
\cr &
\qquad\qquad\quad\thinspace
+e\sqrt2
\permsum{1}{n}
{1\over{2!\thinspace (n-2)!}} \thinspace
\bar\eps^{\dot\alpha\alpha}(n{-}1)\eps_{\alpha\dot\alpha}(n)
\thinspace\PHI\arglt{P}{1}{2}{n{-}2}
\Biggr]
}
\eqlabel\WLrecursionperm
$$
where the symbol ${{\cal{P}}(1\ldots n)}$ denotes the set of permutations
of the momenta $k_1, k_2, \ldots, k_n$.   This form of the
recursion relation is permitted by the Bose symmetry enjoyed
by the photons.  The sources of the two terms in (\WLrecursionperm)
are obvious:  the first term is from the three-point vertex,
while the second term is generated by the seagull vertex.
By definition
$$
\PHI(P)\equiv1.
\eqlabel\PHIP
$$
We denote by $\PHI(1,2,\ldots,n;Q)$
the same current, but with the  $\phi^{+}$ off shell instead.
The two currents are connected in the expected manner:
$$
\PHI(1,2,\ldots,n;Q)=(-1)^n\PHI(Q;1,2,\ldots,n).
\eqlabel\scalarcrossing
$$

Closed form solutions to the recursion relation (\WLrecursionperm)
are known for two special helicity configurations.  If all of the photons
have the same helicity, or all but one of the photons have the same
helicity, then it is possible to choose the gauge momenta  such that
$$
\bar\eps^{\dot\alpha\alpha}(j) \eps_{\alpha\dot\alpha}(\ell) =0
\eqlabel\seagullrid
$$
for any pair of polarization spinors.  The advantage of having
(\seagullrid) hold is the vanishing of all of the seagull
contributions to (\WLrecursionperm), leaving what is effectively
a single-term recursion relation.
Thus, to obtain $\PHI(P;1^{+},\ldots,n^{+})$, we choose
$$
\eps_{\alpha\dot\alpha}(j^{+}) \equiv
{
{ u_{\alpha}(g) \bar u_{\dot\alpha}(k_j) }
\over
{ \braket{j}{g} }
},
\eqlabel\allpluspolarizations
$$
with the same gauge spinor $g$ for each of the photons.  In this
case, the solution to (\WLrecursionperm) is \cite\firstpaper
$$
\PHI(P;1^{+},\ldots,n^{+}) =
(-e\sqrt2)^n
\permsum{1}{n}
{
{\braket{P}{g}}
\over
{\bra{P} 1,\ldots,n \ket{g}}
},
\eqlabel\PHIallplussoln
$$
where
$$
\bra{P} 1,2,\ldots,n \ket{g} \equiv
\braket{P}{1} \braket{1}{2} \cdots \braket{n}{g}.
\eqlabel\stringdef
$$
Some useful properties of this ``string'' of
spinor inner products are given in Appendix \msconvent.
This expression has the proper $n=0$ limit to match smoothly
onto (\PHIP).

Next, consider the case  of one photon with differing helicity.  For
concreteness, let us choose that photon to carry momentum $k_1$.  If
we choose $g=k_1$ in (\allpluspolarizations) for $j=2,\ldots,n$,
and set
$$
\eps_{\alpha\dot\alpha}(1^{-}) \equiv
{
{ u_{\alpha}(k_1) \bar u_{\dot\alpha}(h) }
\over
{ {\braket{1}{h}}^{*} }
},
\eqlabel\oneminuspolarizations
$$
with $h$ an arbitrary null-momentum,   it is not
hard to see that (\seagullrid) still holds.  The solution to
(\WLrecursionperm) for this case reads \cite\firstpaper
$$
\eqalign{
\PHI&(P;1^{-},2^{+},\ldots,n^{+})=
\cr & =
-(-e\sqrt2)^n \permsum{2}{n}
{
{\braket{P}{1}}
\over
{\bra{P} 2,\ldots,n \ket{1}}
}
\Biggl\{
{
{ {\braket{h}{P}}^{*} }
\over
{ { \bra{h}1\ket{P} }^{*} }
} \cr&\qquad\qquad\qquad\qquad
+ (1-{\delta}_{n1})
\sum_{j=2}^n
u^{\alpha}(k_1){{\pole}_{\alpha}}^{\beta}(P,1,2,\ldots,j)
u_{\beta}(k_1)
\Biggr\}.
}
\eqlabel\PHIoneminussoln
$$
Notice that since the first photon is distinguishable by its
helicity, it has been explicitly removed from the permutation sum.
The quantity $\pole$ appearing in (\PHIoneminussoln) is
defined by
$$
{{\pole}_{\alpha}}^{\beta}(P,1,2,\ldots,j) \equiv
{
{(k_{j})_{\alpha\dot\alpha}
[\bar P + \bar \kappa(1,j)]^{\dot\alpha\beta}}
\over
{[P+\kappa(1,j{-}1)]^2
[P+\kappa(1,j)]^2}
}.
\eqlabel\poledef
$$
Some useful properties of $\pole$ are listed in Appendix \msconvent.
Equation (\PHIoneminussoln)
is valid for all $n\ge1$.


\section{The fermion currents}

Next, we consider the spin-$1\over2$ case.
We define the quantity $\positronon(p;1,2,\ldots,n)$ to represent
the sum of all tree-level diagrams consisting of an
unbroken fermion line with $n$ photons attached in all possible
ways.  Again, all momenta flow into the diagram.  The on-shell
positron has momentum $p$, the $n$ on-shell photons have
momenta $k_1,k_2,\ldots,k_n$, and the off-shell electron has
momentum $q= -[p+\kappa(1,n)]$.  Berends and Giele~\cite\BG
obtain  recursion relations for $\positronon$, which may be
cast in the form \cite\firstpaper
$$
\eqalign{
{{\positronon}_{\dot\alpha}}&\argltpos{p}{1}{2}{n}=
\cr & =
-e\sqrt2
\permsum{1}{n} {{1}\over{(n-1)!}}
\positronon_{\dot\beta}\argltpos{p}{1}{2}{n{-}1}
{\bar{\eps}}^{\dot\beta\beta}(n)
{{[p+\kappa(1,n)]_{\beta\dot\alpha}}\over{[p+\kappa(1,n)]^2}},
}
\newlettlabel\spinorperm
$$
$$
\eqalign{
{{\positronon}^{\alpha}}&\argltneg{p}{1}{2}{n}=
\cr & =
-e\sqrt2
\permsum{1}{n} {{1}\over{(n-1)!}}
\positronon^{\beta}\argltneg{p}{1}{2}{n{-}1}
{\bar{\eps}}_{\beta\dot\beta}(n)
{{[\bar p+\bar\kappa(1,n)]^{\dot\beta\alpha}}\over{[p+\kappa(1,n)]^2}}.
}
\lett
$$
In the massless limit considered here, the helicities of all
fermions are conserved; hence the pair of recursion relations.
The zero-photon currents are
$$
\positronon_{\dot\alpha}(p^{+}) \equiv \bar u_{\dot\alpha}(p)
\newlettlabel\spinorZERO
$$
and
$$
\positronon^{\alpha}(p^{-}) \equiv u^{\alpha}(p).
\lett
$$
We may also define currents in which the positron is off shell
and the electron is on shell.  These will be denoted by
$\electronon(1,2,\ldots,n;q)$ and are related to the
$\positronon$'s by
$$
{{\electronon}^{\dot\alpha}}\argrtpos{1}{2}{n}{q}=
(-1)^n{{\positronon}^{\dot\alpha}}\argltpos{q}{1}{2}{n},
\newlettlabel\electrononconnection
$$
$$
{{\electronon}_{\alpha}}\argrtneg{1}{2}{n}{q}=
(-1)^n{{\positronon}_{\alpha}}\argltneg{q}{1}{2}{n},
\lett
$$
as required by charge-conjugation symmetry.

The gauge choice (\allpluspolarizations) allows us
to solve the recursion relations (\spinorperm) for
like-helicity photons, with the results \cite\firstpaper
$$
{\positronon}_{\dot\alpha}(p^{+};1^{+},\ldots,n^{+}) =
(-e\sqrt2)^n \permsum{1}{n}
{
{-u^{\beta}(g)[p+\kappa(1,n)]_{\beta\dot\alpha}}
\over
{ \bra{p} 1,2,\ldots,n \ket{g} }
},
\newlettlabel\spinorallplussoln
$$
$$
{\positronon}^{\alpha}(p^{-};1^{+},\ldots,n^{+}) =
(-e\sqrt2)^n \permsum{1}{n}
{
{u^{\alpha}(p) \varsp \braket{p}{g}}
\over
{ \bra{p} 1,2,\ldots,n \ket{g} }
}.
\lett
$$
Notice that except for  the required spinor structure,
the solutions (\spinorallplussoln) contain the same functional
form as the solution (\PHIallplussoln) for $\PHI(P;1^{+},\ldots,n^{+})$.
That is, currents of differing spins are proportional to each
other, a SUSY-like relation.
The $n=0$ forms of (\spinorallplussoln) match smoothly onto
(\spinorZERO).
\def\RHpositronallplussoln{\spinorallplussoln a}

To obtain solutions when the first photon has negative helicity,
set $g=k_1$ in (\allpluspolarizations), and take (\oneminuspolarizations)
for $\eps_{\alpha\dot\alpha}(1^{-})$.  Then, the following
solutions result:\cite\firstpaper
$$
\eqalign{
{\positronon}_{\dot\alpha}&(p^{+};1^{-},2^{+},\ldots,n^{+}) =
\cr & =
(-e\sqrt2)^n \permsum{2}{n}
{
{u^{\beta}(k_1)
[p+\kappa(1,n)]_{\beta\dot\alpha}}
\over
{\bra{p} 2,\ldots,n \ket{1}}
}
\Biggl\{
{
{ {\braket{h}{p}}^{*} }
\over
{ { \bra{h}1\ket{p} }^{*} }
}
\cr&\qquad\qquad\qquad\quad
+ (1-{\delta}_{n1})
\sum_{j=2}^n
u^{\alpha}(k_1){{\pole}_{\alpha}}^{\beta}(p,1,2,\ldots,j)
u_{\beta}(k_1)
\Biggr\},
}
\newlettlabel\spinoroneminussoln
$$
$$
\eqalign{
{\positronon}^{\alpha}&(p^{-};1^{-},2^{+},\ldots,n^{+}) =
\cr & =
(-e\sqrt2)^n \permsum{2}{n}
{
{-\braket{p}{1}}
\over
{\bra{p} 2,\ldots,n \ket{1}}
}
\Biggl\{
{
{ {\braket{h}{p}}^{*} }
\over
{ { \bra{h}1\ket{p} }^{*} }
}u^{\alpha}(p)
-{
{ u^{\alpha}(k_1)}
\over
{{\braket{p}{1}}^{*}}
}
\cr&\qquad\qquad\qquad\quad
+ (1-{\delta}_{n1}) \braket{p}{1}
\sum_{j=2}^n
u^{\beta}(k_1){{\pole}_{\beta}}^{\alpha}(p,1,2,\ldots,j)
\Biggr\}.
}
\lett
$$
Even though the fermionic case has no seagulls to dispose of, and
hence a simpler recursion relation than  the scalar case, it is
still not possible to solve for currents containing more than
one unlike helicity.


\section{The transverse $W$ currents}

Finally, we consider spin-1 currents.
We define  $\Wnorm(P;1,2,\ldots,n)$ to represent
the sum of all tree-level diagrams consisting of an
unbroken vector line
with $n$ photons attached in all possible ways.  As usual,
all momenta flow into the diagram.  The on-shell $W^{+}$ has momentum $P$,
the $n$ on-shell photons have momenta $k_1,k_2,\ldots,k_n$, and
the off-shell $W^{-}$ has momentum $Q=
-[P+\kappa(1,n)]$.    In Lorentz 4-vector notation the recursion
relation for $\Wnorm$ reads \cite\firstpaper
$$
\eqalign{
\Wnorm^\mu&\arglt{P}{1}{2}{n} =
\cr & =
{-e\over{[P+\kappa(1,n)]^2}}
\Biggl[
\permsum{1}{n}
{1\over{(n-1)!}} \thinspace
\biggl[\eps(n),\Wnorm \arglt{P}{1}{2}{n{-}1}\biggr]^{\mu}
\cr&\quad
+e\permsum{1}{n}
{1\over{2!\thinspace (n-2)!}} \thinspace
\biggl\{ \eps(n{-}1),\Wnorm\arglt{P}{1}{2}{n{-}2},\eps(n)
\biggr\}^{\mu}
\Biggr],
}
\eqlabel\WTrecursionperm
$$
where
$$
\eqalign{
\biggl[\eps(n),{\Wnorm}&\arglt{P}{1}{2}{n{-}1}\biggr]^{\mu}\equiv
\cr & =
2[P+\kappa(1,n{-}1)] \cdot \eps(n)\thinspace\thinspace
{\Wnorm}^{\mu}\arglt{P}{1}{2}{n{-}1} \cr&
- 2k_n \cdot {\Wnorm}\arglt{P}{1}{2}{n{-}1} \thinspace\thinspace
{\eps}^{\mu}(n) \cr&
+ \eps(n)\cdot {\Wnorm}\arglt{P}{1}{2}{n{-}1}\thinspace\thinspace
\biggl[k_n-[P+\kappa(1,n{-}1)] \biggr]^{\mu},
}
\eqlabel\sqbrak
$$
and
$$
\eqalign{
\biggl\{\eps(&n{-}1),\Wnorm\arglt{P}{1}{2}{n{-}2},\eps(n){\biggr\}}^{\mu}
\equiv \cr & =
\eps(n{-}1)\cdot
[\eps(n)\Wnorm^{\mu}\arglt{P}{1}{2}{n{-}2}
-\Wnorm\arglt{P}{1}{2}{n{-}2} \eps^{\mu}(n)]
\cr & \quad
-\eps(n)\cdot
[\Wnorm\arglt{P}{1}{2}{n{-}2}\eps^{\mu}(n{-}1)
-\eps(n{-}1) \Wnorm^{\mu}\arglt{P}{1}{2}{n{-}2}].
}
\eqlabel\curlybrak
$$
The current $W$ is a conserved quantity, satisfying \cite\firstpaper
$$
[P+\kappa(1,n)]\cdot\Wnorm\arglt{P}{1}{2}{n}=0.
\eqlabel\Wcc
$$
The current $W(1,2,\ldots,n;Q)$, with an on-shell $W^{-}$ of momentum
$Q$ and an off-shell $W^{+}$ is related to $W(Q;1,2,\ldots,n)$
by
$$
\Wnorm \argrt{1}{2}{n}{Q}=(-1)^n\Wnorm\arglt{Q}{1}{2}{n}.
\eqlabel\Wcrossing
$$

The recursion relation is soluble if all of the particles have the
same helicity or only one helicity differs.  If all of the
particles have positive helicity, we choose  (\allpluspolarizations)
for the photons and
$$
{\Wms}_{\alpha\dot\alpha}(P^{+})=
{{ u_{\alpha}(g) {\bar u}_{\dot\alpha}(P) }
\over {\braket{P}{g}}}.
\eqlabel\polallplusW
$$
It is not hard to show that with this choice of gauge spinors,
not only does (\seagullrid) hold, but also
$$
\bar\eps^{\dot\alpha\alpha}(j^{+})
{\Wms}_{\alpha\dot\alpha}(P^{+};1^{+},\ldots,n^{+})
=0,
\eqlabel\nice
$$
for all $j$ and $n$.  Because of (\seagullrid) and (\nice),
the seagull contributions represented by the curly bracket
function  (\curlybrak) vanish, and the square bracket function
(\sqbrak) reduces to
$$
\eqalign{
\biggl[\eps(n),{\Wnorm}&\arglt{P}{1}{2}{n{-}1}\biggr]^{\mu}=
\cr & =
2[P+\kappa(1,n{-}1)] \cdot \eps(n)\thinspace
{\Wnorm}^{\mu}\arglt{P}{1}{2}{n{-}1} \cr&
- 2k_n \cdot {\Wnorm}\arglt{P}{1}{2}{n{-}1} \thinspace {\eps}^{\mu}(n).
}
\eqlabel\sqbraknice
$$
These simplifications are sufficient to allow solution of the
recursion relation (\WTrecursionperm), with the result \cite\firstpaper
$$
{\Wms}_{\alpha\dot\alpha}(P^{+};1^{+},\ldots,n^{+}) =
(-e\sqrt2)^n  \permsum{1}{n}
{
{- u_{\alpha}(g)u^{\beta}(g)[P+\kappa(1,n)]_{\beta\dot\alpha} }
\over
{ \braket{P}{g} \varsp \bra{P} 1,2,\ldots,n \ket{g} }
}.
\eqlabel\WTallplussoln
$$
Once more, the same functional form as in the scalar case appears,
with the added spinor structure required to describe a spin-1
particle.  Equation (\WTallplussoln) reduces to (\polallplusW)
for $n=0$.

If we set $g=k_1$ in (\allpluspolarizations) and (\polallplusW),
and use the choice (\oneminuspolarizations) for the first
photon, then we are able to  obtain
${\Wms}(P^{+};1^{-},2^{+},\ldots,n^{+})$.
Because all of the polarization spinors are proportional to
$u_{\alpha}(k_1)$, the key properties (\seagullrid) and (\nice)
still hold, and we may eliminate the seagull contributions
and use  (\sqbraknice) when solving the recursion relation.
The result is not surprising \cite\firstpaper:
$$
\eqalign{
{\Wms}_{\alpha\dot\alpha}&(P^{+};1^{-},2^{+},\ldots,n^{+})=
\cr & =
{ {(-e\sqrt2)^n}\over{\braket{P}{1}} } \permsum{2}{n}
{
{u_{\alpha}(k_1)u^{\beta}(k_1)[P+\kappa(1,n)]_{\beta\dot\alpha}}
\over
{\bra{P} 2,\ldots,n \ket{1}}
}
\Biggl\{
{
{ {\braket{h}{P}}^{*} }
\over
{ { \bra{h}1\ket{P} }^{*} }
}
\cr&\qquad\qquad\qquad\quad
+ (1-{\delta}_{n1})
\sum_{j=2}^n
u^{\gamma}(k_1){{\pole}_{\gamma}}^{\delta}(P,1,2,\ldots,j)
u_{\delta}(k_1)
\Biggr\},
}
\eqlabel\WToneminussoln
$$
valid for $n\ge1$.

Instead of allowing one of the photons to have negative helicity,
we may choose to let
the $W$ have negative helicity.
In this case, we set $g=P$ in (\allpluspolarizations) and write
$$
{\Wms}_{\alpha\dot\alpha}(P^{-})=
{
{ u_{\alpha}(P) {\bar u}_{\dot\alpha}(h) }
\over
{{\braket{P}{h}}^{*}}
}.
\eqlabel\negW
$$
The recursion relation simplifies as before, and we  easily
obtain \cite\firstpaper
$$
\eqalign{
{\Wms}_{\alpha\dot\alpha}(P^{-};1^{+})& =
-e\sqrt2 \thinspace \thinspace
{
{u_{\alpha}(P)\bar u_{\dot\alpha}(k_1)}
\over
{[P+k_1]^2}
}
{
{ {\braket{h}{1}}^{*} }
\over
{ {\braket{P}{h}}^{*} }
},
}
\newlettlabel\negWTallplus
$$
$$
\eqalign{
{\Wms}_{\alpha\dot\alpha}&(P^{-};1^{+},\ldots,n^{+}) =
\cr & =
(-e\sqrt2)^n  \permsum{1}{n}
{
{ {\braket{h}{1}}^{*} }
\over
{ {\braket{P}{h}}^{*} }
}
{
{u_{\alpha}(P)u^{\beta}(P)[P+\kappa(1,n)]_{\beta\dot\alpha} }
\over
{ \braket{P}{1} \bra{P}2,\ldots,n\ket{P} }
}
\cr & \qquad\qquad\qquad\times
\sum_{\ell=2}^n
u^{\gamma}(P) {{\pole}_{\gamma}}^{\delta}(P,1,\ldots,\ell) u_{\delta}(P).
}
\lett
$$
Note that (\negWTallplus b) holds only for $n\ge 2$.  For
$n=1$ we use (\negWTallplus a) and for $n=0$ we use (\negW).
\chapter{QUADRUPLE CURRENT AMPLITUDES}

Since we have already discussed those amplitudes which may be
obtained from the currents taken either
singly or in pairs \cite\firstpaper,
we begin with an examination of those processes which may be
computed from the combination of four currents.  Each process
involves a pair of charged lines, and the corresponding Feynman
diagrams have the basic topology of
Figure \Flabel\quadfig .  Depending upon the
identities of the four currents, variations upon this basic
structure may be possible.  In
particular, note that the $\phi^4$ vertex enters directly when
all four currents are scalars, and the strength of the coupling
$\lambda$ as compared to $e$ becomes an issue when deciding
which contributions are the most important.

The layout of this section is as follows.  First, we will
illustrate the techniques required to compute quadruple current
amplitudes by discussing the process
$$
 e^{+}_{\down}e^{-}_{\up}  \longrightarrow
W^{+}_{\up} W^{-}_{\up} \gamma^{ }_{\down}
\gamma^{ }_{\up} \cdots \gamma^{ }_{\up}.
\eqlabel\quadsample
$$
Then, we will tabulate the various results which may be obtained
using the currents reviewed in the previous section.

\section{The process $e^{+}_{\downarrow}e^{-}_{\uparrow} \longrightarrow
W^{+}_{\uparrow} W^{-}_{\uparrow} \gamma^{ }_{\downarrow}
\gamma^{ }_\uparrow \cdots \gamma^{ }_\uparrow$}

As illustrated in Figure \quadfig, there are three basic
contributions to the process (\quadsample).  The first two
contributions have unbroken $W$ lines with either a photon or
a $Z$-boson connecting the $W$ line to the fermion line, as
illustrated in the upper half of Figure \quadfig.  We will refer
to these as the photon and $Z$ contributions respectively.
The third diagram, which is only present when the fermion line is
left-handed, consists of a  broken $W$ line, with
a neutrino propagating  across the gap, as pictured in the
lower half of Figure \quadfig.   This will be referred to
as the neutrino contribution.

We will choose the convention that all of the momenta flow
into the diagram.  Hence, variations on the basic process
(\quadsample) are easily obtained by crossing.  The
positron will have momentum $p$, the electron momentum $q$,
the $W^{+}$ momentum $P$ and the $W^{-}$ momentum $Q$.
The photons will have momenta labelled by $k_1, k_2, \ldots, k_n$.
The first photon will be the one which is selected to carry
negative helicity; the remaining $n-1$ photons will all have
positive helicity.

\eject

\subsection{The photon  contribution}

Denote by $\amp_{\gamma}$ those contributions to (\quadsample) that
involve the exchange of a photon between the spinor and vector lines.
{}From Figure \quadfig, we see that we have
$$
\eqalign{
\amp_{\gamma}&(p,q;P,Q;1,\ldots,n) =
\cr & =
\permsum{1}{n} \sum_{r=0}^n \sum_{s=r}^n \sum_{t=s}^n
{
{\Wnorm_{\mu}(P;1,\ldots,r)}
\over
{r!}
}
\thinspace
{
{\Wnorm_{\nu}(r{+}1,\ldots,s;Q)}
\over
{(s-r)!}
}
\cr & \qquad\quad\times
(-ie)V^{\mu\nu\lambda}\bigl[P+\kappa(1,r),\kappa(r{+}1,s)+Q,-K\bigr]
\cr & \qquad\quad\times
{
{ -i g_{\lambda\sigma} }
\over
{ K^2 }
}
{
{\positronon(p;s{+}1,\ldots,t)}
\over
{(t-s)!}
}
(-ie\gamma_\sigma)
{
{\electronon(t{+}1,\ldots,n;q)}
\over
{(n-t)!}
}
\cr & +
\permsum{1}{n} \sum_{r=1}^n \sum_{s=r}^n \sum_{t=s}^n
{
{\Wnorm_{\mu}(P;1,\ldots,r{-}1)}
\over
{(r-1)!}
}
\eps_{\rho}(r)
{
{\Wnorm_{\nu}(r{+}1,\ldots,s;Q)}
\over
{(s-r)!}
}
\cr & \qquad\quad\times
(-ie^2)S^{\mu\nu\lambda\rho}
{
{ -i g_{\lambda\sigma} }
\over
{ K^2 }
}
{
{\positronon(p;s{+}1,\ldots,t)}
\over
{(t-s)!}
}
(-ie\gamma_\sigma)
{
{\electronon(t{+}1,\ldots,n;q)}
\over
{(n-t)!}
}
}
\eqlabel\gammastart
$$
where we have used the notations
$$
V^{\mu\nu\lambda}(k_1,k_2,k_3)=g^{\mu\nu}(k_1-k_2)^{\lambda}
                              +g^{\nu\lambda}(k_2-k_3)^{\mu}
                              +g^{\lambda\mu}(k_3-k_1)^{\nu}
\eqlabel\Threepoint
$$
to designate the three-point vertex function and
$$
S^{\mu\nu\lambda\rho}=2g^{\mu\nu}g^{\lambda\rho}
                      -g^{\mu\rho}g^{\nu\lambda}
                      -g^{\mu\lambda}g^{\nu\rho}
\eqlabel\Fourpoint
$$
for the seagull vertex function.  The momentum carried by the
virtual photon is
$$
\eqalign{
K &\equiv P+\kappa(1,s)+Q
\cr &
= -[p+\kappa(s{+}1,n)+q].
}
\eqlabel\Kdef
$$
For the helicity combination of interest, namely all vector bosons
having positive helicity with the exception of a single photon,
we know that each of the currents  $\Wms_{\alpha\dot\alpha}$
are proportional to $u_{\alpha}(k_1)$.  Thus, all  products
of the form $W\negthinspace\negthinspace\cdot W'$ vanish.
Furthermore, since all of the polarization
spinors are also proportional to $u_{\alpha}(k_1)$ as well, their
dot products with any of the $W$'s is also zero.
Since the metric tensor combinations appearing in
(\Fourpoint) inevitably leading to the forms
$W\negthinspace\negthinspace\cdot W'$ or
$\eps\cdot W$, these two important
properties tell us that there are no seagull contributions to
the amplitude. In addition, only two of the three terms appearing
in (\Threepoint) actually contribute.  Thus, (\gammastart)
reduces to
$$
\eqalign{
\amp_{\gamma}&(p,q;P,Q;1,\ldots,n) =
2ie^2 \negthinspace\negthinspace
\permsum{1}{n} \sum_{r=0}^n \sum_{s=r}^n \sum_{t=s}^n
{
{1}
\over
{r!(s{-}r)!(t{-}s)!(n{-}t)!}
}
\thinspace
{ {1}\over{K^2} }
\cr & \negthinspace\times
\biggl\{
\Wnorm_{\mu}(P;1,\ldots,r)
\Wnorm_{\nu}(r{+}1,\ldots,s;Q)
-\Wnorm_{\nu}(P;1,\ldots,r)
\Wnorm_{\mu}(r{+}1,\ldots,s;Q)
\biggr\}
\cr & \negthinspace\times
\positronon(p;s{+}1,\ldots,t) \gamma^{\nu}
\electronon(t{+}1,\ldots,n;q) K^{\mu}
}
\eq
$$
after applying current conservation and rearranging a bit.
Use of the multispinor replacement rules (\msreplacedotprod)
and (\msreplacefermion) plus the Schouten identity (\Shouten)
allows us to simplify further, yielding
$$
\eqalign{
\amp_{\gamma}&(p^{+},q^{-};P,Q;1,\ldots,n) =
\cr & =
-i(-e\sqrt2)^2 \negthinspace\negthinspace
\permsum{1}{n} \sum_{s=0}^n \sum_{r=0}^s \sum_{t=s}^n
{
{1}
\over
{r!(s{-}r)!(t{-}s)!(n{-}t)!}
}
\thinspace
\positronon_{\dot\beta}(p^{+};s{+}1,\ldots,t)
\cr & \qquad\times
{ {\bar K^{\dot\beta\alpha}}\over{K^2} }
\Wms_{\alpha\dot\alpha}(r{+}1,\ldots,s;Q)
{\overline\Wms}^{\dot\alpha\beta}(P;1,\ldots,r)
\electronon_{\beta}(t{+}1,\ldots,n;q^{-}),
}
\eqlabel\gammatemplate
$$
where we have specialized to a left-handed fermion line.

In order to actually work with (\gammatemplate), it is necessary to
explicitly write out the part of the permutation sum involving the
first photon, since for the amplitude we wish to compute that
photon is distinguishable by its negative helicity.  The result
is a sequence of four terms,
corresponding to the possibility that this
photon was radiated by any one of the four charged particles.
At the same time, it is convenient
to use the permutation sum so that the remaining photons radiated
by that particle  are numbered $k_2,\ldots,k_r$.  Thus, we find
the following four contributions to
$\amp_{\gamma}(p^{+},q^{-};P^{+},Q^{+};1^{-},2^{+},\ldots,n^{+})$:
$$
\eqalign{
\amp_{\gamma1}&\equiv
-i(-e\sqrt2)^2 \negthinspace\negthinspace
\permsum{2}{n} \sum_{s=1}^n \sum_{r=1}^s \sum_{t=s}^n
{
{1}
\over
{(r{-}1)!(s{-}r)!(t{-}s)!(n{-}t)!}
}
\cr & \quad\times
\positronon_{\dot\beta}(p^{+};(s{+}1)^{+},\ldots,t^{+})
{ {\bar K^{\dot\beta\alpha}}\over{K^2} }
\Wms_{\alpha\dot\alpha}((r{+}1)^{+},\ldots,s^{+};Q^{+})
\cr & \quad\times
{\overline\Wms}^{\dot\alpha\beta}(P^{+};1^{-},2^{+},\ldots,r^{+})
\electronon_{\beta}((t{+}1)^{+},\ldots,n^{+};q^{-}),
}
\eqlabel\gammaONE
$$
$$
\eqalign{
\amp_{\gamma2}&\equiv
-i(-e\sqrt2)^2 \negthinspace\negthinspace
\permsum{2}{n} \sum_{s=1}^n \sum_{r=1}^s \sum_{t=s}^n
{
{1}
\over
{(r{-}1)!(s{-}r)!(t{-}s)!(n{-}t)!}
}
\cr & \quad\times
\positronon_{\dot\beta}(p^{+};(s{+}1)^{+},\ldots,t^{+})
{ {\bar K^{\dot\beta\alpha}}\over{K^2} }
\Wms_{\alpha\dot\alpha}(1^{-},2^{+},\ldots,r^{+};Q^{+})
\cr & \quad\times
{\overline\Wms}^{\dot\alpha\beta}(P^{+};(r{+}1)^{+},\ldots,s^{+})
\electronon_{\beta}((t{+}1)^{+},\ldots,n^{+};q^{-}),
}
\eqlabel\gammaTWO
$$
$$
\eqalign{
\amp_{\gamma3}&\equiv
i(-e\sqrt2)^2 \negthinspace\negthinspace
\permsum{2}{n} \sum_{s=1}^n \sum_{r=1}^s \sum_{t=s}^n
{
{1}
\over
{(r{-}1)!(s{-}r)!(t{-}s)!(n{-}t)!}
}
\cr & \quad\times
\positronon_{\dot\beta}(p^{+};1^{-},2^{+},\ldots,r^{+})
{ {\bar \curlyK^{\dot\beta\alpha}}\over{\curlyK^2} }
\Wms_{\alpha\dot\alpha}((t{+}1)^{+},\ldots,n^{+};Q^{+})
\cr & \quad\times
{\overline\Wms}^{\dot\alpha\beta}(P^{+};(s{+}1)^{+},\ldots,t^{+})
\electronon_{\beta}((r{+}1)^{+},\ldots,s^{+};q^{-}),
}
\eqlabel\gammaTHREE
$$
$$
\eqalign{
\amp_{\gamma4}&\equiv
i(-e\sqrt2)^2 \negthinspace\negthinspace
\permsum{2}{n} \sum_{s=1}^n \sum_{r=1}^s \sum_{t=s}^n
{
{1}
\over
{(r{-}1)!(s{-}r)!(t{-}s)!(n{-}t)!}
}
\cr & \quad\times
\positronon_{\dot\beta}(p^{+};(r{+}1)^{+},\ldots,s^{+})
{ {\bar \curlyK^{\dot\beta\alpha}}\over{\curlyK^2} }
\Wms_{\alpha\dot\alpha}((t{+}1)^{+},\ldots,n^{+};Q^{+})
\cr & \quad\times
{\overline\Wms}^{\dot\alpha\beta}(P^{+};(s{+}1)^{+},\ldots,t^{+})
\electronon_{\beta}(1^{-},2^{+},\ldots,r^{+};q^{-}).
}
\eqlabel\gammaFOUR
$$
The last two contributions contain the new notation
$$
\curlyK\equiv p+\kappa(1,s)+q,
\eqlabel\DEFcurlyK
$$
which arises because of the different momentum routing
used in this pair of terms.
The actual evaluation of each of the four terms follows essentially
the same procedure.  We will illustrate the method
using $\amp_{\gamma1}$, and then simply quote the results
for the remaining  three contributions.

When we insert (\electrononconnection), (\spinorallplussoln),
(\Wcrossing), (\WTallplussoln), and (\WToneminussoln) for the
currents appearing in (\gammaONE) we obtain
$$
\eqalign{
\amp_{\gamma1}&=
i(-e\sqrt2)^{n+2}
{
{ {\braket{1}{q}}^2 }
\over
{ \bra{P}1\ket{Q} }
}
\permsum{2}{n} \sum_{s=1}^n \sum_{r=1}^s \sum_{t=s}^n
{ {1}\over{K^2} }
\cr & \quad\times
{
{  u^{\delta}(k_1)[\kappa(r{+}1,s)+Q]_{\delta\dot\alpha}
[\bar P + \bar\kappa(1,r)]^{\dot\alpha\eps}u_{\eps}(k_1) }
\over
{ \bra{P}2,\ldots,r\ket{1} \bra{1}r{+}1,\ldots,s\ket{Q} }
}
\cr & \quad\times
{
{  u^{\gamma}(k_1)[p+\kappa(s{+}1,t)]_{\gamma\dot\beta}
\bar K^{\dot\beta\alpha}u_{\alpha}(k_1) }
\over
{ \bra{p}s{+}1,\ldots,t\ket{1} \bra{1}t{+}1,\ldots,n\ket{q} }
}
\cr & \quad\times
\Biggl\{
\linkstar{h}{1}{P}
+(1-\delta_{r1})\sum_{j=2}^r
u^{\omega}(k_1){{\pole}_{\omega}}^{\upsilon}(P,1,2,\ldots,j)
u_{\upsilon}(k_1)
\Biggr\}.
}
\eqlabel\gammaONEstart
$$
We begin by performing the sum on $t$.  In order to write
$$
{
{  1 }
\over
{ \bra{p}s{+}1,\ldots,t\ket{1} \bra{1}t{+}1,\ldots,n\ket{q} }
} =
\link{t}{1}{t{+}1}
{
{  1 }
\over
{ \bra{p}s{+}1,\ldots,n\ket{q} }
}
\eq
$$
for all $t$ in the indicated summation range, we adopt the
device that when $t=s$, we write $p$ and when $t=n+1$, we
write $q$.  These assignments mesh
nicely with the momentum sum in the numerator, which becomes
$$
[p+\kappa(s{+}1,t)]_{\gamma\dot\beta} =
\sum_{v=s}^t (k_{v})_{\gamma\dot\beta}.
\eq
$$
Thus, we write the relevant factors for the sum on $t$ as
$$
\sigma_{\gamma1t} \equiv
\sum_{t=s}^{n} \sum_{v=s}^{t}
\link{t}{1}{t{+}1}
\braket{1}{v}
\bar u_{\dot\beta}(k_v) \bar K^{\dot\beta\alpha} u_{\alpha}(k_1).
\eq
$$
Interchanging the order of the sums and applying (\linkidsummed)
to do the sum on $t$ yields
$$
\eqalign{
\sigma_{\gamma1t} & =
\sum_{v=s}^{n}
\link{v}{1}{q}
\braket{1}{v}
\bar u_{\dot\beta}(k_v) \bar K^{\dot\beta\alpha} u_{\alpha}(k_1)
\cr & =
{ {1}\over{ \braket{1}{q} } }
u^{\gamma}(q)[p+\kappa(s{+}1,n)]_{\gamma\dot\beta}
K^{\dot\beta\alpha} u_{\alpha}(k_1).
}
\eqlabel\sigt
$$
We may use the Weyl equation to extend the sum in the first
factor of (\sigt) to read $p+\kappa(s{+}1,n)+q = -K$:
$$
\eqalign{
\sigma_{\gamma1t} & =
{ {-1}\over{\braket{1}{q}} }
u^{\gamma}(q)
K_{\gamma\dot\beta} \bar K^{\dot\beta\alpha} u_{\alpha}(k_1)
\cr & = K^2,
}
\eqlabel\sigtDONE
$$
where we have applied (\slashsqr) and
(\antisym) to obtain the last line.
Returning to (\gammaONEstart) and using (\slashsqr) to
extend one of the momentum factors from $P+\kappa(1,r)$
to $K$, we have
$$
\eqalign{
\amp_{\gamma1}&=
i(-e\sqrt2)^{n+2}
{
{ {\braket{1}{q}}^2 }
\over
{ \bra{P}1\ket{Q} }
}
\permsum{2}{n} \sum_{s=1}^n \sum_{r=1}^s
{
{  u^{\delta}(k_1)[\kappa(r{+}1,s)+Q]_{\delta\dot\alpha}
\bar K^{\dot\alpha\eps}u_{\eps}(k_1) }
\over
{ \bra{P}2,\ldots,r\ket{1} \bra{1}r{+}1,\ldots,s\ket{Q} }
}
\cr & \quad\qquad\qquad\times
{
{1}
\over
{ \bra{p}s{+}1,\ldots,n \ket{q}}
}
\linkstar{h}{1}{P}
\cr & \quad +
i(-e\sqrt2)^{n+2}
{
{ {\braket{1}{q}}^2 }
\over
{ \bra{P}1\ket{Q} }
}
\permsum{2}{n} \sum_{s=2}^n \sum_{r=2}^s
{
{  u^{\delta}(k_1)[\kappa(r{+}1,s)+Q]_{\delta\dot\alpha}
\bar K^{\dot\alpha\eps}u_{\eps}(k_1) }
\over
{ \bra{P}2,\ldots,r\ket{1} \bra{1}r{+}1,\ldots,s\ket{Q} }
}
\cr & \quad\qquad\qquad\times
{
{1}
\over
{ \bra{p}s{+}1,\ldots,n \ket{q}}
}
\sum_{j=2}^r
u^{\omega}(k_1){{\pole}_{\omega}}^{\upsilon}(P,1,2,\ldots,j)
u_{\upsilon}(k_1).
}
\eqlabel\gammaONEtsummed
$$

Denote the first of the two terms in (\gammaONEtsummed) by
$\amp_{\gamma1a}$.  We may do the sum on $r$ by defining
$r=1$ to mean $P$ and $r=s+1$ to mean $Q$.  Notice that $k_1$
is thus naturally absent from any momentum sums that result. Isolating
the $r$-dependent factors of $\amp_{\gamma1a}$ yields
$$
\sigma_{\gamma1a} \equiv
\sum_{r=1}^s \sum_{w=r+1}^{s+1}
\link{r}{1}{r{+}1}
\braket{1}{w} \bar u_{\dot\alpha}(k_w)
\bar K^{\dot\alpha\eps}u_{\eps}(k_1).
\eq
$$
Interchanging the order of summation and proceeding as before we
find that
$$
\eqalign{
\sigma_{\gamma1a} &=
\sum_{w=2}^{s+1}
\link{P}{1}{w}
\braket{1}{w} \bar u_{\dot\alpha}(k_w)
\bar K^{\dot\alpha\eps}u_{\eps}(k_1)
\cr & =
{ {1}\over{\braket{P}{1}} }
u^{\delta}(P)[\kappa(2,s)+Q]_{\delta\dot\alpha}
\bar K^{\dot\alpha\eps}u_{\eps}(k_1)
\cr &
= [P+\kappa(2,s)+Q]^2.
}
\eq
$$
Hence,
$$
\eqalign{
\amp_{\gamma1a}&=
i(-e\sqrt2)^{n+2}
{
{ {\braket{1}{q}}^2 }
\over
{ \bra{P}1\ket{Q} }
}
\cr & \quad\times
\permsum{2}{n} \sum_{s=1}^n
{
{ [P+\kappa(2,s)+Q]^2 }
\over
{ \bra{P}2,\ldots,s\ket{Q}
\bra{p}s{+}1,\ldots,n \ket{q} }
}
\linkstar{h}{1}{P}.
}
\eqlabel\gammaONEA
$$

The additional summation appearing in the second term
of (\gammaONEtsummed)
makes the evaluation of the sums there somewhat more complex.
We may join the denominator strings involving $r$ using the
same ground rules outlined in the previous paragraph.  Thus, we
are led to consider
$$
\sigma_{\gamma1b} \equiv
\sum_{r=j}^s \sum_{w=r+1}^{s+1}
\link{r}{1}{r{+}1}
\braket{1}{w} \bar u_{\dot\alpha}(k_w)
\bar K^{\dot\alpha\eps}u_{\eps}(k_1).
\eq
$$
The (unwritten) sum on $j$ has been moved to the
left of the sum on $r$ and now ranges from 2 to $s$.
We begin the evaluation as before:
$$
\eqalign{
\sigma_{\gamma1b} &=
\sum_{w=j+1}^{s+1}
\link{j}{1}{w}
\braket{1}{w} \bar u_{\dot\alpha}(k_w)
\bar K^{\dot\alpha\eps}u_{\eps}(k_1)
\cr & =
{ {-1}\over{\braket{1}{j}} }
u^{\delta}(k_j)[\kappa(j{+}1,s)+Q]_{\delta\dot\alpha}
\bar K^{\dot\alpha\eps}u_{\eps}(k_1).
}
\eq
$$
At this stage, we transpose the order of multiplication, the
shifting of three contractions producing a net sign change.
In addition, we write $\kappa(j{+}1,s)+Q =
K-[P+\kappa(1,j)]$ to obtain
$$
\sigma_{\gamma1b} =
K^2 -
{ {1}\over{\braket{1}{j}} }
u^{\eps}(k_1)K_{\eps\dot\alpha}
[\bar P + \bar\kappa(1,j)]^{\dot\alpha\delta}
\bar u_{\delta}(k_j).
\eqlabel\sigmaoneb
$$
Insertion of (\sigmaoneb) into the second term of (\gammaONEtsummed)
produces
$$
\eqalign{
\amp_{\gamma1b}&=
i(-e\sqrt2)^{n+2}
{
{ {\braket{1}{q}}^2 }
\over
{ \bra{P}1\ket{Q} }
}
\permsum{2}{n} \sum_{s=2}^n
{
{ K^2 }
\over
{ \bra{P}2,\ldots,s\ket{Q} \bra{p}s{+}1,\ldots,n \ket{q}}
}
\cr & \quad\qquad\qquad\times
\sum_{j=2}^s
u^{\omega}(k_1){{\pole}_{\omega}}^{\upsilon}(P,1,2,\ldots,j)
u_{\upsilon}(k_1)
\cr & -
i(-e\sqrt2)^{n+2}
{
{ {\braket{1}{q}}^2 }
\over
{ \bra{P}1\ket{Q} }
}
\permsum{2}{n} \sum_{s=2}^n
{
{  1  }
\over
{ \bra{P}2,\ldots,s\ket{Q}\bra{p}s{+}1,\ldots,n \ket{q} }
}
\cr & \quad\qquad\qquad\times
\sum_{j=2}^s
u^{\eps}(k_1)K_{\eps\dot\alpha}
[\bar P + \bar\kappa(1,j)]^{\dot\alpha\delta}
{{\pole}_{\delta}}^{\upsilon}(P,1,2,\ldots,j)
u_{\upsilon}(k_1)
}
\eqlabel\gammaONEBstart
$$
where we have used the fact that
$u^{\omega}(k_1){{\pole}_{\omega}}^{\upsilon}(P,1,2,\ldots,j)
u_{\upsilon}(k_1)$ contains a factor of $\braket{1}{j}$ to
obtain the second term.

It is possible to do the sum on $j$ appearing in the second term
($\equiv\amp_{\gamma1b2}$)
of (\gammaONEBstart) with the help of (\splitid).  Thus, we write
$$
\eqalign{
\amp_{\gamma1b2}&=
-i(-e\sqrt2)^{n+2}
{
{ {\braket{1}{q}}^2 }
\over
{ \bra{P}1\ket{Q} }
}
\permsum{2}{n} \sum_{s=1}^n
{
{  u^{\eps}(k_1)K_{\eps\dot\alpha}  }
\over
{ \bra{P}2,\ldots,s\ket{Q}\bra{p}s{+}1,\ldots,n \ket{q} }
}
\cr & \quad\qquad\qquad\times
\Biggl[
{
{ [\bar P+\bar k_1]^{\dot\alpha\upsilon} u_{\upsilon}(k_1) }
\over
{ [P+k_1]^2 }
} - {
{ [\bar P+\bar\kappa(1,s)]^{\dot\alpha\upsilon} u_{\upsilon}(k_1) }
\over
{ [P+\kappa(1,s)]^2 }
}
\Biggr],
}
\eqlabel\gammaONEBtwostart
$$
where we have used the fact that the quantity in square brackets
vanishes for $s=1$ to extend the range of the summation.  We
may use the Weyl equation and (\slashsqr) to rewrite
(\gammaONEBtwostart) as
$$
\eqalign{
\amp_{\gamma1b2}&=
-i(-e\sqrt2)^{n+2}
{
{ {\braket{1}{q}}^2 }
\over
{ \bra{P}1\ket{Q} }
}
\cr & \qquad\qquad\times
\permsum{2}{n} \sum_{s=1}^n
{
{  u^{\eps}(k_1)K_{\eps\dot\alpha}  \bar u^{\dot\alpha}(P) }
\over
{ \bra{P}2,\ldots,s\ket{Q}\bra{p}s{+}1,\ldots,n \ket{q} }
}
{ {1}\over {{\braket{P}{1}}^{*}} }
\cr & \quad
+i(-e\sqrt2)^{n+2}
{
{ {\braket{1}{q}}^2 }
\over
{ \bra{P}1\ket{Q} }
}
\permsum{2}{n} \sum_{s=1}^n
{
{  1  }
\over
{ \bra{P}2,\ldots,s\ket{Q}\bra{p}s{+}1,\ldots,n \ket{q} }
}
\cr & \qquad\qquad\times
{
{ u^{\eps}(k_1)Q_{\eps\dot\alpha}
[\bar P+\bar\kappa(1,s)]^{\dot\alpha\upsilon} u_{\upsilon}(k_1) }
\over
{ [P+\kappa(1,s)]^2 }
}.
}
\eqlabel\gammaONEBtwo
$$
Notice that the last factor in (\gammaONEBtwo) is just
$K^2 u^{\eps}(k_1){{\pole}_{\eps}}^{\upsilon}(P,1,2,\ldots,s,Q)
u_{\upsilon}(k_1)$.  Hence, we may combine (\gammaONEBtwo)
with (\gammaONEBstart) to form
$$
\eqalign{
\amp_{\gamma1b}&=
i(-e\sqrt2)^{n+2}
{
{ {\braket{1}{q}}^2 }
\over
{ \bra{P}1\ket{Q} }
}
\permsum{2}{n} \sum_{s=1}^n
{
{ [P+\kappa(1,s)+Q]^2 }
\over
{ \bra{P}2,\ldots,s\ket{Q} \bra{p}s{+}1,\ldots,n \ket{q}}
}
\cr & \quad\qquad\qquad\times
\sum_{j=2}^{s+1}
u^{\omega}(k_1){{\pole}_{\omega}}^{\upsilon}(P,1,2,\ldots,j)
u_{\upsilon}(k_1)
{\biggl\vert}_{j=s+1\equiv Q}
\cr & -
i(-e\sqrt2)^{n+2}
{
{ {\braket{1}{q}}^2 }
\over
{ \bra{P}1\ket{Q} }
}
\permsum{2}{n} \sum_{s=1}^n
{
{ \bar u_{\dot\alpha}(P)
[\bar P+\bar\kappa(1,s)+\bar Q]^{\dot\alpha\eps}
u_{\eps}(k_1)  }
\over
{ \bra{P}2,\ldots,s\ket{Q}\bra{p}s{+}1,\ldots,n \ket{q} }
}
{ {1}\over {{\braket{P}{1}}^{*}} },
}
\eqlabel\gammaONEB
$$
where the notation $j=s+1\equiv Q$ is used to remind us to write $Q$
for $s+1$ in the sum on $j$.

Applying the same procedure to $\amp_{\gamma2}$ yields the contributions
$$
\eqalign{
\amp_{\gamma2a}&=
-i(-e\sqrt2)^{n+2}
{
{ {\braket{1}{q}}^2 }
\over
{ \bra{P}1\ket{Q} }
}
\cr & \quad\times
\permsum{2}{n} \sum_{s=1}^n
{
{ [P+\kappa(2,s)+Q]^2 }
\over
{ \bra{P}2,\ldots,s\ket{Q}
\bra{p}s{+}1,\ldots,n \ket{q} }
}
\linkstar{h}{1}{Q}
}
\eqlabel\gammaTWOA
$$
and
$$
\eqalign{
\amp_{\gamma2b}&=
-i(-e\sqrt2)^{n+2}
{
{ {\braket{1}{q}}^2 }
\over
{ \bra{P}1\ket{Q} }
}
\permsum{2}{n} \sum_{s=1}^n
{
{ [P+\kappa(1,s)+Q]^2 }
\over
{ \bra{P}s,s{-}1,\ldots,2\ket{Q} \bra{p}s{+}1,\ldots,n \ket{q}}
}
\cr & \quad\qquad\qquad\times
\sum_{j=2}^{s+1}
u^{\omega}(k_1){{\pole}_{\omega}}^{\upsilon}(Q,1,2,\ldots,j)
u_{\upsilon}(k_1)
{\biggl\vert}_{j=s+1\equiv P}
\cr & +
i(-e\sqrt2)^{n+2}
{
{ {\braket{1}{q}}^2 }
\over
{ \bra{P}1\ket{Q} }
}
\permsum{2}{n} \sum_{s=1}^n
{
{ \bar u_{\dot\alpha}(Q)
[\bar P+\bar\kappa(1,s)+\bar Q]^{\dot\alpha\eps}
u_{\eps}(k_1)  }
\over
{ \bra{P}2,\ldots,s\ket{Q}\bra{p}s{+}1,\ldots,n \ket{q} }
}
{ {1}\over {{\braket{Q}{1}}^{*}} }.
}
\eqlabel\gammaTWOB
$$
The structure $\bra{P}s,s{-}1,\ldots,2\ket{Q}$ appearing in
(\gammaTWOB) is a consequence of (\Wcrossing) and the desire
to write the current in a form containing $\pole(Q,1,2,\ldots,j)$.
Other forms of this term
are possible, but this is the most convenient.

The remaining contributions, $\amp_{\gamma3}$ and
$\amp_{\gamma4}$, turn out to be so nearly identical to
$\amp_{\gamma1}$ and $\amp_{\gamma2}$ respectively that
it would be redundant to write them down.  These terms
may be obtained by exchanging $p\leftrightarrow P$ and
$q\leftrightarrow Q$ in (\gammaONEA), (\gammaONEB),
(\gammaTWOA), and (\gammaTWOB) {\it inside the permutation sums only.}
The factor ${\braket{1}{q}}^2  { \bra{P}1\ket{Q} }^{-1}$
appearing outside the permutation sums remains unchanged.  This
close relation, which becomes apparent after the explicit expressions
for the currents are inserted into (\gammaTHREE) and (\gammaFOUR),
is a reflection  of the SUSY-like relationships between the
currents.

At this stage, we may combine some of the
fragments that do not depend on $\pole$.
Application of (\linkidnosum) to  the sum of (\gammaONEA) and
(\gammaTWOA) gives
$$
\eqalign{
\amp_{\gamma12a}&\equiv
-i(-e\sqrt2)^{n+2}
{
{ {\braket{1}{q}}^2 }
\over
{ \bra{P}1\ket{Q} }
}
\cr & \quad\times
\permsum{2}{n} \sum_{s=1}^n
{
{ [P+\kappa(2,s)+Q]^2 }
\over
{ \bra{P}2,\ldots,s\ket{Q}
\bra{p}s{+}1,\ldots,n \ket{q} }
}
\linkstar{P}{1}{Q}.
}
\eqlabel\gammaONETWOA
$$
Likewise, the second term of (\gammaONEB) may be added to
the second term of (\gammaTWOB) to produce
$$
\eqalign{
\amp_{\gamma12b2}&\equiv
-i(-e\sqrt2)^{n+2}
{
{ {\braket{1}{q}}^2 }
\over
{ \bra{P}1\ket{Q} }
}
\permsum{2}{n} \sum_{s=1}^n
\Biggl[
{
{\bar u_{\dot\alpha}(P)}
\over
{ {\braket{P}{1}}^{*} }
} - {
{\bar u_{\dot\alpha}(Q)}
\over
{ {\braket{Q}{1}}^{*} }
}
\Biggr]
\cr & \qquad\times
{
{ [\bar P+\bar\kappa(1,s)+\bar Q]^{\dot\alpha\eps}u_{\eps}(k_1) }
\over
{ \bra{P}2,\ldots,s\ket{Q}
\bra{p}s{+}1,\ldots,n \ket{q} }
}
\cr & =
-i(-e\sqrt2)^{n+2}
{
{ {\braket{1}{q}}^2 }
\over
{ \bra{P}1\ket{Q} }
}
\permsum{2}{n} \sum_{s=1}^n
\linkstar{P}{1}{Q}
\cr & \qquad\times
{
{ \bar u_{\dot\alpha}(k_1)
[\bar P+\bar\kappa(1,s)+\bar Q]^{\dot\alpha\eps}u_{\eps}(k_1) }
\over
{ \bra{P}2,\ldots,s\ket{Q}
\bra{p}s{+}1,\ldots,n \ket{q} }
}.
}
\eqlabel\gammaONETWOBtwo
$$
Since
$$
\bar u_{\dot\alpha}(k_1)
[\bar P+\bar\kappa(1,s)+\bar Q]^{\dot\alpha\eps}u_{\eps}(k_1)
= 2k_1\cdot[P+\kappa(1,s)+Q],
\eq
$$
we see that (\gammaONETWOBtwo) is precisely what must be added
to  (\gammaONETWOA)
to extend  its numerator from $[P+\kappa(2,s)+Q]^2$ to
$[P+\kappa(1,s)+Q]^2$.
Thus,
$$
\eqalign{
\amp_{\gamma12a}+\amp_{\gamma12b2} &=
-i(-e\sqrt2)^{n+2}
{
{ {\braket{1}{q}}^2 }
\over
{ \bra{P}1\ket{Q} }
}
\cr & \quad\times
\permsum{2}{n} \sum_{s=1}^n
{
{ [P+\kappa(1,s)+Q]^2 }
\over
{ \bra{P}2,\ldots,s\ket{Q}
\bra{p}s{+}1,\ldots,n \ket{q} }
}
\linkstar{P}{1}{Q}.
}
\eqlabel\gammafrags
$$
We now present the final result for the photon exchange graphs.  There
are a total of six terms:  (\gammafrags), the first term of
(\gammaONEB), the first term of (\gammaTWOB), and the counterparts
to these three terms generated
from $\amp_{\gamma3}$ and $\amp_{\gamma4}$.  Hence,
$$
\eqalign{
\amp_{\gamma}&(p^{+},q^{-};P^{+},Q^{+};1^{-},2^{+},\ldots,n^{+})=
\cr & =
i(-e\sqrt2)^{n+2}
{
{ {\braket{1}{q}}^2 }
\over
{ \bra{P}1\ket{Q} }
}
\permsum{2}{n} \sum_{s=1}^n
{
{ [P+\kappa(1,s)+Q]^2 }
\over
{ \bra{P}2,\ldots,s\ket{Q} \bra{p}s{+}1,\ldots,n \ket{q}}
}
\cr & \quad\qquad\qquad\times
\sum_{j=2}^{s+1}
u^{\omega}(k_1){{\pole}_{\omega}}^{\upsilon}(P,1,2,\ldots,j)
u_{\upsilon}(k_1)
{\biggl\vert}_{j=s+1\equiv Q}
\cr & \quad
-i(-e\sqrt2)^{n+2}
{
{ {\braket{1}{q}}^2 }
\over
{ \bra{P}1\ket{Q} }
}
\permsum{2}{n} \sum_{s=1}^n
{
{ [P+\kappa(1,s)+Q]^2 }
\over
{ \bra{P}s,s{-}1,\ldots,2\ket{Q} \bra{p}s{+}1,\ldots,n \ket{q}}
}
\cr & \quad\qquad\qquad\times
\sum_{j=2}^{s+1}
u^{\omega}(k_1){{\pole}_{\omega}}^{\upsilon}(Q,1,2,\ldots,j)
u_{\upsilon}(k_1)
{\biggl\vert}_{j=s+1\equiv P}
\cr & \quad
+i(-e\sqrt2)^{n+2}
{
{ {\braket{1}{q}}^2 }
\over
{ \bra{P}1\ket{Q} }
}
\permsum{2}{n} \sum_{s=1}^n
{
{ [p+\kappa(1,s)+q]^2 }
\over
{ \bra{p}2,\ldots,s\ket{q} \bra{P}s{+}1,\ldots,n \ket{Q}}
}
\cr & \quad\qquad\qquad\times
\sum_{j=2}^{s+1}
u^{\omega}(k_1){{\pole}_{\omega}}^{\upsilon}(p,1,2,\ldots,j)
u_{\upsilon}(k_1)
{\biggl\vert}_{j=s+1\equiv q}
\cr & \quad
-i(-e\sqrt2)^{n+2}
{
{ {\braket{1}{q}}^2 }
\over
{ \bra{P}1\ket{Q} }
}
\permsum{2}{n} \sum_{s=1}^n
{
{ [p+\kappa(1,s)+q]^2 }
\over
{ \bra{p}s,s{-}1,\ldots,2\ket{q} \bra{P}s{+}1,\ldots,n \ket{Q}}
}
\cr & \quad\qquad\qquad\times
\sum_{j=2}^{s+1}
u^{\omega}(k_1){{\pole}_{\omega}}^{\upsilon}(q,1,2,\ldots,j)
u_{\upsilon}(k_1)
{\biggl\vert}_{j=s+1\equiv p}
\cr & \quad
-i(-e\sqrt2)^{n+2}
{
{ {\braket{1}{q}}^2 }
\over
{ \bra{P}1\ket{Q} }
}
\cr & \quad\qquad\qquad\times
\permsum{2}{n} \sum_{s=1}^n
{
{ [P+\kappa(1,s)+Q]^2 }
\over
{ \bra{P}2,\ldots,s\ket{Q}
\bra{p}s{+}1,\ldots,n \ket{q} }
}
\linkstar{P}{1}{Q}
\cr & \quad
-i(-e\sqrt2)^{n+2}
{
{ {\braket{1}{q}}^2 }
\over
{ \bra{P}1\ket{Q} }
}
\cr & \quad\qquad\qquad\times
\permsum{2}{n} \sum_{s=1}^n
{
{ [p+\kappa(1,s)+q]^2 }
\over
{ \bra{p}2,\ldots,s\ket{q}
\bra{P}s{+}1,\ldots,n \ket{Q} }
}
\linkstar{p}{1}{q}.
}
\eqlabel\Mgammadone
$$
\vfill\eject


\subsection{The $Z$  contribution}

Given that, in the massless limit, the $Z$-boson is almost a photon
with ``odd'' couplings, it is not hard to see how the inclusion
of $Z$-exchange modifies (\Mgammadone).  Emission of a photon from
the $W$ line occurs at a vertex
$ig \sin \theta_W V^{\mu\nu\lambda}$,
while $Z$-emission involves $ig \cos \theta_W V^{\mu\nu\lambda}$.
The same ratio of coupling constants holds between the corresponding
pair of seagull vertices.  The fermion-photon vertex is
$$
-ie\gamma^{\mu} = -ig  \gamma^{\mu}
\biggl[ {1\over2}(1-\gamma_5)\sin \theta_W
+ {1\over2}(1+\gamma_5)\sin \theta_W\biggr],
\eq
$$
while the fermion-Z vertex is
$$
{ {ig}\over{4 \cos\theta_W} }
\gamma^{\mu}(-1+4 \sin^2\theta_W + \gamma_5)
=-ig  \gamma^{\mu}
\biggl[ {1\over2}(1-\gamma_5) { {\cos2\theta_W}\over{2\cos\theta_W} }
+  {1\over2}(1+\gamma_5)
{ {-\sin^2\theta_W}\over{\cos\theta_W} } \biggr].
\eq
$$
Thus, we see that if we replace the exchanged photon by an exchanged
$Z$, we simply replace
$$
(-e\sqrt2)^{n+2}= 2 g^2 \sin^2\theta_W (-e\sqrt2)^{n}\longrightarrow
-2 g^2 \sin^2\theta_W (-e\sqrt2)^{n}
\eqlabel\Zreplacementright
$$
if the electron is right-handed and
$$
 2 g^2 \sin^2\theta_W (-e\sqrt2)^{n}\longrightarrow
 g^2 \cos2\theta_W (-e\sqrt2)^{n}
\eqlabel\Zreplacementleft
$$
if the electron is left-handed.  The consequence of
(\Zreplacementright) is the exact cancellation of the tree-level
photon-exchange diagrams with the $Z$-exchange diagrams
for $e_{\up} \bar e_{\down} \rightarrow
W^{+}W^{-}\gamma\cdots\gamma$
in the high-energy limit.  Since the right-handed neutrinos
which would be required for the third type of diagram
do not exist within the Standard Model, this means that
$$
\amp(p^{-},q^{+};P,Q;1,\ldots,n) = 0
\eqlabel\neatresult
$$
to tree level in the high-energy limit independent of the helicity
combination of the (transverse) vector bosons.
This is a reflection of a number of
properties of the high-energy limit of the  Standard Model.
First, the $W$-boson has only left-handed couplings to fermions.
Thus, a right-handed fermion cannot radiate the $W's$ directly.
Second, the photon and $Z$ exchanges between
a $W$ line and a right-handed
fermion line conspire to cancel.  Finally, helicity conservation for
what are effectively massless fermions means that the emission of any
number of photons in any helicity configuration cannot
change a right-handed fermion line into a left-handed one.
Hence, we have (\neatresult).

For the left-handed fermions we have been considering in detail,
the net effect when the photon and $Z$ contributions are summed
is to make the replacement
$$
(-e\sqrt2)^{n+2} \longrightarrow g^2(-e\sqrt2)^n
\eqlabel\Mzeedone
$$
in (\Mgammadone).

\subsection{The neutrino  contribution}

For left-handed fermions, there is also a contribution in which
the $W$'s are radiated directly by the fermions, with a neutrino
propagating in the gap.
{}From Figure \quadfig, we see that this contribution is
$$
\eqalign{
\amp_{\nu}&(p^{+},q^{-};P,Q;1,\ldots,n)=
\cr & =
\permsum{1}{n} \sum_{r=0}^n \sum_{s=r}^n \sum_{t=s}^n
{
{\positronon(p^{+};s{+}1,\ldots,t) }
\over
{(t-s)!}
}
{{ig}\over{\sqrt2}}
{
{\thinspace{\not{\negthinspace\negthinspace W}}(t{+}1,\ldots,n;Q)}
\over
{(n-t)!}
}
{1\over2}(1-\gamma_5)
\cr & \qquad\times
{
{ i[\slash{P}+\slash\kappa(1,s)+\slash{q}] }
\over
{ [P+\kappa(1,s)+q]^2 }
}
{{ig}\over{\sqrt2}}
{
{\thinspace{\not{\negthinspace\negthinspace W}}(P;1,\ldots,r)}
\over
{r!}
}
{1\over2}(1-\gamma_5)
{
{\electronon(r{+}1,\ldots,s;q^{-}) }
\over
{(s-r)!}
}
}
\eq
$$
Applying (\msreplacefermion) to obtain the multispinor form produces
$$
\eqalign{
\amp_{\nu}&(p^{+},q^{-};P,Q;1,\ldots,n)=
\cr & = -ig^2 \negthinspace\negthinspace
\permsum{1}{n} \sum_{r=0}^n \sum_{s=r}^n \sum_{t=s}^n
{ {1}\over{r!(s{-}r)!(t{-}s)!(n{-}t)!} }
{\positronon_{\dot\alpha}(p^{+};s{+}1,\ldots,t) }
\cr & \qquad\times
\Wbar^{\dot\alpha\alpha}(t{+}1,\ldots,n;Q)
{
{ [P+\kappa(1,s)+q]_{\alpha\dot\beta} }
\over
{ [P+\kappa(1,s)+q]^2 }
}
\Wbar^{\dot\beta\beta}(P;1,\ldots,r)
\cr & \qquad\times
{\electronon_{\beta}(r{+}1,\ldots,s;q^{-}) }
}
\eqlabel\neutrinostart
$$
Evaluation of (\neutrinostart) for the case of all like helicities
except for the first photon is exactly analogous to the evaluation
of the photon-exchange case already discussed.   Hence, we will only
present the result:
$$
\eqalign{
\amp_{\nu}&(p^{+},q^{-};P^{+},Q^{+};1^{-},2^{+},\ldots,n^{+})=
\cr & =
ig^2(-e\sqrt2)^{n}
{
{ {\braket{1}{q}}^2 }
\over
{ \bra{P}1\ket{Q} }
}
\permsum{2}{n} \sum_{s=1}^n
{
{ [P+\kappa(1,s)+q]^2 }
\over
{ \bra{P}2,\ldots,s\ket{q} \bra{p}s{+}1,\ldots,n \ket{Q}}
}
\cr & \quad\qquad\qquad\times
\sum_{j=2}^{s+1}
u^{\omega}(k_1){{\pole}_{\omega}}^{\upsilon}(P,1,2,\ldots,j)
u_{\upsilon}(k_1)
{\biggl\vert}_{j=s+1\equiv q}
\cr & \quad
-ig^2(-e\sqrt2)^{n}
{
{ {\braket{1}{q}}^2 }
\over
{ \bra{P}1\ket{Q} }
}
\permsum{2}{n} \sum_{s=1}^n
{
{ [P+\kappa(1,s)+q]^2 }
\over
{ \bra{P}s,s{-}1,\ldots,2\ket{q} \bra{p}s{+}1,\ldots,n \ket{Q}}
}
\cr & \quad\qquad\qquad\times
\sum_{j=2}^{s+1}
u^{\omega}(k_1){{\pole}_{\omega}}^{\upsilon}(q,1,2,\ldots,j)
u_{\upsilon}(k_1)
{\biggl\vert}_{j=s+1\equiv P}
\cr & \quad
+ig^2(-e\sqrt2)^{n}
{
{ {\braket{1}{q}}^2 }
\over
{ \bra{P}1\ket{Q} }
}
\permsum{2}{n} \sum_{s=1}^n
{
{ [p+\kappa(1,s)+Q]^2 }
\over
{ \bra{p}2,\ldots,s\ket{Q} \bra{P}s{+}1,\ldots,n \ket{q}}
}
\cr & \quad\qquad\qquad\times
\sum_{j=2}^{s+1}
u^{\omega}(k_1){{\pole}_{\omega}}^{\upsilon}(p,1,2,\ldots,j)
u_{\upsilon}(k_1)
{\biggl\vert}_{j=s+1\equiv Q}
\cr & \quad
-ig^2(-e\sqrt2)^{n}
{
{ {\braket{1}{q}}^2 }
\over
{ \bra{P}1\ket{Q} }
}
\permsum{2}{n} \sum_{s=1}^n
{
{ [p+\kappa(1,s)+Q]^2 }
\over
{ \bra{p}s,s{-}1,\ldots,2\ket{Q} \bra{P}s{+}1,\ldots,n \ket{q}}
}
\cr & \quad\qquad\qquad\times
\sum_{j=2}^{s+1}
u^{\omega}(k_1){{\pole}_{\omega}}^{\upsilon}(Q,1,2,\ldots,j)
u_{\upsilon}(k_1)
{\biggl\vert}_{j=s+1\equiv p}
\cr & \quad
-ig^2(-e\sqrt2)^{n}
{
{ {\braket{1}{q}}^2 }
\over
{ \bra{P}1\ket{Q} }
} \negthinspace
\permsum{2}{n} \sum_{s=1}^n
{
{ [P+\kappa(1,s)+q]^2 }
\over
{ \bra{P}2,\ldots,s\ket{q}
\bra{p}s{+}1,\ldots,n \ket{Q} }
}
\linkstar{P}{1}{q}
\cr & \quad
-ig^2(-e\sqrt2)^{n}
{
{ {\braket{1}{q}}^2 }
\over
{ \bra{P}1\ket{Q} }
} \negthinspace
\permsum{2}{n} \sum_{s=1}^n
{
{ [p+\kappa(1,s)+Q]^2 }
\over
{ \bra{p}2,\ldots,s\ket{Q}
\bra{P}s{+}1,\ldots,n \ket{q} }
}
\linkstar{p}{1}{Q}.
}
\eqlabel\Mneutrinodone
$$


\subsection{Cross-channel identities}

The contributions represented by (\Mgammadone) and (\Mneutrinodone)
look quite similar to each other.  Since they contain the
same combination of coupling
constants (recall (\Mzeedone)!), one is led to investigate whether
or not they may be re-written in such a way so as to be
combined into a simpler form.  This is indeed the case, as
is shown in Appendix \Xid.  The results relevant to the present
discussion are (see (\XidsimpleA), (\XidsimpleB), (\XidmessyA),
and (\XidmessyB)):
$$
\eqalign{
\permsum{2}{n} \sum_{s=1}^n  &
{
{ [P+\kappa(1,s)+q]^2 }
\over
{ \bra{P}2,\ldots,s\ket{q}
\bra{p}s{+}1,\ldots,n \ket{Q} }
}
\cr & =
-\permsum{2}{n} \sum_{s=1}^n
{
{ [P+\kappa(1,s)+Q]^2 }
\over
{ \bra{P}2,\ldots,s \ket{Q}
\bra{p}s{+}1,\ldots,n\ket{q} }
}
\cr & \quad +
\permsum{2}{n} \sum_{s=1}^n
{
{ \braket{P}{1} }
\over
{ \braket{P}{q} }
}
{
{ \bar u_{\dot\alpha}(k_1)
[\bar P + \bar\kappa(1,s)+\bar Q]^{\dot\alpha\beta}u_{\beta}(q) }
\over
{ \bra{P}2,\ldots,s \ket{Q}
\bra{p}s{+}1,\ldots,n\ket{q} }
}
}
\eqlabel\XingONE
$$
and
$$
\eqalign{
\permsum{2}{n}& \sum_{s=1}^n
{
{ [P+\kappa(1,s)+q]^2 }
\over
{ \bra{P}2,\ldots,s\ket{q} \bra{p}s{+}1,\ldots,n \ket{Q}}
}
\cr & \quad\qquad\qquad\times
\sum_{j=2}^{s+1}
u^{\gamma}(k_1){{\pole}_{\gamma}}^{\delta}(P,1,2,\ldots,j)
u_{\delta}(k_1)
{\biggl\vert}_{j=s+1\equiv q}=
\cr & =
-\negthinspace\negthinspace
\permsum{2}{n} \sum_{s=1}^n
{
{ [P + \kappa(1,s) + Q ]^2  }
\over
{ \bra{P}2,\ldots,s \ket{Q} \bra{p}s{+}1,\ldots,n\ket{q}}
}
\cr & \quad\qquad\qquad\times
\sum_{j=2}^{s+1}
u^{\alpha}(k_1)
{{\pole}_{\alpha}}^{\delta}(P,1,2,\ldots,j)
u_{\delta}(k_1)
{\biggl\vert}_{j=s+1\equiv Q}
\cr & \quad
+ \negthinspace\negthinspace
\permsum{2}{n} \sum_{s=1}^n
\invlink{q}{1}{Q}
\Biggl[
{
{ 1  }
\over
{ \bra{P}2,\ldots,s \ket{q} \bra{p}s{+}1,\ldots,n\ket{Q} }
}
\cr & \qquad\qquad\qquad\qquad\qquad
-
{
{ 1  }
\over
{ \bra{P}2,\ldots,s \ket{Q} \bra{p}s{+}1,\ldots,n\ket{q} }
}
\Biggr]
\cr & \quad
+\negthinspace\negthinspace
\permsum{2}{n} \sum_{s=1}^n
{
{ \braket{P}{1} }
\over
{ \braket{P}{q} }
}
{
{ \bar u_{\dot\alpha}(P)
[\bar P + \bar\kappa(1,s) +\bar Q]^{\dot\alpha\beta}
u_{\beta}(q)  }
\over
{ \bra{P}2,\ldots,s \ket{Q} \bra{p}s{+}1,\ldots,n\ket{q}}
}
{
{1}
\over
{ { \braket{P}{1} }^{*} }
}.
}
\eqlabel\XingTWO
$$
In addition to (\XingONE) and (\XingTWO), we require variants
of these identities
that are  easily obtained by permuting among $P$, $Q$, $p$, and
$q$.  We apply the appropriate form of (\XingONE) or (\XingTWO)
to each of the six terms appearing in (\Mneutrinodone), and
combine them with (\Mgammadone).  Every one of the double-sum
terms in (\Mneutrinodone) ({\it i.e.}{ }those containing $\pole$)
produce a contribution that cancels one of the double-sum
terms in (\Mgammadone).  In addition, each of the single-sum
terms in (\Mneutrinodone) produces a piece that combines
logically with a corresponding piece in (\Mgammadone)
via (\linkidnosum).  The result of (carefully) doing all of this
algebra may be organized into the following ten contributions:
$$
\eqalign{
&\amp_{\rm A}=
{\cal C} \negthinspace\negthinspace
\permsum{2}{n} \sum_{s=1}^n
\invlink{q}{1}{Q}
\cr & \quad\times
\Biggl[
{
{1}
\over
{ \bra{P}2,\ldots,s\ket{q} \bra{p}s{+}1,\ldots,n\ket{Q} }
}
-
{
{1}
\over
{ \bra{P}2,\ldots,s\ket{Q} \bra{p}s{+}1,\ldots,n\ket{q} }
}
\cr & \qquad
-
{
{1}
\over
{ \bra{P}s{+}1,\ldots,n\ket{q} \bra{p}2,\ldots,s\ket{Q} }
}
+
{
{1}
\over
{ \bra{P}s{+}1,\ldots,n\ket{Q} \bra{p}2,\ldots,s\ket{q} }
}
\Biggr],
}
\eqlabel\EINS
$$
$$
\eqalign{
&\amp_{\rm B}=
{\cal C} \negthinspace\negthinspace
\permsum{2}{n} \sum_{s=1}^n
\invlink{P}{1}{p}
\cr & \quad\times
\Biggl[
{
{1}
\over
{ \bra{P}2,\ldots,s\ket{q} \bra{p}s{+}1,\ldots,n\ket{Q} }
}
-
{
{1}
\over
{ \bra{p}2,\ldots,s\ket{q} \bra{P}s{+}1,\ldots,n\ket{Q} }
}
\cr & \qquad
-
{
{1}
\over
{ \bra{P}s{+}1,\ldots,n\ket{q} \bra{p}2,\ldots,s\ket{Q} }
}
+
{
{1}
\over
{ \bra{p}s{+}1,\ldots,n\ket{q} \bra{P}2,\ldots,s\ket{Q} }
}
\Biggr],
}
\eqlabel\ZWEI
$$
$$
\eqalign{
&\amp_{\rm C}=
{\cal C} \negthinspace\negthinspace
\permsum{2}{n} \sum_{s=1}^n
{
{ \braket{P}{1} }
\over
{ \braket{P}{q} }
}
{
{ \bar u_{\dot\alpha}(P) \bar K^{\dot\alpha\beta}
u_{\beta}(q)  }
\over
{ \bra{P}2,\ldots,s \ket{Q} \bra{p}s{+}1,\ldots,n\ket{q}}
}
{
{1}
\over
{ { \braket{P}{1} }^{*} }
},
}
\eqlabel\DREI
$$
$$
\eqalign{
&\amp_{\rm D}=
{\cal C} \negthinspace\negthinspace
\permsum{2}{n} \sum_{s=1}^n
{
{ \braket{p}{1} }
\over
{ \braket{p}{Q} }
}
{
{ \bar u_{\dot\alpha}(p) \bar \curlyK^{\dot\alpha\beta}
u_{\beta}(Q)  }
\over
{ \bra{p}2,\ldots,s \ket{q} \bra{P}s{+}1,\ldots,n\ket{Q}}
}
{
{1}
\over
{ { \braket{p}{1} }^{*} }
},
}
\eqlabel\VIER
$$
$$
\eqalign{
&\amp_{\rm E}=
-{\cal C} \negthinspace\negthinspace
\permsum{2}{n} \sum_{s=1}^n
{
{ \braket{Q}{1} }
\over
{ \braket{p}{Q} }
}
{
{ \bar u_{\dot\alpha}(Q) \bar K^{\dot\alpha\beta}
u_{\beta}(p)  }
\over
{ \bra{P}2,\ldots,s \ket{Q} \bra{p}s{+}1,\ldots,n\ket{q}}
}
{
{1}
\over
{ { \braket{1}{Q} }^{*} }
},
}
\eqlabel\FUNF
$$
$$
\eqalign{
&\amp_{\rm F}=
-{\cal C} \negthinspace\negthinspace
\permsum{2}{n} \sum_{s=1}^n
{
{ \braket{q}{1} }
\over
{ \braket{P}{q} }
}
{
{ \bar u_{\dot\alpha}(q) \bar \curlyK^{\dot\alpha\beta}
u_{\beta}(P)  }
\over
{ \bra{p}2,\ldots,s \ket{q} \bra{P}s{+}1,\ldots,n\ket{Q}}
}
{
{1}
\over
{ { \braket{1}{q} }^{*} }
},
}
\eqlabel\SECHS
$$
$$
\eqalign{
&\amp_{\rm G}=
{\cal C} \negthinspace\negthinspace
\permsum{2}{n} \sum_{s=1}^n
{
{K^2}
\over
{\bra{P}2,\ldots,s \ket{Q} \bra{p}s{+}1,\ldots,n\ket{q}}
}
\linkstar{Q}{1}{q},
}
\eqlabel\SIEBEN
$$
$$
\eqalign{
&\amp_{\rm H}=
{\cal C} \negthinspace\negthinspace
\permsum{2}{n} \sum_{s=1}^n
{
{\curlyK^2}
\over
{\bra{p}2,\ldots,s \ket{q} \bra{P}s{+}1,\ldots,n\ket{Q}}
}
\linkstar{q}{1}{Q},
}
\eqlabel\ACHT
$$
$$
\eqalign{
&\amp_{\rm I}=
-{\cal C} \negthinspace\negthinspace
\permsum{2}{n} \sum_{s=1}^n
{
{ \braket{P}{1} }
\over
{ \braket{P}{q} }
}
{
{ \bar u_{\dot\alpha}(k_1) \bar K^{\dot\alpha\beta}
u_{\beta}(q)  }
\over
{ \bra{P}2,\ldots,s \ket{Q} \bra{p}s{+}1,\ldots,n\ket{q}}
}
\linkstar{P}{1}{q},
}
\eqlabel\NEUN
$$
$$
\eqalign{
&\amp_{\rm J}=
-{\cal C} \negthinspace\negthinspace
\permsum{2}{n} \sum_{s=1}^n
{
{ \braket{p}{1} }
\over
{ \braket{p}{Q} }
}
{
{ \bar u_{\dot\alpha}(k_1) \bar\curlyK^{\dot\alpha\beta}
u_{\beta}(Q)  }
\over
{ \bra{p}2,\ldots,s \ket{q} \bra{P}s{+}1,\ldots,n\ket{Q}}
}
\linkstar{p}{1}{Q},
}
\eqlabel\ZEHN
$$
where we use the abbreviation
$$
{\cal C} \equiv ig^2(-e\sqrt2)^{n}
{
{ {\braket{1}{q}}^2 }
\over
{ \bra{P}1\ket{Q} }
}.
\eqlabel\CDEFN
$$
Recall that $K$ and $\curlyK$ are as given in (\Kdef) and
(\DEFcurlyK) respectively.
At first glance, it seems like the trade of
(\Mgammadone) and (\Mneutrinodone)
for (\EINS)--(\ZEHN) has produced more terms than it has
cancelled.  We shall now demonstrate how to combine this
bewildering array of terms into a simple result.

The first observation
to make is that $\amp_{\rm A}$ and $\amp_{\rm B}$ both vanish.
This is readily seen by noting that the momenta of the second
two terms in the square brackets may be relabelled  using
$$
\eqalign{
\{s{+}1,\ldots,n\} &\longrightarrow \{2,\ldots,n{-}s{+}1\}
\cr
\{2,\ldots,s\} & \longrightarrow \{n{-}s{+}2,\ldots,n\},
}
\eq
$$
followed by the variable change $s'=n-s+1$.  If we subsequently
drop the primes, the net effect is
$$
\{2,\ldots,s\} \longleftrightarrow \{s{+}1,\ldots,n\}.
\eqlabel\KEYrelabel
$$
Upon applying (\KEYrelabel) to the appropriate parts of (\EINS)
and (\ZWEI), we see that both contributions are identically zero.

The next step is to combine $\amp_{\rm C}$ with $\amp_{\rm I}$.
The factors which differ between these two terms read
$$
{
{\bar u_{\dot\alpha}(P)}
\over
{ {\braket{P}{1}}^{*} }
}
-
{
{{\braket{P}{q}}^{*}\thinspace\bar u_{\dot\alpha}(k_1)}
\over
{ {\braket{P}{1}}^{*} {\braket{1}{q}}^{*} }
}
= -
{
{\bar u_{\dot\alpha}(q)}
\over
{ {\braket{1}{q}}^{*} }
},
\eq
$$
where we have applied (\fierz).  Thus, the sum of (\DREI)
and (\NEUN) is
$$
\eqalign{
&\amp_{\rm CI}=
-{\cal C} \negthinspace\negthinspace
\permsum{2}{n} \sum_{s=1}^n
{
{ \braket{P}{1} }
\over
{ \braket{P}{q} }
}
{
{ \bar u_{\dot\alpha}(q) \bar K^{\dot\alpha\beta}
u_{\beta}(q)  }
\over
{ \bra{P}2,\ldots,s \ket{Q} \bra{p}s{+}1,\ldots,n\ket{q}}
}
{
{1}
\over
{ { \braket{1}{q} }^{*} }
}.
}
\eqlabel\DREIplusNEUN
$$
The combination of $\amp_{\rm D}$ with $\amp_{\rm J}$ proceeds
in analogous fashion to produce
$$
\eqalign{
&\amp_{\rm DJ}=
-{\cal C} \negthinspace\negthinspace
\permsum{2}{n} \sum_{s=1}^n
{
{ \braket{p}{1} }
\over
{ \braket{p}{Q} }
}
{
{ \bar u_{\dot\alpha}(Q) \bar \curlyK^{\dot\alpha\beta}
u_{\beta}(Q)  }
\over
{ \bra{p}2,\ldots,s \ket{q} \bra{P}s{+}1,\ldots,n\ket{Q}}
}
{
{1}
\over
{ { \braket{1}{Q} }^{*} }
}.
}
\eqlabel\VIERplusZEHN
$$

In order to obtain the same denominator strings in all of the
contributions, we apply the momentum relabelling given
by (\KEYrelabel).  The action of (\KEYrelabel) on $\curlyK$
is
$$
\eqalign{
\curlyK = p+\kappa(1,s)+q &\longrightarrow p+k_1+\kappa(s{+}1,n)+q
\cr & \qquad = k_1-P-\kappa(1,s)-Q
\cr & \qquad = k_1-K.
}
\eqlabel\action
$$
If we relabel $\amp_{\rm H}$  in this manner
and combine it with $\amp_{\rm G}$,
the result is
$$
\eqalign{
&\amp_{\rm GH}=
{\cal C} \negthinspace\negthinspace
\permsum{2}{n} \sum_{s=1}^n
{
{2k_1\cdot K}
\over
{\bra{P}2,\ldots,s \ket{Q} \bra{p}s{+}1,\ldots,n\ket{q}}
}
\linkstar{Q}{1}{q}.
}
\eqlabel\SIEBENplusACHT
$$
We will set this contribution aside for cancellation.

The result of using (\KEYrelabel) to get all of the remaining
contributions on a common denominator is
$$
\eqalign{
&\amp_{\rm CDEFIJ}=
{\cal C} \negthinspace\negthinspace
\permsum{2}{n} \sum_{s=1}^n
{
{1}
\over
{ \bra{P}2,\ldots,s\ket{Q} \bra{p}s{+}1,\ldots,n\ket{q} }
}
\cr &  \enspace
\times
\Biggl\{
-
{ {1}\over{{\braket{1}{q}}^{*}}  }
{ {\braket{P}{1}} \over {\braket{P}{q}} }
\bar u_{\dot\alpha}(q)\bar K^{\dot\alpha\alpha} u_{\alpha}(q)
+
{ {1}\over{{\braket{1}{Q}}^{*}}  }
{ {\braket{p}{1}} \over {\braket{p}{Q}} }
\bar u_{\dot\alpha}(Q)[\bar K- \bar k_1]^{\dot\alpha\alpha}
u_{\alpha}(Q)
\cr & \enspace\qquad\negthinspace
-
{ {1}\over{{\braket{1}{Q}}^{*}}  }
{ {\braket{Q}{1}} \over {\braket{p}{Q}} }
\bar u_{\dot\alpha}(Q)\bar K^{\dot\alpha\alpha} u_{\alpha}(p)
+
{ {1}\over{{\braket{1}{q}}^{*}}  }
{ {\braket{q}{1}} \over {\braket{P}{q}} }
\bar u_{\dot\alpha}(q)[\bar K- \bar k_1]^{\dot\alpha\alpha}
u_{\alpha}(P)
\Biggr\}.
}
\eqlabel\yoyo
$$
Equation (\yoyo) consists of the contributions from
(\FUNF), (\SECHS), (\DREIplusNEUN), and (\VIERplusZEHN).
Denoting the portion of (\yoyo) in curly brackets by ${\cal R}$,
we have, after grouping the terms appropriately,
$$
\eqalign{
{\cal R}&=
{ {1}\over{{\braket{1}{q}}^{*}}  }
{
{\bar u_{\dot\alpha}(q)\bar K^{\dot\alpha\alpha}}
\over
{\braket{P}{q}}
}
\biggl[
-u_{\alpha}(q)\thinspace\braket{P}{1}
+u_{\alpha}(P)\thinspace\braket{q}{1}
\biggr]
\cr & \quad
+
{ {1}\over{{\braket{1}{Q}}^{*}}  }
{
{\bar u_{\dot\alpha}(Q) \bar K^{\dot\alpha\alpha} }
\over
{\braket{p}{Q}}
}
\biggl[
 u_{\alpha}(Q)\braket{p}{1}
- u_{\alpha}(p)\braket{Q}{1}
\biggr]
\cr & \quad
-
{ {\braket{p}{1}} \over {\braket{p}{Q}} }
\braket{1}{Q}
-
{ {\braket{q}{1}} \over {\braket{P}{q}} }
\braket{1}{P}.
}
\eq
$$
The terms in square brackets are simplified using (\fierz), yielding
$$
\eqalign{
{\cal R}&=
\Biggl[
-{ {\bar u_{\dot\alpha}(q)}\over{{\braket{1}{q}}^{*}}  }
+{ {\bar u_{\dot\alpha}(Q)}\over{{\braket{1}{Q}}^{*}}  }
\Biggr]
\bar K^{\dot\alpha\alpha}u_{\alpha}(k_1)
-\invlink{p}{1}{Q}-\invlink{q}{1}{P},
}
\eq
$$
where we have again regrouped.  One final application of (\fierz)
produces
$$
{\cal R}=
-2k_1\cdot K \linkstar{Q}{1}{q}
+\invlink{P}{1}{q} - \invlink{p}{1}{Q}.
\eqlabel\doneatlast
$$
Notice that the first term in (\doneatlast), when inserted
back into (\yoyo), exactly cancels (\SIEBENplusACHT).
Thus, the only surviving contributions to the total amplitude
come from the second two terms of (\doneatlast), leading to
the result
$$
\eqalign{
\amp&(p^{+},q^{-};P^{+},Q^{+};1^{-},2^{+},\ldots,n^{+})=
ig^2(-e\sqrt2)^n
{
{ {\braket{1}{q}}^2 }
\over
{ \bra{P}1\ket{Q} }
}
\cr & \times
\Biggl[ \invlink{P}{1}{q}-\invlink{p}{1}{Q} \Biggr]
\permsum{2}{n} \sum_{s=1}^n
{
{1}
\over
{ \bra{P}2,\ldots,s\ket{Q} \bra{p} s{+}1,\ldots,n\ket{q} }
}.
}
\eq
$$

The only technique required to do the remaining quadruple current
amplitudes which was not illustrated by this example is the
disposal of non-vanishing seagull terms.  However, the
procedure is very much like the seagull-disposal method
described in reference \cite\firstpaper, and so we will not
repeat it here.

\section{Summary of quadruple current amplitudes}

We now summarize the results of those processes we are able
to compute from four currents.  They fall naturally into two
groupings, as described below.  In order to list such a large
variety of processes in as compact a form as possible, we adopt
a ``standard'' process, namely
$$
P \thinspace Q \thinspace p \thinspace q
\longrightarrow
\gamma \thinspace \gamma \cdots \gamma.
\eqlabel\STANDARD
$$
In (\STANDARD), $P$ and $p$ denote positively-charged particles
of momenta $P$ and $p$ respectively, while $Q$ and $q$ are
negatively-charged particles of momenta $Q$ and $q$.  The
$n$ photons have momenta $k_1,\ldots,k_n$.  All momenta are
directed inward, so each amplitude listed below can be made
to describe many different processes via  the crossing relations.
The helicity labels appearing in the tables of results in this
section reflect this convention.

The first  group of amplitudes all involve $n$ like-helicity
photons.  They  have the generic form
$$
\eqalign{
\amp(P,Q;&p,q;1^{+},\ldots,n^{+}) =
\cr & =
ig^2 (-e\sqrt2)^n
\permsum{1}{n} \sum_{s=0}^n
{
{ f(P,Q,p,q) }
\over
{ \bra{P}1,\ldots,s\ket{Q} \bra{p}s{+}1,\ldots,n\ket{q} }
},
}
\eqlabel\quadGROUPone
$$
where $f(P,Q,p,q)$ is a scalar function that depends upon
the spins and helicities of the charged particles.
Table\Tlabel\GROUPoneQUADS{ } lists the values of $f$
appropriate for the various combinations of four currents
in this grouping.

Certain aspects of the entries in Table  \GROUPoneQUADS{ }
deserve special mention.  Notice that although the amplitudes
for transverse $W$ production from ``wrong''-helicity
fermions vanish in accordance with (\neatresult),
the same is not true for  longitudinal $W$
production.  Both types of fermion pairs may produce
longitudinal $W$'s in the high-energy limit.

Many of the entries contain functions of the weak-mixing angle
$\theta_W$.  This quantity enters in because of interference
between photon and $Z$ exchanges.  The entry involving four
longitudinal $W$'s contains two terms, one of which was
generated from the $\phi^4$ coupling of the Higgs sector.
Obviously, the (still unknown) size of $\lambda$ is crucial
in determining the importance of this contribution.

\noadvancetable{
{\begintable
$P$  \vb $Q$  \vb $p$  \vb $q$  \| $f(P,Q,p,q)$ \crthick
$W_{\up}^{+}$ \vb $W_{\down}^{-}$ \vb $\bar e_{\up}$ \vb $e_{\down}$\|
$-{\braket{Q}{q}}^3 {\braket{P}{q}}^{-1}$ \cr
$W_{\down}^{+}$ \vb $W_{\up}^{-}$ \vb $\bar e_{\up}$ \vb $e_{\down}$\|
$ {\braket{P}{q}}^2 \braket{P}{p} {\braket{p}{Q}}^{-1} $ \cr
$W_{\up}^{+}$ \vb $W_{\down}^{-}$ \vb $\bar e_{\down}$ \vb $e_{\up}$\|
$0$ \cr
$W_{\down}^{+}$ \vb $W_{\up}^{-}$ \vb $\bar e_{\down}$ \vb $e_{\up}$\|
$0$ \cr
$W_{\up}^{+}$ \vb $W_{\down}^{-}$ \vb $W_L^{+}$ \vb $W_L^{-}$\|
$-\braket{p}{Q} {\braket{Q}{q}}^2 {\braket{P}{q}}^{-1} $  \cr
$W_{\down}^{+}$ \vb $W_{\up}^{-}$ \vb $W_L^{+}$ \vb $W_L^{-}$\|
$-{\braket{P}{p}}^2 \braket{P}{q} {\braket{p}{Q}}^{-1} $ \cr
$\bar e_{\up}$ \vb $e_{\down}$ \vb $\bar\mu_{\up}$ \vb $\mu_{\down}$ \|
$-{1\over2} \sec^2\theta_W {\braket{Q}{q}}^2$ \cr
$\bar e_{\up}$ \vb $e_{\down}$ \vb $\bar e_{\up}$ \vb $\ e_{\down}$ \|
${1\over2} \sec^2\theta_W
\braket{P}{p} {\braket{Q}{q}}^3
{\braket{P}{q}}^{-1} {\braket{p}{Q}}^{-1} $ \cr
$\bar e_{\up}$ \vb $e_{\down}$ \vb $\bar\mu_{\down}$ \vb $\mu_{\up}$ \|
$ \tan^2\theta_W {\braket{p}{Q}}^2$ \cr
$\bar e_{\down}$ \vb $e_{\up}$ \vb $\bar\mu_{\down}$ \vb $\mu_{\up}$ \|
$-2 \tan^2\theta_W {\braket{P}{p}}^2$ \cr
$\bar e_{\down}$ \vb $e_{\up}$ \vb $\bar e_{\down}$ \vb $ e_{\up}$ \|
$2\tan^2\theta_W
{\braket{P}{p}}^3 \braket{Q}{q}
{\braket{P}{q}}^{-1} {\braket{p}{Q}}^{-1} $ \cr
$W_L^{+}$ \vb $W_L^{-}$ \vb $\bar e_{\up}$ \vb $e_{\down}$ \|
${1\over2} \sec^2\theta_W \braket{P}{q} \braket{Q}{q}$ \cr
$W_L^{+}$ \vb $W_L^{-}$ \vb $\bar e_{\down}$ \vb $e_{\up}$ \|
$- \tan^2\theta_W \braket{p}{P} \braket{p}{Q}$ \cr
$W_{L}^{+}$ \vb $W_{L}^{-}$ \vb $W_L^{+}$ \vb $W_L^{-}$\|
$-{1\over2}\sec^2\theta_W
\braket{P}{p} \braket{Q}{q}
-4\lambda g^{-2} \braket{P}{Q} \braket{p}{q}
$
\endtable}
}
{Group one  quadruple current amplitude
helicity functions}

For processes involving two fermion lines, the question
of indistinguishability
is especially interesting, since there is more than one
type of charged fermion in the standard model.
Extra graphs are possible when
both fermion lines have both the same helicity and the same identity:
the two fermions (or the two antifermions) may be exchanged.
When these graphs are included (with the relative minus sign
dictated by Fermi statistics), the entries for $e\bar e e\bar e$
listed result, and differ from the entries for $e\bar e\mu\bar\mu$.
On the other hand, if the two fermion lines have
different helicities, the additional graphs do not
exist.  In this case, $e \bar e e\bar e$ and $e\bar e\mu\bar\mu$
give the same results.

We now turn to the results for the second group of amplitudes, all
of which have a single negative-helicity photon and $n-1$
positive-helicity photons.  Most of the amplitudes in this group
have a very complicated structure, as evidenced by the general
form
$$
\eqalign{
&\amp(P,Q;p,q;1^{-},2^{+},\ldots,n^{+}) =
ig^2 (-e\sqrt2)^n
\cr & \times
\Biggl\{
\permsum{2}{n} \sum_{s=1}^n \sum_{j=2}^{s+1}
{
{ {{F_1}^{\alpha}}_{\beta}(P,Q,p,q,1)
{{\pole}_{\alpha}}^{\beta}(P,1,2,\ldots,j) }
\over
{ \bra{P}2,\ldots,s\ket{Q} \bra{p}s{+}1,\ldots,n\ket{q} }
}
{\Biggl\vert}_{j=s+1\equiv Q}
\cr & \enspace\thinspace +
\permsum{2}{n} \sum_{s=1}^n \sum_{j=2}^{s+1}
{
{ {{F_2}^{\alpha}}_{\beta}(P,Q,p,q,1)
{{\pole}_{\alpha}}^{\beta}(Q,1,2,\ldots,j) }
\over
{ \bra{P}s,s{-}1,\ldots,2\ket{Q} \bra{p}s{+}1,\ldots,n\ket{q} }
}
{\Biggl\vert}_{j=s+1\equiv P}
\cr & \enspace\thinspace +
\permsum{2}{n} \sum_{s=1}^n \sum_{j=2}^{s+1}
{
{ {{F_3}^{\alpha}}_{\beta}(P,Q,p,q,1)
{{\pole}_{\alpha}}^{\beta}(p,1,2,\ldots,j) }
\over
{ \bra{p}2,\ldots,s\ket{q} \bra{P}s{+}1,\ldots,n\ket{Q} }
}
{\Biggl\vert}_{j=s+1\equiv q}
\cr & \enspace\thinspace +
\permsum{2}{n} \sum_{s=1}^n \sum_{j=2}^{s+1}
{
{ {{F_4}^{\alpha}}_{\beta}(P,Q,p,q,1)
{{\pole}_{\alpha}}^{\beta}(q,1,2,\ldots,j) }
\over
{ \bra{p}s,s{-}1,\ldots,2\ket{q} \bra{P}s{+}1,\ldots,n\ket{Q} }
}
{\Biggl\vert}_{j=s+1\equiv p}
\cr & \enspace\thinspace +
\permsum{2}{n} \sum_{s=1}^n
{
{ f_5(P,Q,p,q,1) }
\over
{ \bra{P}2,\ldots,s\ket{Q} \bra{p}s{+}1,\ldots,n\ket{q} }
}
\Biggl[
\linkstar{P}{1}{Q} + \linkstar{p}{1}{q}
\Biggr]
\cr & \enspace\thinspace +
\permsum{2}{n} \sum_{s=1}^n
{
{ f_6(P,Q,p,q,1) }
\over
{ \bra{P}2,\ldots,s\ket{Q} \bra{p}s{+}1,\ldots,n\ket{q} }
}
{
{1}
\over
{ [P+\kappa(1,s)+Q]^2 }
}
\cr & \enspace\thinspace +
\permsum{2}{n} \sum_{s=1}^n
{
{ f_7(P,Q,p,q,1) }
\over
{ \bra{p}2,\ldots,s\ket{q} \bra{P}s{+}1,\ldots,n\ket{Q} }
}
{
{1}
\over
{ [p+\kappa(1,s)+q]^2 }
}
\cr & \enspace\thinspace +
\permsum{2}{n} \sum_{s=1}^n
{
{ f_8(P,Q,p,q,1) }
\over
{ \bra{P}2,\ldots,s\ket{Q} \bra{p}s{+}1,\ldots,n\ket{q} }
}
\Biggr\}.
}
\eqlabel\quadGROUPtwo
$$
Most of the structure in (\quadGROUPtwo) is simply the reflection
that the negative helicity photon can be radiated from
by one of the four charged particles.  The functions and momenta
appropriate to various processes are summarized in
Table\Tlabel\GROUPtwoQUADS.

\vfill\eject
\ \ \
\vfill\eject
\ \ \
\vfill\eject
\ \ \
\vfill\eject

\continuetable{
{\begintable
$P$  \vb $Q$  \vb $p$  \vb $q$  \|
$f_8(P,Q,p,q,1)$  \crthick
$W^{+}_{\up}$ \vb $W^{-}_{\up}$ \vb $\bar e_{\up}$ \vb $e_{\down}$ \|
${\braket{1}{q}}^2
\biggl[ \braket{1}{q} {\braket{P}{q}}^{-1} {\braket{1}{Q}}^{-1}
      - \braket{p}{1} {\braket{p}{Q}}^{-1} {\braket{P}{1}}^{-1} \biggr] $
\cr
$W^{+}_{\up}$ \vb $W^{-}_{\up}$ \vb $W_L^{+}$ \vb $W_L^{-}$ \|
$\bra{p}1\ket{q}
\biggl[ \braket{1}{q} {\braket{P}{q}}^{-1} {\braket{1}{Q}}^{-1}
      - \braket{p}{1} {\braket{p}{Q}}^{-1} {\braket{P}{1}}^{-1} \biggr] $
\cr
$\bar e_{\up}$ \vb $e_{\down}$ \vb $\bar\mu_{\up}$ \vb $\mu_{\down}$ \|
$-{1\over2}\sec^2\theta_W
\braket{Q}{q} \braket{1}{Q} \braket{1}{q}
\biggl[ (k_1+q)^{-2} - (k_1+Q)^{-2} \biggr] $
\cr
$\bar e_{\down}$ \vb $e_{\up}$ \vb $\bar\mu_{\down}$ \vb $\mu_{\up}$ \|
$2\tan^2\theta_W
\braket{P}{p} \braket{P}{1} \braket{p}{1}
\biggl[ (P+k_1)^2 - (p+k_1)^2 \biggr] $
\cr
$\bar e_{\up}$ \vb $e_{\down}$ \vb $\bar\mu_{\down}$ \vb $\mu_{\up}$ \|
$\tan^2\theta_W
\braket{p}{Q} \bra{p}1\ket{Q}
\biggl[ (p+k_1)^{-2} + (k_1+Q)^{-2} \biggr] $
\cr
$W_L^{+}$ \vb $W_L^{-}$ \vb $\bar e_{\up}$ \vb $ e_{\down}$ \|
${1\over2}\sec^2\theta_W
\bra{P}q,1\ket{Q} (k_1+q)^{-2} $
\cr
$W_L^{+}$ \vb $W_L^{-}$ \vb $\bar e_{\down}$ \vb $ e_{\up}$ \|
$\tan^2\theta_W
\bra{P}p,1\ket{Q} (p+k_1)^{-2} $
\cr
$W_L^{+}$ \vb $W_L^{-}$ \vb $W_L^{+}$ \vb $W_L^{-}$ \|
$ 0 $\cr
$W_L^{+}$ \vb $W_L^{-}$ \vb $W_L^{+}$ \vb $W_L^{-}$ \|
$ \thinspace\thinspace 0 $
\endtable}
}
{Group two  quadruple current amplitude helicity functions}

Table \GROUPtwoQUADS{ } requires a few comments.  First,
the only four-fermion amplitudes given involve $e \bar e \mu\bar\mu$,
because no additional simplifications were found in the
$e \bar e e \bar e$ case.  Hence, to obtain the result for
two identical fermion lines with the same helicity,
proceed as follows.  First, write down the contribution
implied by  the appropriate entry in Table \GROUPtwoQUADS.
Then,
subtract the same quantity with $Q \leftrightarrow q$,
to account for the indistinguishability.
The amplitude
involving two fermion lines of opposing helicities receives no
such additional contribution.
\eject

The process with four $W_L$'s has been given two entries in
Table \GROUPtwoQUADS.  The first one is the sum of photon
and $Z$ exchanges.
To it should be added another contribution
obtained by exchanging $Q$ and $q$,
since the $W_L$'s are indistinguishable.
The other entry is the result from the $\phi^4$ coupling.
Since the form of
this vertex automatically accounts for the indistinguishability
of the $W_L$'s, this contribution is complete as it stands
and requires no $Q \leftrightarrow q$ addition.

\chapter{CURRENTS WITH TWO OFF-SHELL PARTICLES}

The quadruple current amplitudes discussed in the previous section
all share a common feature.  The particle which is propagating
between the two charged lines is neutral.  It cannot radiate
photons.   Suppose we change the situation, and consider
processes where the virtual particle has a non-zero electric
charge.  Examples include:
$$
e \thinspace \bar\nu \longrightarrow
W^{-}_L \thinspace H \thinspace \gamma \ldots \gamma,
\eqlabel\scalarexample
$$
which contains a propagating $\phi^{\pm}$,
$$
W^{+}\thinspace W^{-} \longrightarrow
\nu \thinspace \bar\nu \thinspace \gamma \ldots \gamma,
\eqlabel\spinorexample
$$
which contains a propagating $e^{\pm}$, and
$$
e \thinspace \bar e \longrightarrow
\nu \thinspace \bar\nu \thinspace \gamma \ldots \gamma,
\eqlabel\vectorexample
$$
which contains a propagating $W^{\pm}$.  In every case, we
require a current which has {\it both} ends of the charged
line off shell.  In this section we will examine the situation
with respect to the computability of such amplitudes.  We will
find that, although we can obtain the actual double-off-shell
current only for a scalar line, we can obtain the combination
of factors required to compute all three of the above
types of amplitudes.


\section{The double-off-shell scalar current}

We define the  current $\PHImod(\Poff;1,\ldots,n;\Qoff)$
as consisting of a charged scalar line with $n$ photons
attached all possible ways.  All momenta are directed
inward.  The off-shell $\phi^{+}$ has momentum
$\Poff$, $\Poff^2 \ne 0$.  The off-shell $\phi^{-}$ has
momentum $\Qoff$, $\Qoff^2 \ne 0$.  The argument list
of $\PHImod$ is overspecified in that
$$
\Poff+\kappa(1,n)+\Qoff = 0,
\eqlabel\overspecify
$$
allowing us to always eliminate one of the momenta from the
result.  We define the zero-photon current by
$$
\PHImod(\Poff;\Qoff) = { {i}\over{\Poff^2} }
                     = { {i}\over{\Qoff^2} },
\eqlabel\ZEROphimod
$$
that is, just a propagator for the scalar particle.  A moment's
reflection upon the derivation\cite{\firstpaper}
of the recursion relation for
the scalar current $\PHI(P;1,\ldots,n)$ will
reveal that nowhere was the fact that $P^2=0$ used in the
development.  Hence, we have the same recursion formula as
(\WLrecursionperm), but with $\PHI$ replaced by $\PHImod$,
and seeded by (\ZEROphimod) instead of (\PHIP).

Furthermore, if we make the gauge choice
indicated by (\allpluspolarizations),
we are able to solve this recursion relation in the case of
all like-helicity photons.  Since the seagull contributions all vanish
in this gauge ({\it cf.}{ }equation (\seagullrid)), the recursion
relation for $\PHImod$ becomes
$$
\eqalign{
&\PHImod(\Poff;1^{+},\ldots,n^{+};\Qoff)=
\cr& \enspace =
{{-e\sqrt2}\over{[\Poff+\kappa(1,n)]^2}}
\permsum{1}{n} \negthinspace
{1\over{(n-1)!}}
\bar\eps^{\dot\alpha\alpha}(n) [\Poff{+}\kappa(1,n)]_{\alpha\dot\alpha}
\thinspace
\PHImod(\Poff;1^{+},\ldots,(n{-}1)^{+};\Qoff)
\cr & \enspace =
-e\sqrt2
\permsum{1}{n} \negthinspace
{1\over{(n-1)!}}
{
{\bar u_{\dot\alpha}(k_n)
[\bar\Poff{+}\bar\kappa(1,n)]^{\dot\alpha\alpha}
u_{\alpha}(g) }
\over
{ \braket{n}{g} \thinspace [\Poff+\kappa(1,n)]^2 }
}
\PHImod(\Poff;1^{+},\ldots,(n{-}1)^{+};\Qoff)
}
\eqlabel\herewego
$$
We may iterate (\herewego) until we reach $\PHImod(\Poff;\Qoff)$.
The result is
$$
\PHImod(\Poff;1^{+},\ldots,n^{+};\Qoff)=
(-e\sqrt2)^n \permsum{1}{n}
\PHImod(\Poff;\Qoff) \prod_{\ell=1}^n
{
{\bar u_{\dot\alpha}(k_\ell)
[\bar\Poff{+}\bar\kappa(1,\ell)]^{\dot\alpha\alpha}
u_{\alpha}(g) }
\over
{ \braket{\ell}{g} \thinspace [\Poff+\kappa(1,\ell)]^2 }
}.
\eqlabel\iterated
$$
We recognize that (\iterated) contains a factor of
$\Xi(1,n)$, as defined in Appendix \msconvent.  Thus, we
apply (\prodtosum) to obtain
$$
\eqalign{
\PHImod&(\Poff;1^{+},\ldots,n^{+};\Qoff)=
\cr & =
(-e\sqrt2)^n \permsum{1}{n}
{
{\Poff^2 \thinspace \PHImod(\Poff;\Qoff)}
\over
{\bra{g} 1,\ldots,n \ket{g}}
}
\sum_{\ell=1}^n
u^{\alpha}(g)
{{\pole}_{\alpha}}^{\beta}(\Poff,1,2,\ldots,\ell)
u_{\beta}(g).
}
\eqlabel\pointofdeparture
$$
Inserting (\ZEROphimod), we find that
$$
\eqalign{
\PHImod&(\Poff;1^{+},\ldots,n^{+};\Qoff)=
\cr & =
i(-e\sqrt2)^n \permsum{1}{n}
{
{1}
\over
{\bra{g} 1,\ldots,n \ket{g}}
}
\sum_{\ell=1}^n
u^{\alpha}(g)
{{\pole}_{\alpha}}^{\beta}(\Poff,1,2,\ldots,\ell)
u_{\beta}(g).
}
\eqlabel\PHImodallplus
$$
Equation (\PHImodallplus) is valid for $n\ge 1$.  The limit onto
$n=0$ is not smooth, and we must treat that case separately.

We note  that the
derivation for $\PHImod(\Poff;1^{+},\ldots,n^{+};\Qoff)$
matches the derivation
of $\PHI(P;1^{+},\ldots,n^{+})$ until the point at which
the zero particle current is introduced.  It is at this stage
where $P^2=0$ was used to eliminate the sum over the various
$\pole$'s in the latter case.  One might hope, in
light of this similarity, that it would be possible to compute
$\PHImod(\Poff;1^{-},2^{+},\ldots,n^{+};\Qoff)$, a current
with one unlike helicity.  Unfortunately,
because of the extra complications
introduced by the form of (\PHImodallplus) as compared to
(\PHIallplussoln), we have been unable to obtain a
simplified expression in this case.


\section{The double-off-shell spinor current}

We define the current $\Psi_{\alpha\dot\alpha}(\Poff;1,\ldots,n;\Qoff)$
as consisting of a charged spinor line with $n$ photons
attached all possible ways.  All momenta are directed
inward.  The off-shell $e^{+}$ has momentum
$\Poff$, $\Poff^2 \ne 0$.  The off-shell $e^{-}$ has
momentum $\Qoff$, $\Qoff^2 \ne 0$.  We choose the line to be
left-handed, as we are interested in diagrams that couple this
current to the $W$ of the standard model.  A right-handed current
could also be defined analogously.

The zero-photon current is given by a propagator for the fermion:
$$
\Psi_{\alpha\dot\alpha}(\Poff;\Qoff) =
{
{ -i\Poff_{\alpha\dot\alpha} }
\over
{ \Poff^2 }
}
=
{
{ i\Qoff_{\alpha\dot\alpha} }
\over
{ \Qoff^2 }
}.
\eqlabel\PSIzero
$$
The only place in which the on-shell condition
was used in the derivation
of the recursion for ${{\positronon}_{\dot\alpha}}\argltpos{p}{1}{2}{n}$
is in the form of the zero photon current---that is, a massless
spinor was used for the on-shell particle.  If we remove it from
(\spinorperm{a}), we obtain the recursion relation for
$\Psi_{\alpha\dot\alpha}(\Poff;1,\ldots,n;\Qoff)$:
$$
\eqalign{
\Psi_{\alpha\dot\alpha}&(\Poff;1,\ldots,n;\Qoff)=
\cr & =
-e\sqrt2
\permsum{1}{n} {{1}\over{(n-1)!}}
\Psi_{\alpha\dot\alpha}(\Poff;1,\ldots,n{-}1;\Qoff)
{\bar{\eps}}^{\dot\beta\beta}(n)
{{[\Poff+\kappa(1,n)]_{\beta\dot\alpha}}\over{[\Poff+\kappa(1,n)]^2}}.
}
\eqlabel\PSIrecursion
$$

In the case of all like-helicity photons, we are able to solve
(\PSIrecursion) for the combination
$u^{\alpha}(g)\Psi_{\alpha\dot\alpha}(\Poff;1^{+},\ldots,n^{+};\Qoff)$,
where $g$ is the gauge momentum of the photons.
Fortunately, for the processes under consideration in
this paper, this is sufficient.  We see this by noting
that typically, one of the transverse $W$ currents
is contracted into the undotted index of $\Psi$.  Since all of
the $\Wms$'s are proportional to $u(g)$ (we may have to choose
a specific value for $g$, however), the above combination
is all that is required.

Contracting $u^{\alpha}(g)$ into both sides of (\PSIrecursion)
and employing (\allpluspolarizations) for the polarization
spinors, we have
$$
\eqalign{
u^{\alpha}&(g)\Psi_{\alpha\dot\alpha}(\Poff;1^{+},\ldots,n^{+};\Qoff)=
\cr & =
-e\sqrt2  \negthinspace\negthinspace
\permsum{1}{n}  {{1}\over{(n-1)!}}
u^{\alpha}(g)
\Psi_{\alpha\dot\alpha}(\Poff;1^{+},\ldots,(n{-}1)^{+};\Qoff)
\cr & \qquad\times
\bar u^{\dot\alpha}(k_n)
{{u^{\beta}(g)[\Poff+\kappa(1,n)]_{\beta\dot\alpha}}
\over{\braket{n}{g}\thinspace[\Poff+\kappa(1,n)]^2}}.
}
\eqlabel\obvious
$$
{}From the form of (\obvious), is is obvious that $u^{\alpha}(g)
\Psi_{\alpha\dot\alpha}$ satisfies the following ansatz:
$$
u^{\alpha}(g)\Psi_{\alpha\dot\alpha}(\Poff;1^{+},\ldots,n^{+};\Qoff)=
u^{\alpha}(g)[\Poff+\kappa(1,n)]_{\alpha\dot\alpha}
\thinspace {\cal Y}(\Poff;1^{+},\ldots,n^{+};\Qoff),
\eqlabel\PSIansatz
$$
where, according to (\PSIzero), we have
$$
{\cal Y}(\Poff;\Qoff) = { {-i}\over{\Poff^2} }.
\eqlabel\Yzero
$$
If we insert (\PSIansatz) into (\obvious), we obtain
$$
\eqalign{
{\cal Y}&(\Poff;1^{+},\ldots,n^{+};\Qoff)=
\cr & =
-e\sqrt2  \negthinspace\negthinspace
\permsum{1}{n}  {{1}\over{(n-1)!}}
{
{\bar u_{\dot\alpha}(k_n)
[\bar\Poff+\bar\kappa(1,n)]^{\dot\alpha\beta}
u_{\beta}(g)}
\over
{\braket{n}{g}\thinspace[\Poff+\kappa(1,n)]^2}
}
{\cal Y}(\Poff;1^{+},\ldots,(n{-}1)^{+};\Qoff).
}
\eqlabel\Yrecurse
$$
A comparison of (\Yrecurse) with (\herewego) reveals that
$\PHImod$ and ${\cal Y}$ satisfy the same recursion relation.
Hence, the two solutions are proportional, differing only
by the ratio of the zero-photon currents.  Thus, we find
another example of a SUSY-like relationship between
currents of differing spins.
Hence, we have
$$
\eqalign{
u^{\alpha}&(g)\Psi_{\alpha\dot\alpha}(\Poff;1^{+},\ldots,n^{+};\Qoff)=
\cr & =
-i(-e\sqrt2)^n \permsum{1}{n}
{
{u^{\alpha}(g)[\Poff+\kappa(1,n)]_{\alpha\dot\alpha}}
\over
{\bra{g} 1,\ldots,n \ket{g}}
}
\sum_{\ell=1}^n
u^{\alpha}(g)
{{\pole}_{\alpha}}^{\beta}(\Poff,1,2,\ldots,\ell)
u_{\beta}(g),
}
\eqlabel\PSIallplus
$$
valid for $n\ge1$.  As in the scalar case, $n=0$ remains separate.


\section{The modified vector current}

The case of a vector line with both ends off shell is much more
difficult to solve in general.
This is because the vector current with one off-shell $W$ is
a conserved current, while the vector current with two off-shell
$W$'s is not.  Let us denote this current by
${I_{\mu\nu}}(\Poff;1,\ldots,n;\Qoff)$.   The $\mu$ index is
associated with the incoming $W^{+}$ of momentum $\Poff$,
while the $\nu$ index belongs to the incoming $W^{-}$ of
momentum $\Qoff$.  Since the
zero photon current is just a propagator, we have simply
$$
I_{\mu\nu}(\Poff;\Qoff)=
-{
{ig_{\mu\nu}}
\over
{\Poff^2}
}
=
-{
{ig_{\mu\nu}}
\over
{\Qoff^2}
}.
\eqlabel\newstarter
$$
It is straightforward to repeat the derivation of the
transverse $W$ current given in reference \cite\firstpaper\
using (\newstarter) as the starting point, and avoiding the
use of current conservation.  Hence, we shall immediately present
the result:
$$
\eqalign{
I_{\mu\nu}&(\Poff;1,2,\ldots,n;\Qoff) =
\cr & =
{-e\over{[\Poff+\kappa(1,n)]^2}}
\Biggl[
\permsum{1}{n}
{1\over{(n-1)!}} \thinspace
\biggl[\negthinspace\negthinspace\biggl[\eps(n),
I(\Poff,1,\ldots,n{-}1;\Qoff)
\biggr]\negthinspace\negthinspace\biggr]_{\mu\nu}
\cr&\quad
+e\permsum{1}{n}
{1\over{2!\thinspace (n-2)!}} \thinspace
\biggl\{\negthinspace\negthinspace\negthinspace\biggl\{
\eps(n{-}1),
I(\Poff,1,\ldots,n{-}2;\Qoff),\eps(n)
\biggr\}\negthinspace\negthinspace\negthinspace\biggr\}_{\mu\nu}
\Biggr],
}
\eqlabel\WTdbloffrecursionperm
$$
where
$$
\eqalign{
{\biggl[\negthinspace\negthinspace\biggl[}
\eps(n),I&(\Poff,1,\ldots,n{-}1;\Qoff)
\biggr]\negthinspace\negthinspace\biggr]_{\mu\nu}\equiv
\cr & =
I_{\mu\nu}(\Poff,1,\ldots,n{-}1;\Qoff) \thinspace\thinspace
\eps(n)\cdot\biggl\{2[\Poff+\kappa(1,n{-}1)]+k_n\biggr\}
\cr&\quad
-I_{\mu\xi}(\Poff;1,\ldots,n{-}1;\Qoff)
\biggl\{ [\Poff+\kappa(1,n{-}1)]+2k_n \biggr\}^{\xi}
\thinspace\thinspace \eps_{\nu}(n)
\cr&\quad
+ I_{\mu\xi}(\Poff;1,\ldots,n{-}1;\Qoff)
\eps^{\xi}(n)
\biggl\{k_n-[P+\kappa(1,n{-}1)]\biggr\}^{\nu},
}
\eqlabel\newsqbrak
$$
and
$$
\eqalign{
\biggl\{\negthinspace\negthinspace\negthinspace\biggl\{
\eps(&n{-}1),I(\Poff;1,\ldots,n{-}2;\Qoff),\eps(n)
\biggr\}\negthinspace\negthinspace\negthinspace{\biggr\}}_{\mu\nu}
\equiv \cr & =
\eps^{\xi}(n{-}1)
[\eps_{\xi}(n)I_{\mu\nu}(\Poff;1,\ldots,n{-}2;\Qoff)
-I_{\mu\xi}(\Poff;1,\ldots,n{-}2;\Qoff)\eps_{\nu}(n)]
\cr & \quad
-\eps^{\xi}(n)
[I_{\mu\xi}(\Poff;1,\ldots,n{-}2;\Qoff)\eps_{\nu}(n{-}1)
-\eps_{\xi}(n{-}1)I_{\mu\nu}(\Poff;1,\ldots,n{-}2;\Qoff) ].
}
\eqlabel\newcurlybrak
$$
Equations (\newsqbrak) and (\newcurlybrak) reduce to
(\sqbrak) and (\curlybrak) respectively if we contract
in a  transverse polarization vector $W^{\mu}(\Poff)$,
multiply by $i\Poff^2$ to remove the propagator for the
$W^{+}$, and let $\Poff^2\rightarrow0$.

As written, the recursion relation for the double off-shell
$W$ current is prohibitively difficult to solve.
Instead, let us define a ``modified'' transverse $W$ current,
$\Wmod(P^{*};1,\ldots,n)$, in the spirit of reference
[\ref{G. Mahlon, T.--M. Yan and C. Dunn,
Cornell preprint CLNS 91/1120, 1992.}
\refname\secondpaper{\kern-.6em}{].}  This current differs from the
usual transverse $W$ current in that $P^2 \ne 0$, and
$$
\Wmod_{\alpha\dot\alpha}(P^{*}) \equiv
u_{\alpha}(g)  u^{\beta}(g)P_{\beta\dot\alpha}
\eqlabel\WMODzero
$$
replaces $\Wms_{\alpha\dot\alpha}(P)$.
The form of (\WMODzero) is such that
$$
\bar P^{\dot\alpha\alpha} \Wmod_{\alpha\dot\alpha}(P^{*}) =0,
\eqlabel\metoo
$$
just as if $\Wmod(P^{*})$ were a real
polarization spinor.  Consequently,
$\Wmod(P^{*};1,\ldots,n)$ is a conserved current,
just like $\Wms(P;1,\ldots,n)$.  Furthermore,
since  $\Wmod(P^{*})$ is proportional to $u(g)$,
$$
\bar \eps^{\dot\alpha\alpha}(j^{+})
\Wmod_{\alpha\dot\alpha}(P^{*}) = 0
\eqlabel\SEAgullRID
$$
for the gauge choice (\allpluspolarizations).  As a result of
(\metoo) and (\SEAgullRID), $\Wmod(P^{*};1^{+},\ldots,n^{+})$
satisfies the same simplified form of the recursion relation
as $\Wms(P^{+};1^{+},\ldots,n^{+})$, namely \cite{\firstpaper}
$$
\eqalign{
{\Wmod}_{\alpha\dot\alpha}(P^{*};1^{+},\ldots,n^{+})& =
-e\sqrt2
\permsum{1}{n}   { 1\over{(n-1)!} }
{
{[P+\kappa(1,n)]_{\beta\dot\beta}}
\over
{[P+\kappa(1,n)]^2}
}
\cr & \quad\times
\Biggl[
{\bar\eps}^{\dot\beta\beta}(n^{+})
{\Wmod}_{\alpha\dot\alpha}\bigl(P^{*};1^{+},\ldots,(n{-}1)^{+}\bigr)
\cr &  \quad\quad
-
{\Wmodbar}^{\dot\beta\beta}\bigl(P^{*};1^{+},\ldots,(n{-}1)^{+}\bigr)
{\eps}_{\alpha\dot\alpha}(n^{+})
\Biggr].
}
\eqlabel\WMODallplusstart
$$
The form of (\WMODzero) is consistent with the following ansatz
for the spinor structure of  $\Wmod(P^{*};1^{+},\ldots,n^{+})$:
$$
\Wmod(P^{*};1^{+},\ldots,n^{+}) =
u_{\alpha}(g)u^{\beta}(g)[P+\kappa(1,n)]_{\beta\dot\alpha}
\thinspace\zhe(P^{*};1^{+},\ldots,n^{+}),
\eqlabel\WMODansatz
$$
with
$$
\zhe(P^{*}) = 1.
\eqlabel\zhezero
$$
The correctness of this ansatz is not immediately obvious, but it
is easily proven inductively.
Assume that the ansatz is true for the $(n-1)$-particle current.
Then, the $n$-particle current is given by
$$
\eqalign{
{\Wmod}_{\alpha\dot\alpha}&(P^{*};1^{+},\ldots,n^{+})=
-e\sqrt2 \permsum{1}{n}  { 1\over{(n-1)!} }
{
{\zhe(P^{*};1^{+},\ldots,(n{-}1)^{+})}
\over
{\braket{n}{g}\varsp[P+\kappa(1,n)]^2}
} \cr&
\times
\Biggl\{ u^{\beta}(g)[P+\kappa(1,n)]_{\beta\dot\beta}
{\bar u}^{\dot\beta}(k_n) u_{\alpha}(g)u^{\gamma}(g)
[P+\kappa(1,n{-}1)]_{\gamma\dot\alpha}
\cr &\qquad
+u^{\beta}(g)[P+\kappa(1,n)]_{\beta\dot\beta}
[\bar P+\bar \kappa(1,n{-}1)]^{\dot\beta\gamma}u_{\gamma}(g)
u_{\alpha}(g){\bar u}_{\dot\alpha}(k_n) \Biggr\}.
}
\eqlabel\midstep
$$
We simplify the quantity in curly brackets may be simplified using
(\slashsqr), obtaining
$$
\eqalign{&
u_{\alpha}(g)
\bigl\{
u^{\beta}(g)
[ P+\kappa(1,n)]_{\beta\dot\beta} {\bar u}^{\dot\beta}(k_n)
u^{\gamma}(g)[P+\kappa(1,n{-}1)]_{\gamma\dot\alpha}
\cr & \qquad
+ u^{\beta}(g)k_{n\beta\dot\beta}
[\bar P+\bar \kappa(1,n)]^{\dot\beta\gamma}
u_{\gamma}(g) {\bar u}_{\dot\alpha}(k_n) \bigr\}
\cr &
= u_{\alpha}(g)
{\bar u}_{\dot\beta}(k_n)
[\bar P + \bar \kappa(1,n)]^{\dot\beta\gamma}
u_{\gamma}(g)
\cr & \qquad\times
\bigl\{
u^{\beta}(g)[P+\kappa(1,n{-}1)]_{\beta\dot\alpha}
+ u^{\beta}(g)k_{n\beta\dot\alpha} \bigr\}
\cr &
= u_{\alpha}(g)u^{\beta}(g)
[P+\kappa(1,n)]_{\beta\dot\alpha}\thinspace
{\bar u}_{\dot\beta}(k_n)
[\bar P + \bar \kappa(1,n)]^{\dot\beta\gamma}
u_{\gamma}(g).
}
\eqlabel\SQbrakets
$$
If we insert (\SQbrakets) into  (\midstep), we see that the following
recursion relation for $\zhe(P^{+};1^{+},\ldots,n^{+})$ must hold:
$$
\eqalign{
\zhe&(P^{+};1^{+},\ldots,n^{+}) =
\cr & =
-e \sqrt2 \permsum{1}{n}  { 1\over{(n-1)!} }
{
{ {\bar u}_{\dot\beta}(k_n)[P+\kappa(1,n)]^{\dot\beta\beta}u_{\beta}(g)}
\over
{\braket{n}{g}\varsp [P+\kappa(1,n)]^2}
}
\zhe\bigl(P^{+};1^{+},\ldots,(n{-}1)^{+}\bigr).
}
\eqlabel\zherecurs
$$
When (\zherecurs) is compared to (\herewego), we see
that $\zhe$ also satisfies the
same recursion relation  as $\PHImod$.  Thus, we immediately
write down
$$
\eqalign{
\zhe&(P^{*};1^{+},\ldots,n^{+})=
\cr & =
(-e\sqrt2)^n \permsum{1}{n}
{
{P^2}
\over
{\bra{g} 1,\ldots,n \ket{g}}
}
\sum_{\ell=1}^n
u^{\alpha}(g)
{{\pole}_{\alpha}}^{\beta}(\Poff,1,2,\ldots,\ell)
u_{\beta}(g).
}
\eqlabel\zheallplus
$$
In the next section we will see how $\Wmod(P^{*};1,\ldots,n)$
enters into amplitudes that, on the surface at least, would seem to
require knowledge of the full double-off-shell vector current.
\chapter{TRIPLE CURRENT AMPLITUDES}

In this section we will examine the amplitudes which may
be computed from a combination of three currents.  Because
one of these currents must have two off-shell particles, we
are limited to those cases with like-helicity photons.
The process we will illustrate the computational methods
with is
$$
\bar e_{\up} \thinspace \nu_{\down} \longrightarrow
W_L \thinspace H \thinspace \gamma_{\up} \cdots \gamma_{\up}.
\eqlabel\tripleexample
$$
We have selected (\tripleexample) since it demonstrates
how the modified vector current $\Wmod$ fits into the
picture.  It is straightforward to apply $\PHImod$ and
$\Psi$ to the amplitudes in which they appear.


\section{The process $\bar e_{\uparrow} \thinspace \nu_{\downarrow}
\longrightarrow W_L \thinspace H \thinspace
\gamma_{\uparrow} \cdots \gamma_{\uparrow}$}

Figure \Flabel\tripfig\ illustrates the Feynman diagrams describing the
process (\tripleexample) and indicates the momentum routing
that has been chosen in order to evaluate them.
According to Figure \tripfig, we have
\def \p{p}   
\def \q{q}   
\def \P{H}   
\def \Q{\nu} 
$$
\eqalign{
\amp&(\p,\q;\P,\Q;1,\ldots,n) =
\cr & =
{
{-g^2}
\over
{2\sqrt2}
}
\permsum{1}{n} \sum_{s=0}^n \sum_{t=s}^n
{
{1}
\over
{s!(t{-}s)!(n{-}t)!}
}
\positronon(\p;1,\ldots,s) \thinspace
\gamma_{\lambda} \thinspace {1\over2}(1-\gamma_5) \thinspace u(\Q)
\cr & \qquad\qquad\quad\times
\PHI(t{+}1,\ldots,n;\q) \thinspace
\biggl[ \P - \{\q+\kappa(t{+}1,n)\}\biggr]_{\mu}
I^{\mu\lambda}(\Poff;s{+}1,\ldots,t;\Qoff)
\cr & \enspace
-{
{g^2e}
\over
{2\sqrt2}
}
\permsum{1}{n} \sum_{s=0}^{n-1} \sum_{t=s}^{n-1}
{
{1}
\over
{s!(t{-}s)!(n{-}t{-}1)!}
}
\positronon(\p;1,\ldots,s) \thinspace
\gamma_{\lambda} \thinspace {1\over2}(1-\gamma_5) \thinspace u(\Q)
\cr & \qquad\qquad\quad\times
\PHI(t{+}2,\ldots,n;\q) \thinspace
\eps_{\mu}(t{+}1)
I^{\mu\lambda}(\Poff;s{+}1,\ldots,t;\Qoff).
}
\eqlabel\tripstart
$$
In (\tripstart) the positron has momentum $\p$, the
$W_L^{-}$ has momentum $\q$, the Higgs boson has
momentum $\P$, and the neutrino has momentum $\Q$.
All of these momenta are directed into the diagram.
We also have defined
$$
\Poff \equiv \p + \kappa(1,s) + \Q
\eqlabel\curlyPdef
$$
and
$$
\Qoff \equiv \P + \kappa(t{+}1,n) + \q.
\eqlabel\curlyQdef
$$
Note the presence of
$I^{\mu\lambda}(\Poff;s{+}1,\ldots,t;\Qoff)$,
the vector current with both $W$'s off shell.
The momenta $\Poff$ and $\Qoff$ are directed towards the center
of this current.  Thus, $\Poff$ is the momentum of the off-shell
$W^{+}$, while $\Qoff$ is the momentum of the off-shell $W^{-}$.

We now demonstrate how to connect $I^{\mu\lambda}$ to
$\Wmod_{\alpha\dot\alpha}$.  One way to obtain the current
with two off-shell $W$'s is to begin with a current with
just one off-shell $W$, remove the polarization vector of the
other $W$ by differentiation, and supply a propagator for
the newly off-shell particle.  Hence, we formally write
$$
I^{\mu\lambda}(\Poff;s{+}1,\ldots,t;\Qoff) =
{ {-i}\over{\Poff^2} }
{
{ \partial }
\over
{ \partial\eps_{\lambda}(\Poff) }
}
\Wnorm^{\mu}(\Poff;s{+}1,\ldots,t).
\eqlabel\doubleoffW
$$
Applying (\doubleoffW) to (\tripstart) and translating
to multispinor form via (\msreplacedotprod) and
(\msreplacefermion), we have
$$
\eqalign{
\amp&(\p^{+},\q^{0};\P^{0},\Q^{-};1^{+},\ldots,n^{+}) =
\cr & =
{
{ig^2}
\over
{2\sqrt2}
}
\permsum{1}{n} \sum_{s=0}^n \sum_{t=s}^n
{
{1}
\over
{s!(t{-}s)!(n{-}t)!}
}
\positronon_{\dot\beta}(\p^{+};1^{+},\ldots,s^{+}) \thinspace
u_{\beta}(\Q)
\cr & \qquad\qquad\quad\times
\PHI((t{+}1)^{+},\ldots,n^{+};\q) \thinspace
\biggl[ \P - \{\q+\kappa(t{+}1,n)\}\biggr]_{\alpha\dot\alpha}
\cr & \qquad\qquad\quad\times
{ {1}\over{\Poff^2} }
{
{ \partial }
\over
{ \partial\eps_{\beta\dot\beta}(\Poff) }
}
\Wbar^{\dot\alpha\alpha}(\Poff;(s{+}1)^{+},\ldots,t^{+})
\cr & \enspace
+{
{ig^2e}
\over
{2}
}
\permsum{1}{n} \sum_{s=0}^{n-1} \sum_{t=s}^{n-1}
{
{1}
\over
{s!(t{-}s)!(n{-}t{-}1)!}
}
\positronon_{\dot\beta}(\p^{+};1^{+},\ldots,s^{+}) \thinspace
u_{\beta}(\Q)
\cr & \qquad\qquad\quad\times
\PHI((t{+}2)^{+},\ldots,n^{+};\q) \thinspace
\eps_{\alpha\dot\alpha}((t{+}1)^{+})
\cr & \qquad\qquad\quad\times
{ {1}\over{\Poff^2} }
{
{ \partial }
\over
{ \partial\eps_{\beta\dot\beta}(\Poff) }
}
\Wbar^{\dot\alpha\alpha}(\Poff;(s{+}1)^{+},\ldots,t^{+}).
}
\eqlabel\readytoIDwmod
$$
In (\readytoIDwmod) we have specialized to the helicity configuration
we are able to compute.   In both terms of (\readytoIDwmod) we
are instructed to remove the polarization spinor for the $W^{+}$
and replace it by
$\positronon_{\dot\beta}(\p^{+};1^{+},\ldots,s^{+})
u_{\beta}(\Q)$.  But, according to (\RHpositronallplussoln),
we may write
$$
{\positronon}_{\dot\beta}(\p^{+};1^{+},\ldots,s^{+}) =
u^{\gamma}(g)[\p+\kappa(1,s)]_{\gamma\dot\beta}
Y(\p^{+};1^{+},\ldots,s^{+})
\eq
$$
where
$$
Y(\p^{+};1^{+},\ldots,s^{+}) \equiv
\permsum{1}{s}
{
{-(-e\sqrt2)^{s}}
\over
{ \bra{p} 1,\ldots,s \ket{g} }
}.
\eqlabel\YDEF
$$
Thus, the spinor structure of what we replace the polarization
spinor with is
$$
u^{\gamma}(\Q)[\p+\kappa(1,s)]_{\gamma\dot\beta}u_{\beta}(\Q)
= u_{\beta}(\Q) u^{\gamma}(\Q) \Poff_{\gamma\dot\beta},
\eq
$$
where we have chosen $g=\Q$ and  applied the Weyl equation
along with (\curlyPdef) to obtain a form that matches
(\WMODzero). Consequently, (\readytoIDwmod) becomes
$$
\eqalign{
\amp&(\p^{+},\q^{0};\P^{0},\Q^{-};1^{+},\ldots,n^{+}) =
\cr & =
{
{ig^2}
\over
{2\sqrt2}
}
\permsum{1}{n} \sum_{s=0}^n \sum_{t=s}^n
{
{1}
\over
{s!(t{-}s)!(n{-}t)!}
}
{ {1}\over{\Poff^2} }
Y(\p^{+};1^{+},\ldots,s^{+})
\cr & \qquad\qquad\quad\times
\biggl[ \P - \{\q+\kappa(t{+}1,n)\}\biggr]_{\alpha\dot\alpha}
\Wmodbar^{\dot\alpha\alpha}(\Poff^{*};(s{+}1)^{+},\ldots,t^{+})
\cr & \qquad\qquad\quad\times
\PHI((t{+}1)^{+},\ldots,n^{+};\q)
\cr & \enspace
+{
{ig^2e}
\over
{2}
}
\permsum{1}{n} \sum_{s=0}^{n-1} \sum_{t=s}^{n-1}
{
{1}
\over
{s!(t{-}s)!(n{-}t{-}1)!}
}
{ {1}\over{\Poff^2} }
Y(\p^{+};1^{+},\ldots,s^{+})
\cr & \qquad\qquad\quad\times
\Wmodbar^{\dot\alpha\alpha}(\Poff^{*};(s{+}1)^{+},\ldots,t^{+})
\eps_{\alpha\dot\alpha}((t{+}1)^{+})
\cr & \qquad\qquad\quad\times
\PHI((t{+}2)^{+},\ldots,n^{+};\q).
}
\eqlabel\readytostart
$$

Let us insert (\WMODansatz) to incorporate the spinor
structure of $\Wmodbar^{\dot\alpha\alpha}$.
Because of (\allpluspolarizations),
the second term in (\readytostart) vanishes.  This leaves just
$$
\eqalign{
\amp&(\p^{+},\q^{0};\P^{0},\Q^{-};1^{+},\ldots,n^{+}) =
\cr & =
{
{ig^2}
\over
{2\sqrt2}
}
\permsum{1}{n} \sum_{s=0}^n \sum_{t=s}^n
{
{1}
\over
{s!(t{-}s)!(n{-}t)!}
}
{ {1}\over{\Poff^2} }
Y(\p^{+};1^{+},\ldots,s^{+})
\cr & \qquad\qquad\quad\times
\zhe(\Poff^{*};(s{+}1)^{+},\ldots,t^{+})
\PHI((t{+}1)^{+},\ldots,n^{+};\q)
\cr & \qquad\qquad\quad\times
\biggl[\bar \P - \{\bar\q+\bar\kappa(t{+}1,n)\}\biggr]^{\alpha\dot\alpha}
u_{\alpha}(\Q) u^{\beta}(\Q) [\Poff+\kappa(s{+}1,t)]_{\beta\dot\alpha}.
}
\eqlabel\zhepluggedin
$$
Let us examine the last line of factors appearing in (\zhepluggedin),
which we will call $\chi$.  Applying (\curlyPdef), we find that
$$
\eqalign{
\chi &= u^{\beta}(\Q)[ \p+\kappa(1,t)+\Q ]_{\beta\dot\alpha}
[2\bar\P-\{\bar\P+\bar\kappa(t{+}1,n)+\bar\q\}]^{\dot\alpha\alpha}
u_{\alpha}(\Q)
\cr & =
-u^{\beta}(\Q)[ \P+\kappa(t{+}1,n)+\q ]_{\beta\dot\alpha}
[2\bar\P-\{\bar\P+\bar\kappa(t{+}1,n)+\bar\q\}]^{\dot\alpha\alpha}
u_{\alpha}(\Q),
}
\eq
$$
where we have used momentum conservation in the second line.
Because of (\slashsqr) and the antisymmetry of the spinor
product, this reduces to
$$
\chi =- 2 u^{\beta}(\Q)[\P+\kappa(t{+}1,n)+\q  ]_{\beta\dot\alpha}
{\bar\P}^{\dot\alpha\alpha}
u_{\alpha}(\Q).
\eqlabel\CHI
$$

At this stage, we insert (\scalarcrossing)
and (\PHIallplussoln) for $\PHI$,
(\zhezero) and (\zheallplus) for $\zhe$,
(\YDEF) for $Y$, and (\CHI) for $\chi$ back into (\zhepluggedin).
Because of the special form of $\zhe$ when $t=s$,
we obtain two separate contributions.  After collecting
related factors, these two terms are
$$
\amp_1 \equiv
{ {ig^2}\over{\sqrt2} }
(-e\sqrt2)^n \negthinspace\negthinspace\negthinspace
\permsum{1}{n} \sum_{s=0}^n
{
{ \braket{\P}{\Q} \braket{\Q}{\q}
\bar u_{\dot\alpha}(\P)
[\bar\P+\bar\kappa(s{+}1,n)+\bar\q]^{\dot\alpha\alpha}
u_{\alpha}(\Q) }
\over
{ \bra{\p}1,\ldots,s\ket{\Q} \bra{\Q}s{+}1,\ldots,n\ket{\q}
[\p+\kappa(1,s)+\Q]^2 }
}
\eqlabel\firstterm
$$
and
$$
\eqalign{
\amp_2 & \equiv
{ {ig^2}\over{\sqrt2} }
(-e\sqrt2)^n \negthinspace\negthinspace\negthinspace
\permsum{1}{n} \sum_{\ell=1}^n \sum_{t=\ell}^n \sum_{s=0}^{\ell-1}
{
{ \bar u_{\dot\alpha}(\P)
[\bar\kappa(t{+}1,n)+\bar\q]^{\dot\alpha\alpha}
u_{\alpha}(\Q) }
\over
{ \bra{\p}1,\ldots,s\ket{\Q} \bra{\Q}s{+}1,\ldots,t\ket{\Q}  }
}
\cr & \qquad\qquad\qquad\times
{
{\braket{\P}{\Q} \braket{\Q}{\q}}
\over
{\bra{\Q}t{+}1,\ldots,n\ket{\q}}
}
u^{\gamma}(\Q){{\pole}_{\gamma}}^{\delta}(\Poff,s{+}1,\ldots,\ell)
u_{\delta}(\Q).
}
\eqlabel\secondterm
$$
Note that we have reordered the sums appearing in (\secondterm)
so that the sum on $\ell$, which came from $\zhe$, is to
be done last.

Since it is not immediately obvious how to simplify $\amp_1$,
we begin with $\amp_2$.  Since $\Poff$ is never the
last argument of $\pole$, we have
$$
\pole(\Poff,s{+}1,\ldots,\ell)=\pole(\p,\Q,1,\ldots,\ell),
\eqlabel\poleprettyup
$$
where we have applied (\curlyPdef) and exploited the
definition  (\poledef) for $\pole$.  Thus, the $s$-dependence
of (\secondterm) may be summarized in
$$
\sum_{s=0}^{\ell-1} \link{s}{\Q}{s{+}1}
= \link{\p}{\Q}{\ell}.
\eqlabel\doSUMonS
$$
Application of (\doSUMonS) to (\secondterm) produces
$$
\eqalign{
\amp_2 & =
{ {ig^2}\over{\sqrt2} }
(-e\sqrt2)^n \negthinspace\negthinspace\negthinspace
\permsum{1}{n} \sum_{\ell=1}^n \sum_{t=\ell}^n
{
{ \bar u_{\dot\alpha}(\P)
[\bar\kappa(t{+}1,n)+\bar\q]^{\dot\alpha\alpha}
u_{\alpha}(\Q) }
\over
{ \bra{\p}1,\ldots,t\ket{\Q} \bra{\Q}t{+}1,\ldots,n\ket{\q} }
}
\cr & \qquad\qquad\qquad\times
{
{\braket{\P}{\Q} \braket{\Q}{\q}}
\over
{\braket{\p}{\Q}}
}
u^{\gamma}(\p){{\pole}_{\gamma}}^{\delta}(\p,\Q,1,\ldots,\ell)
u_{\delta}(\Q).
}
\eqlabel\nomoreS
$$
The sum on $t$ appearing in (\nomoreS) is not much harder,
although we do have to deal with the $t$-dependence of the
numerator:
$$
\eqalign{
\sum_{t=\ell}^n \sum_{u=t+1}^{n+1}
\link{t}{\Q}{t{+}1}
{\braket{u}{\P}}^{*}
\braket{u}{\Q}
& =
\sum_{u=\ell+1}^{n+1}
\link{\ell}{\Q}{u}
{\braket{u}{\P}}^{*}
\braket{u}{\Q}
\cr & =
{ {-1}\over{\braket{\Q}{\ell}} }
\bar u_{\dot\alpha}(\P)
[\bar\kappa(\ell{+}1,n)+\bar\q]^{\dot\alpha\alpha}
u_{\alpha}(k_{\ell}).
}
\eq
$$
Thus,
$$
\eqalign{
\amp_2 & =
{ -{ig^2}\over{\sqrt2} }
(-e\sqrt2)^n \negthinspace\negthinspace\negthinspace
\permsum{1}{n} \sum_{\ell=1}^n
{
{\braket{\P}{\Q} \braket{\Q}{\q}}
\over
{\braket{\p}{\Q}}
}
{
{\bar u_{\dot\alpha}(\P)
[\bar\kappa(\ell{+}1,n)+\bar\q]^{\dot\alpha\alpha} }
\over
{ \bra{\p}1,\ldots,n\ket{\q} }
}
\cr & \qquad\qquad\qquad\times
{
{u_{\alpha}(k_\ell)}
\over
{ \braket{\Q}{\ell} }
}
u^{\gamma}(\p){{\pole}_{\gamma}}^{\delta}(\p,\Q,1,\ldots,\ell)
u_{\delta}(\Q).
}
\eqlabel\nomoreTeither
$$

At this stage, we apply (\reverseid) to turn
$u^{\gamma}(\p){{\pole}_{\gamma}}^{\delta}(\p,\Q,1,\ldots,\ell)
u_{\delta}(\Q)$ into
$u^{\gamma}(\Q){{\pole}_{\gamma}}^{\delta}(\p,\Q,1,\ldots,\ell)
u_{\delta}(\p)$.  The result of this manipulation is
$$
\eqalign{
\amp_{2 {\rm A}} & \equiv
{ -{ig^2}\over{\sqrt2} }
(-e\sqrt2)^n \negthinspace\negthinspace\negthinspace
\permsum{1}{n} \sum_{\ell=1}^n
{
{\braket{\P}{\Q} \braket{\Q}{\q} }
\over
{ \braket{\p}{\Q} }
}
{
{1}
\over
{ \bra{\p}1,\ldots,n\ket{\q} }
}
\cr & \qquad\qquad\times
\bar u_{\dot\alpha}(\P)
[\bar\P+\bar\kappa(\ell{+}1,n)+\bar\q]^{\dot\alpha\alpha}
{{\pole}_{\alpha}}^{\delta}(\p,\Q,1,\ldots,\ell)
u_{\delta}(\p),
}
\eqlabel\firstfrag
$$
$$
\eqalign{
\amp_{2 {\rm B}} & \equiv
{ -{ig^2}\over{\sqrt2} }
(-e\sqrt2)^n \negthinspace\negthinspace\negthinspace
\permsum{1}{n} \sum_{\ell=1}^n
{\braket{\P}{\Q} \braket{\Q}{\q}}
{
{\bar u_{\dot\alpha}(\P)
[\bar\kappa(\ell{+}1,n)+\bar\q]^{\dot\alpha\alpha} }
\over
{ \bra{\p}1,\ldots,n\ket{\q} }
}
\cr & \qquad\qquad\times
{
{u_{\alpha}(k_\ell)}
\over
{ \braket{\Q}{\ell} }
}
\Biggl[
{
{1}
\over
{ [\p+\Q+\kappa(1,\ell{-}1)]^2 }
}
-
{
{1}
\over
{ [\p+\Q+\kappa(1,\ell)]^2 }
}
\Biggr].
}
\eqlabel\secondfrag
$$
We have written the contributions from the three terms of (\reverseid)
as two separate pieces since we will deal with them differently.

We begin by considering $\amp_{2 {\rm A}}$.  Momentum conservation
allows us to replace
$[\bar\P+\bar\kappa(\ell{+}1,n)+\bar\q]^{\dot\alpha\alpha}$
by
$-[\bar\p+\bar\Q+\bar\kappa(1,\ell)]^{\dot\alpha\alpha}$.
Hence, we may apply (\splitid) and perform the sum on $\ell$,
giving
$$
\eqalign{
\amp_{2 {\rm A}} & =
{ {ig^2}\over{\sqrt2} }
(-e\sqrt2)^n \negthinspace\negthinspace\negthinspace
\permsum{1}{n}
{
{\braket{\P}{\Q} \braket{\Q}{\q} }
\over
{ \braket{\p}{\Q} }
}
{
{1}
\over
{ \bra{\p}1,\ldots,n\ket{\q} }
}
\cr & \qquad\qquad\times
\bar u_{\dot\alpha}(\P)
\Biggl[
{
{[\bar\p+\bar\Q]^{\dot\alpha\delta}}
\over
{ [\p+\Q]^2 }
}
-
{
{[\bar\p+\bar\Q+\bar\kappa(1,n)]^{\dot\alpha\delta}}
\over
{ [\p+\Q+\kappa(1,n)]^2 }
}
\Biggr]
u_{\delta}(\p).
}
\eq
$$
Momentum conservation plus straightforward spinor algebra produces
$$
\eqalign{
\amp_{2 {\rm A}} & =
{ {ig^2}\over{\sqrt2} }
(-e\sqrt2)^n \negthinspace\negthinspace\negthinspace
\permsum{1}{n}
{
{\braket{\P}{\Q} \braket{\Q}{\q} }
\over
{ \braket{\p}{\Q} }
}
{
{ {\braket{\P}{\Q}}^{*} }
\over
{ {\braket{\p}{\Q}}^{*} }
}
{
{1}
\over
{ \bra{\p}1,\ldots,n\ket{\q} }
}
\cr & \quad
+{ {ig^2}\over{\sqrt2} }
(-e\sqrt2)^n \negthinspace\negthinspace\negthinspace
\permsum{1}{n}
{
{\braket{\P}{\Q} \braket{\Q}{\q} }
\over
{ \braket{\p}{\Q}\braket{\P}{\q} }
}
{
{\braket{p}{q}}
\over
{ \bra{\p}1,\ldots,n\ket{\q} }
}.
}
\eqlabel\firstfragdone
$$

We now return to $\amp_{2 {\rm B}}$.  The form of this contribution
is such that we hope to combine it  with $\amp_1$ in some
manner.  To this end, we shift the sum on $\ell$ by 1 in the
first of the two contributions to (\secondfrag):
$$
\eqalign{
\amp_{2 {\rm B}} & =
{ -{ig^2}\over{\sqrt2} }
(-e\sqrt2)^n \negthinspace\negthinspace\negthinspace
\permsum{1}{n} \sum_{\ell=0}^{n-1}
{
{\braket{\P}{\Q} \braket{\Q}{\q}}
\over
{\bra{\p}1,\ldots,n\ket{\q} }
}
{
{\bar u_{\dot\alpha}(\P)
[\bar\kappa(\ell{+}2,n)+\bar\q]^{\dot\alpha\alpha} }
\over
{[\p+\Q+\kappa(1,\ell)]^2  }
}
{
{u_{\alpha}(k_{\ell+1})}
\over
{ \braket{\Q}{\ell{+}1} }
}
\cr & \quad
+{ {ig^2}\over{\sqrt2} }
(-e\sqrt2)^n \negthinspace\negthinspace\negthinspace
\permsum{1}{n} \sum_{\ell=1}^n
{
{\braket{\P}{\Q} \braket{\Q}{\q}}
\over
{\bra{\p}1,\ldots,n\ket{\q} }
}
{
{\bar u_{\dot\alpha}(\P)
[\bar\kappa(\ell{+}1,n)+\bar\q]^{\dot\alpha\alpha} }
\over
{[\p+\Q+\kappa(1,\ell)]^2  }
}
{
{u_{\alpha}(k_{\ell})}
\over
{ \braket{\Q}{\ell} }
}.
}
\eqlabel\ShIfTeD
$$
The Weyl equation allows us to extend the momentum sum appearing
in the first term of (\ShIfTeD) to $\kappa(\ell{+}1,n)$, matching
the second term.  We now extend both sums to the range
$\ell\in [0,n]$, and compensate.  Note that we choose to let
$k_0\equiv \p$ and $k_{n+1}\equiv \q$
respectively.
Thus, we obtain
$$
\eqalign{
\amp_{2 {\rm B}} & =
{ -{ig^2}\over{\sqrt2} }
(-e\sqrt2)^n \negthinspace\negthinspace\negthinspace
\permsum{1}{n} \sum_{\ell=0}^{n}
{
{\braket{\P}{\Q} \braket{\Q}{\q}}
\over
{\bra{\p}1,\ldots,n\ket{\q} }
}
\cr & \qquad\qquad\times
{
{\bar u_{\dot\alpha}(\P)
[\bar\kappa(\ell{+}1,n)+\bar\q]^{\dot\alpha\alpha} }
\over
{[\p+\Q+\kappa(1,\ell)]^2  }
}
\Biggl[
{
{u_{\alpha}(k_{\ell+1})}
\over
{ \braket{\Q}{\ell{+}1} }
}
-
{
{u_{\alpha}(k_{\ell})}
\over
{ \braket{\Q}{\ell} }
}
\Biggr]
\cr & \quad
+{ {ig^2}\over{\sqrt2} }
(-e\sqrt2)^n \negthinspace\negthinspace\negthinspace
\permsum{1}{n}
{
{\braket{\P}{\Q} \braket{\Q}{\q}}
\over
{\bra{\p}1,\ldots,n\ket{\q} }
}
{
{\bar u_{\dot\alpha}(\P)
\bar\q^{\dot\alpha\alpha} }
\over
{[\p+\Q+\kappa(1,n)]^2  }
}
{
{u_{\alpha}(\q)}
\over
{ \braket{\Q}{\q} }
}
\cr & \quad
-{ {ig^2}\over{\sqrt2} }
(-e\sqrt2)^n \negthinspace\negthinspace\negthinspace
\permsum{1}{n}
{
{\braket{\P}{\Q} \braket{\Q}{\q}}
\over
{\bra{\p}1,\ldots,n\ket{\q} }
}
{
{\bar u_{\dot\alpha}(\P)
[\bar\kappa(1,n)+\bar\q]^{\dot\alpha\alpha} }
\over
{[\p+\Q]^2  }
}
{
{u_{\alpha}(p)}
\over
{ \braket{\Q}{\p} }
}.
}
\eqlabel\notmuchmore
$$
The first term of (\notmuchmore) contains
$$
{
{u_{\alpha}(k_{\ell+1})}
\over
{ \braket{\Q}{\ell{+}1} }
}
-
{
{u_{\alpha}(k_{\ell})}
\over
{ \braket{\Q}{\ell} }
} =
u_{\alpha}(\Q) \link{\ell}{\Q}{\ell{+}1},
\eq
$$
where we have applied (\fierz).  The second term vanishes because
of the Weyl equation.  We use momentum conservation and a
little bit of spinor algebra to simplify the third term.
The result of these manipulations is
$$
\eqalign{
\amp_{2 {\rm B}} & =
{ -{ig^2}\over{\sqrt2} }
(-e\sqrt2)^n \negthinspace\negthinspace\negthinspace
\permsum{1}{n} \sum_{\ell=0}^{n}
{
{\braket{\P}{\Q} \braket{\Q}{\q}
\bar u_{\dot\alpha}(\P)
[\bar\P+\bar\kappa(\ell{+}1,n)+\bar\q]^{\dot\alpha\alpha}
u_{\alpha}(\Q) }
\over
{\bra{\p}1,\ldots,\ell\ket{\Q}
\bra{\Q}\ell{+}1,\ldots,n\ket{\q}
[\p+\kappa(1,\ell)+\Q]^2 }
}
\cr & \quad
-{ {ig^2}\over{\sqrt2} }
(-e\sqrt2)^n \negthinspace\negthinspace\negthinspace
\permsum{1}{n}
{
{\braket{\P}{\Q} \braket{\Q}{\q} }
\over
{ \braket{\p}{\Q} }
}
{
{ {\braket{\P}{\Q}}^{*} }
\over
{ {\braket{\p}{\Q}}^{*} }
}
{
{1}
\over
{ \bra{\p}1,\ldots,n\ket{\q} }
}.
}
\eqlabel\secondfragdone
$$
The first term of (\secondfragdone) exactly cancels the
contribution from $\amp_1$ given in (\firstterm).
The second term of (\secondfragdone) disposes of the
first term of (\firstfragdone), leaving the second
term from $\amp_{2 {\rm A}}$ as the sole surviving contribution.
Therefore, we have
$$
\amp(\p,\q;\P,\Q;1,\ldots,n) =
{ {ig^2}\over{\sqrt2} }
(-e\sqrt2)^n \negthinspace\negthinspace\negthinspace
\permsum{1}{n}
{
{\braket{\P}{\Q} \braket{\Q}{\q} }
\over
{ \braket{\p}{\Q}\braket{\P}{\q} }
}
{
{\braket{p}{q}}
\over
{ \bra{\p}1,\ldots,n\ket{\q} }
}.
\eqlabel\tripalmostdone
$$
We may apply (\dopermsum) to write this in the form
$$
\amp(\p,\q;\P,\Q;1,\ldots,n) =
{ {ig^2}\over{\sqrt2} }
(-e\sqrt2)^n
{
{\braket{\P}{\Q} \braket{\Q}{\q} }
\over
{ \braket{\p}{\Q}\braket{\P}{\q} }
}
{
{ {\braket{\p}{\q}}^n }
\over
{\prod\limits_{j=1}^n \bra{\p} j \ket{\q} }
}.
\eqlabel\tripdone
$$
\section{Summary of triple current amplitudes}

We now summarize the results of those processes we are able
to compute from three currents.  These are limited to the
production of like-helicity photons since the double-off-shell
currents are limited to that case.  We adopt the ``standard''
process to be
$$
P \thinspace Q \thinspace N_1 \thinspace N_2
\longrightarrow
\gamma\thinspace\gamma\cdots\gamma.
\eqlabel\triplestandard
$$
In (\triplestandard), $P$ is a positively-charged particle of
momentum $P$, $Q$ is a negatively-charged particle of momentum $Q$,
$N_1$ and $N_2$ are neutral particles of momenta $N_1$ and $N_2$.
The $n$ photons have momenta $k_1,\ldots,k_n$.  The results
presented below have all momenta directed into the diagrams.
The helicity labels applied to quantities in this section
reflect this convention.
Amplitudes for variants on the above process are easily obtained
from crossing symmetry.

It should be noted that not every diagram contributing to
the processes listed in this section contains three
currents.  In particular, the amplitudes involving a pair
of transversely-polarized $W$'s have at least two types
of contributions.  First, there is the three-current type
of diagram where the $W_T$ line is broken, with a double-off-shell
fermion or scalar current in the intervening space.  The
$W$'s can annihilate into a $Z$, however, and the $Z$ can
then pair-produce the fermion-antifermion or scalar pair.
Such diagrams contain only two currents.
In the neutral scalar case, there is also the possibility
of a seagull vertex that joins two $W$'s and two $\phi$'s
directly.  For the cases considered here, this vertex always
vanishes, since it contains $\Wms_{\alpha\dot\alpha}
\Wbar^{\dot\alpha\alpha}=0$.

The triple current amplitudes may be cast into the form
$$
\amp(P,Q;N_1,N_2;1^{+},\ldots,n^{+}) =
{ {ig^2}\over2 }
(-e\sqrt2)^n
F(P,Q,N_1,N_2)
{
{ {\braket{P}{Q}}^{n-1} }
\over
{\prod\limits_{j=1}^n \bra{P} j \ket{Q} }
},
\eqlabel\tripform
$$
where the scalar function $F(P,Q,N_1,N_2)$ depends on
the identities and helicities of the four particles.
The values of these helicity functions are listed in
Table\Tlabel\TRIPLES.

\noadvancetable{
{\begintable
$P$  \vb $Q$  \vb $N_1$  \vb $N_2$  \| $F(P,Q,N_1,N_2)$ \crthick
$W_{\up}^{+}$ \vb $W_{\down}^{-}$ \vb $\bar \nu_{\up}$ \vb $\nu_{\down}$\|
$2{\braket{N_1}{Q}}^2 \braket{N_2}{Q}
{\braket{N_1}{N_2}}^{-1} {\braket{P}{N_2}}^{-1}$  \cr
$W_{\down}^{+}$ \vb $W_{\up}^{-}$ \vb $\bar \nu_{\up}$ \vb $\nu_{\down}$\|
$-2{\braket{N_1}{P}}^3
{\braket{N_1}{N_2}}^{-1} {\braket{N_1}{Q}}^{-1}$ \cr
$W_{\up}^{+}$ \vb $W_{\down}^{-}$ \vb $H$ \vb $H$\|
$  \braket{P}{Q}
{ \bra{N_1}Q\ket{N_2} }
{ \bra{N_1}P\ket{N_2} }^{-1}
$ \cr
$W_{\up}^{+}$ \vb $W_{\down}^{-}$ \vb $Z_L$ \vb $H$\|
$ -i
{ \bra{N_1}Q\ket{N_2} } { \braket{N_1}{N_2} }^{-1}
\sum\limits_{\ell=1}^2
\braket{N_\ell}{Q} {\braket{P}{N_\ell}}^{-1}
$ \cr
$W_{\up}^{+}$ \vb $W_{\down}^{-}$ \vb $Z_L$ \vb $Z_L$\|
$  \braket{P}{Q}
{ \bra{N_1}Q\ket{N_2} }
{ \bra{N_1}P\ket{N_2} }^{-1}
$ \cr
$W_{\down}^{+}$ \vb $W_{\up}^{-}$ \vb $H$ \vb $H$\|
$  \braket{P}{Q}
{ \bra{N_1}P\ket{N_2} }
{ \bra{N_1}Q\ket{N_2} }^{-1}
$ \cr
$W_{\down}^{+}$ \vb $W_{\up}^{-}$ \vb $Z_L$ \vb $H$\|
$-i
\bra{N_1}P\ket{N_2} { \braket{N_1}{N_2} }^{-1}
\sum\limits_{\ell=1}^2
\braket{P}{N_\ell} {\braket{N_\ell}{Q}}^{-1}
$ \cr
$W_{\down}^{+}$ \vb $W_{\up}^{-}$ \vb $Z_L$ \vb $Z_L$\|
$  \braket{P}{Q}
{ \bra{N_1}P\ket{N_2} }
{ \bra{N_1}Q\ket{N_2} }^{-1}
$  \cr
$\bar e_{\up}$ \vb $e_{\down}$ \vb $\bar \nu_{\up}$ \vb $\nu_{\down}$\|
$- 2{\braket{N_2}{Q}}^2 \braket{P}{Q}
{\braket{P}{N_2}}^{-1} {\braket{N_1}{Q}}^{-1}$ \cr
$\bar e_{\up}$ \vb $W_L^{-}$ \vb $H$ \vb $\nu_{\down}$\|
$\sqrt2 {\braket{N_1}{N_2}} \braket{N_2}{Q} \braket{P}{Q}
{\braket{P}{N_2}}^{-1} {\braket{N_1}{Q}}^{-1} $ \cr
$\bar e_{\up}$ \vb $W_L^{-}$ \vb $Z_L$ \vb $\nu_{\down}$\|
$i\sqrt2 {\braket{N_1}{N_2}} \braket{N_2}{Q} \braket{P}{Q}
{\braket{P}{N_2}}^{-1} {\braket{N_1}{Q}}^{-1} $\cr
$W_L^{+}$ \vb $e_{\down}$ \vb $\bar \nu_{\up}$ \vb $H$\|
$\sqrt2 {\braket{P}{Q}}^2 \braket{Q}{N_2}
{\braket{P}{N_2}}^{-1} {\braket{N_1}{Q}}^{-1} $ \cr
$W_L^{+}$ \vb $e_{\down}$ \vb $\bar \nu_{\up}$ \vb $Z_L$\|
$-i\sqrt2 {\braket{P}{Q}}^2 \braket{Q}{N_2}
{\braket{P}{N_2}}^{-1} {\braket{N_1}{Q}}^{-1} $
\endtable}
}{Triple current amplitude helicity functions}

Not listed in Table \TRIPLES{ } are the combinations involving
two positive-helicity $W$'s.  These amplitudes vanish for
all-positive-helicity photon production.  Although the
amplitudes involving two {\it negative}-helicity $W$'s do
not vanish, we are unable to compute them from the available
currents.  Such amplitudes would require the simultaneous use of
$\Wms(P^{-};1^{+},\ldots,s^{+})$ and
$\Wms((t{+}1)^{+},\ldots,n^{+};Q^{-})$.  The known solution
for the first of these two
quantities requires that the gauge momentum be $g\equiv P$, while
the latter requires that $g\equiv Q$.

The amplitudes involving a pair of neutral scalars forms an
interesting series.  While Bose symmetry dictates that the
amplitude for
the production of two $H$'s or two $Z_L$'s  be symmetric under
the interchange of the momenta of the two scalars, no such
requirement holds  for the production of a one $H$ plus one $Z_L$.
Instead, the form of the Feynman rules forces the amplitude
to be {\it antisymmetric} under the interchange of $H$ and $Z_L$.
This may be traced back to the identifications of the Higgs
with the $\phi_1$ and the $Z_L$ with the $\phi_2$ of the
unbroken theory.  Only the combinations $\phi_1 \pm i\phi_2$
appear as part of the
original complex scalar doublet of hypercharge $+1$.
The extra $i$ in front of the $\phi_2$ appears in the vertices,
and leads to the antisymmetry  between $H$ and $Z_L$ just mentioned.
It is also the source of the differing relative signs between
the last two pairs of amplitudes.

Conspicuously absent from Table \TRIPLES{ } is
$$
W_L^{+} \thinspace W_L^{-} \thinspace H \thinspace H
\longrightarrow \gamma\gamma\cdots\gamma
\eqlabel\notthere
$$
as well as its relatives involving $Z_L$'s.
The reason for this is the presence of diagrams containing
the double-off-shell vector current, but contracted into
scalar vertices at both ends.  In the mixed case ($W_LHe\nu$),
it was always possible to avoid the complications caused by
the seagulls on the scalar line by making the fermion line
into the effective polarization spinor, and using the
form of the modified $W$ current to eliminate the seagulls.
If both lines are scalars, however, it is not apparent where to
begin.  In order to show that the modified $W$ current
even appears in the amplitude, it seems that one must eliminate
the seagulls first.  But eliminating the seagulls requires
knowledge of the current tying the two scalar lines together!
Hence, we have been unable to obtain expressions for the
process (\notthere).

\chapter{CONCLUSION}

We have seen how to combine the currents of
reference \cite\firstpaper\ in groups of four to produce
helicity amplitudes
for processes containing two charged lines.  An important example
of this type of process involves the production of a $W^{+}W^{-}$ pair
from an $e\bar e$ pair.  We are able to obtain expressions for
amplitudes containing photons of all the same helicity or one
differing helicity.  The latter type of
amplitude cannot be  easily obtained
from within the $U(N)$ framework of reference \cite\DY.
We have considered currents which have both ends of the charged line
off shell.  We have found that except for the case of the
double off-shell scalar
current, we cannot solve for these quantities directly.  We are able,
however, to find expressions for the combination of the spinor and
vector double off-shell currents with some other suitable factors.
These combinations occur naturally in amplitudes containing
a pair of Higgs particles, transversely-polarized $Z$-bosons, or
neutrinos.  We are limited to all like-helicity photons in
this case.

All of the cautions mentioned in the conclusion to
reference \cite\firstpaper\ apply to the results obtained here.
In particular, it is potentially difficult to square most
of the amplitudes in this paper.  A notable exception is
the set of triple current amplitudes, which may be squared
trivially.
A second potential difficulty involves the finite masses of
the particles.
In those special regions of phase space where
some of the invariants formed from pairs of momenta are
of  the same order as the neglected masses, the corrections
to the amplitudes presented here are potentially large.
A further problem lies in the fact that
all of the amplitudes presented here are
infrared divergent.  In principle, the satisfactory
treatment of the divergences involves a knowledge of loop diagrams.
Finally,  current experimental capabilities preclude the measurement
of helicity-projected amplitudes.  Thus, it would be desirable
to obtain a complete set of helicity amplitudes, in order to
be able to sum over helicities.
In spite of the difficulties, however, from
the large variety of processes  for which we
have obtained amplitudes, it is clear that the combination of
multispinors, the equivalence theorem, and recursion relations
forms a powerful tool in the
study of the high energy limit of the Standard Model.

\bigskip\bigskip
\noindent
{\chapterfont Acknowledgements:}

\medskip
\noindent
I would like to thank Professor Tung--Mow Yan for
the many useful discussions and helpful suggestions
during the course of this project.

\medskip
\noindent
This work was supported in part by the National Science
Foundation.

\appendix{MULTISPINOR CONVENTIONS}

Below we list the important results of application of
Weyl-van der Waerden spinor calculus to gauge theories.
Readers interested
in the details should refer to references \cite\DY\ and
[\ref{For a brief introduction to properties of two-component
Weyl-van der Waerden spinors, see, for example, M. F. Sohnius,
Phys. Reports {\bf 128}, 39 (1985).}].

We use the Weyl basis
$$
\gamma^\mu =
\pmatrix{  0      & \sigma^\mu \cr
         \bar\sigma^\mu &      0   \cr},
\eqlabel\weylbasis
$$
for the Dirac matrices.  In (\weylbasis), $\sigma^\mu$
and $\bar\sigma^\mu$ refer to the convenient Lorentz-covariant
grouping of the $2\times2$ Pauli matrices plus the unit matrix:
$$
\sigma^{\mu} \equiv (1, \vec\sigma),
\newlettlabel\sigmamus
$$
$$
\bar\sigma^{\mu} \equiv (1, -\vec\sigma),
\lett
$$
and satisfy the anticommutators
$$
(\bar\sigma^{\mu})^{\dot\alpha\beta}
(\sigma^{\nu})_{\beta\dot\beta}
+
(\bar\sigma^{\nu})^{\dot\alpha\beta}
(\sigma^{\mu})_{\beta\dot\beta}
=
2 g^{\mu\nu} \delta_{\dot\beta}^{\dot\alpha},
\newlettlabel\sigmaanticommutator
$$
$$
(\sigma^{\mu})_{\alpha\dot\beta}
(\bar\sigma^{\nu})^{\dot\beta\beta}
+
(\sigma^{\nu})_{\alpha\dot\beta}
(\bar\sigma^{\mu})^{\dot\beta\beta}
=
2 g^{\mu\nu} \delta_{\alpha}^{\beta}.
\lett
$$

To each Lorentz 4-vector there corresponds a rank two multispinor,
formed from the contraction of the 4-vector with $\sigma^{\mu}$:
$$
\Wms_{\alpha\dot\beta} =
{1\over{\sqrt2}}
\sigma^{\mu}_{\alpha\dot\beta} W_{\mu},
\newlettlabel\msvector
$$
$$
{\overline\Wms}^{\dot\alpha\beta} =
{1\over{\sqrt2}}
\bar\sigma_{\mu}^{\dot\alpha\beta} W^{\mu}.
\lett
$$
For the purposes of normalization, it is convenient to use a
different convention when converting momenta:
$$
k_{\alpha\dot\beta} =
\sigma^{\mu}_{\alpha\dot\beta} k_{\mu},
\newlettlabel\msmomenta
$$
$$
{\bar k}^{\dot\alpha\beta} =
\bar\sigma_{\mu}^{\dot\alpha\beta} k^{\mu}.
\lett
$$
Useful consequences of (\msmomenta) and (\sigmaanticommutator) are
$$
\bar k^{\dot\alpha\beta} k_{\beta\dot\beta}
= k^2 \delta_{\dot\beta}^{\dot\alpha},
\newlettlabel\slashsquaredidentity
$$
$$
k_{\alpha\dot\beta} \bar k^{\dot\beta\beta}
= k^2 \delta_{\alpha}^{\beta}.
\lett
$$

The spinor indices may be raised and lowered using the
2-component antisymmetric tensor:
$$
u^{\alpha} = \vareps^{\alpha\beta}u_{\beta},
\newlettlabel\raiseandlower
$$
$$
\bar v^{\dot\alpha} = \vareps^{\dot\alpha\dot\beta}\bar v_{\dot\beta},
\lett
$$
$$
\vareps^{\alpha\beta}=\vareps_{\alpha\beta},
\lett
$$
$$
\vareps^{\dot\alpha\dot\beta}=\vareps_{\dot\alpha\dot\beta},
\lett
$$
$$
\vareps_{12}=\vareps_{\dot1\dot2}=1.
\lett
$$
Many useful relations may be easily proven from the Schouten identity
$$
\delta_{\gamma}^{\alpha} \delta_{\delta}^{\beta}
-\delta_{\delta}^{\alpha} \delta_{\gamma}^{\beta}
+\vareps^{\alpha\beta} \vareps_{\gamma\delta}=0,
\eqlabel\SCHOUTEN
$$
the generator of 2-component Fierz transformations.

We denote by $u(k)$ and $\bar u(k)$ the solutions to the
2-component Weyl equations:
$$
\bar k^{\dot\alpha\beta} u_{\beta}(k) = 0,
\newlettlabel\Weylequations
$$
$$
\bar u_{\dot\beta}(k) \bar k^{\dot\beta\alpha} = 0.
\lett
$$
These two spinors are related by complex conjugation
$$
\bar u_{\dot\alpha}(k) = \bigl[u_{\alpha}(k)\bigr]^{*},
\eqlabel\complexconj
$$
and have the normalization
$$
u_{\alpha}(k) \bar u_{\dot\alpha}(k) = k_{\alpha\dot\alpha}.
\eqlabel\spinornormalization
$$
It is useful to define a scalar product
$$
\braket{1}{2} \equiv u^{\alpha}(k_1) u_{\alpha}(k_2),
\eqlabel\scalarproduct
$$
which has two elementary properties
$$
\braket{1}{2} = - \braket{2}{1},
\newlettlabel\antisymmetry
$$
$$
\braket{1}{2} {\braket{1}{2}}^{*} =2k_1\cdot k_2.
\lett
$$
Contraction of $u_{\alpha}(k_1)u_{\beta}(k_2)u^{\gamma}(k_3)
u^{\delta}(k_4)$ into (\SCHOUTEN) produces the extremely
useful relation
$$
\braket{1}{2}\braket{3}{4}
+\braket{1}{3}\braket{4}{2}
+\braket{1}{4}\braket{2}{3}
=0.
\eqlabel\veryhelpful
$$
A second relation of great utility may be derived from (\veryhelpful):
$$
{
{ \braket{1}{2} }
\over
{ \braket{1}{P} \braket{P}{2} }
}
+
{
{ \braket{2}{3} }
\over
{ \braket{2}{P} \braket{P}{3} }
}
=
{
{ \braket{1}{3} }
\over
{ \braket{1}{P} \braket{P}{3} }
}.
\eqlabel\LINK
$$
Equation (\LINK) may be used to demonstrate that
$$
\sum_{j=\ell}^{m-1}
{
{ \braket{j}{j{+}1} }
\over
{ \braket{j}{P} \braket{P}{j{+}1} }
}
=
{
{ \braket{\ell}{m} }
\over
{ \braket{\ell}{P} \braket{P}{m} }
}.
\eqlabel\SUMLINK
$$

An important structure built from the inner products is
$$
\bra{p} 1,2,\ldots,n \ket{q} \equiv
\braket{p}{1} \braket{1}{2} \cdots \braket{n}{q}.
\eqlabel\stringdef
$$
We note the following basic properties
of  $\bra{p} 1,2,\ldots,n \ket{q}$:
$$
\bra{p}\thinspace\ket{q} \equiv \braket{p}{q}
\newlett
$$
$$
\bra{p} 1,2,\ldots,j{-}1 \ket{j} \bra{j} j{+}1,j{+}2,\ldots,n \ket{q}
= \bra{p} 1,2,\ldots,n \ket{q}
\lett
$$
$$
\bra{q} n,n{-}1,\ldots,1 \ket{p} =
(-1)^{n-1} \bra{p} 1,2,\ldots,n \ket{q}.
\lett
$$
The inverse of this ``string'' of spinor inner products is
easy to sum over permutations: [\ref{M. Mangano,
Nucl. Phys. {\bf B309}, 461 (1988).}]
$$
\permsum{1}{n}
{
{1}
\over
{\bra{p} 1,2,\ldots,n \ket{q} }
}
=
{
{{\braket{p}{q}}^{n-1}}
\over
{\prod\limits_{j=1}^n \bra{p} j \ket{q} }
}.
\eq
$$

Helicities $\pm1$ for massless vector bosons may be described by
$$
\eps_{\alpha\dot\alpha}(k^{+}) \equiv
{
{ u_{\alpha}(q) \bar u_{\dot\alpha}(k) }
\over
{ \braket{k}{q} }
},
\newlettlabel\polarizations
$$
$$
\eps_{\alpha\dot\alpha}(k^{-}) \equiv
{
{ u_{\alpha}(k) \bar u_{\dot\alpha}(q) }
\over
{ {\braket{k}{q}}^{*} }
},
\lett
$$
where $q$ is any null-vector such that $k\cdot q\ne0$.  As the
choice of $q$ does not affect any physics result, we will refer
to $u(q)$ and $\bar u(q)$ as gauge spinors.  The  corresponding
polarization vectors $\eps^{\mu}(k)$ defined through (\msvector)
differ from the ``standard'' polarization vectors
$$
\vareps_{0}^{\mu}(k^{\pm}) =
\left(
0, \mp{1\over{\sqrt2}},{{-i}\over{\sqrt2}},0
\right),
\newlettlabel\standardpolarizations
$$
$$
k^{\mu}=(k,0,0,k),
\lett
$$
by a $q$-dependent phase and  gauge transformation  \cite\DY.

To save accounting for a large number of indices, an efficient
method is to initially write quantities in the usual formalism
and then convert to multispinor notation at a later stage using
the substitutions
$$
k\cdot k' =
{1\over2} \bar k^{\dot\alpha\alpha}{k'}_{\alpha\dot\alpha} =
{1\over2} k_{\alpha\dot\alpha} {\bar k}^{\prime\dot\alpha\alpha},
\newlettlabel\conversiondot
$$
$$
k\cdot\eps(k') =
{1\over{\sqrt2}} \bar k^{\dot\alpha\alpha}\eps_{\alpha\dot\alpha}(k') =
{1\over{\sqrt2}} k_{\alpha\dot\alpha}\bar\eps^{\dot\alpha\alpha}(k'),
\lett
$$
$$
\eps(k)\cdot\eps(k') =
\bar\eps^{\dot\alpha\alpha}(k)\eps_{\alpha\dot\alpha}(k') =
\eps_{\alpha\dot\alpha}(k)\bar\eps^{\dot\alpha\alpha}(k'),
\lett
$$
for Lorentz dot products and
$$
{1\over2}(1-\gamma_5)\psi \longrightarrow \psi_{\alpha},
\newlettlabel\conversionfermion
$$
$$
{1\over2}(1+\gamma_5)\psi \longrightarrow \psi^{\dot\alpha},
\lett
$$
$$
{1\over2}(1-\gamma_5)\thinspace{\not{\negthinspace\negthinspace W}}
{1\over2}(1+\gamma_5)
\longrightarrow \sqrt2 \thinspace {{\cal{W}}}_{\alpha\dot\alpha},
\lett
$$
$$
{1\over2}(1+\gamma_5)\thinspace{\not{\negthinspace\negthinspace W}}
{1\over2}(1-\gamma_5)
\longrightarrow \sqrt2 \thinspace
{{\overline{{\cal{W}}}}}^{\dot\alpha\alpha},
\lett
$$
$$
{1\over2}(1-\gamma_5)\slash{k}{1\over2}(1+\gamma_5)
\longrightarrow k_{\alpha\dot\alpha},
\lett
$$
$$
{1\over2}(1+\gamma_5)\slash{k}{1\over2}(1-\gamma_5)
\longrightarrow \bar k^{\dot\alpha\alpha},
\lett
$$
in strings of Dirac matrices.  Note the unequal treatments of momenta
versus other 4-vectors caused by the conventions (\msvector)
and (\msmomenta).

The following combination of factors forms the key building block
in the solutions to the recursion relations:
$$
\Xi(j,n)\equiv \permsum{j}{n} \prod_{\ell=j}^{n}
{
{ {\bar u}_{\dot\beta}(k_{\ell})
[\bar P + \bar \kappa(1,\ell)]^{\dot\beta\beta} u_{\beta}(g) }
\over
{ \braket{\ell}{g} \varsp [P+\kappa(1,\ell)]^2 }
}.
\eqlabel\XIdef
$$
In reference \firstpaper, we show that this expression reduces
to
$$
\Xi(j,n) =
\permsum{j}{n}
{
{[P+\kappa(1,j{-}1)]^2}
\over
{\bra{g} j,\ldots,n \ket{g}}
}
\sum_{\ell=j}^n
u^{\alpha}(g)
{{\pole}_{\alpha}}^{\beta}(P,1,2,\ldots,\ell)
u_{\beta}(g),
\eqlabel\XIidappend
$$
where
$$
{{\pole}_{\alpha}}^{\beta}(P,1,2,\ldots,\ell) \equiv
{
{(k_{\ell})_{\alpha\dot\alpha}
[\bar P + \bar \kappa(1,\ell)]^{\dot\alpha\beta}}
\over
{[P+\kappa(1,\ell{-}1)]^2
[P+\kappa(1,\ell)]^2}
}.
\eqlabel\poledef
$$
The factor $\pole$ has a pair of useful properties:
$$
[\bar P + \bar\kappa(1,j)]^{\dot\beta\alpha}
{\pole_{\alpha}}^{\beta}(P,1,\ldots,j)
=
{
{ [\bar P + \bar\kappa(1,j{-}1)]^{\dot\beta\beta} }
\over
{ [ P +  \kappa(1,j{-}1)]^2 }
} - {
{ [\bar P + \bar\kappa(1,j)]^{\dot\beta\beta} }
\over
{ [ P +  \kappa(1,j)]^2 }
}   ,
\eqlabel\splitid
$$

$$
\eqalign{
u^{\alpha}(g)
{{\pole}_{\alpha}}^{\beta}(P,1,2,\ldots,j)
u_{\beta}(h)&=
u^{\alpha}(h){\pole_{\alpha}}^{\beta}(P,1,\ldots,j)u_{\beta}(g)
\cr &
+
\braket{g}{h}
\biggl[
{
{1}
\over
{[P+\kappa(1,j{-}1)]^2}
}
-
{
{1}
\over
{[P+\kappa(1,j)]^2}
}
\biggr].
}
\eqlabel\REVERSE
$$
Both (\splitid) and (\REVERSE) contain combinations of terms which are
trivially summed over $j$.


\appendix{Cross-channel identities}

In many of the quadruple current amplitudes, there are
structures which result from particles propagating
in two different channels. They are characterized by
denominators that look like
$$
{\cal D}_1 = \bra{P}2,\ldots,s\ket{q}
           \bra{p}s{+}1,\ldots,n\ket{Q}
\eq
$$
as opposed to
$$
{\cal D}_2 = \bra{P}2,\ldots,s\ket{Q}
           \bra{p}s{+}1,\ldots,n\ket{q}.
\eq
$$
The key to the generation of identities that these two forms
is the observation that the  way in which we
labelled the photon momenta on the diagrams was dictated by convenience.
Any other labelling scheme is physically equivalent.
Hence, we may turn one type of term into the other simply
by ``undoing'' one of the sums tying the legs together, relabelling
the momenta, and redoing the sum.   Unfortunately, it does
not seem possible to start with the diagrams written in a form which
leads to the same denominators in both cases.
The gauge momentum  $g$ appears naturally as the momentum that
``links'' the denominator strings.  Unfortunately, the
linking momentum required to derive these identities is
different. It is this conflict that motivated the approach outlined
below.

A simple example  of this type of identity
may be generated from the fifth term of (\Mneutrinodone).
Let us define
$$
\eqalign{
\Lambda &\equiv
\permsum{2}{n} \sum_{s=1}^n
{
{ [P+\kappa(1,s)+q]^2 }
\over
{ \bra{P}2,\ldots,s\ket{q}
\bra{p}s{+}1,\ldots,n \ket{Q} }
}
\cr & =
\permsum{2}{n} \sum_{s=1}^n
{
{ [p+\kappa(s{+}1,n)+Q]^2 }
\over
{ \bra{P}2,\ldots,s\ket{q}
\bra{p}s{+}1,\ldots,n \ket{Q} }
}.
}
\eqlabel\firstidentitystart
$$
Since we wish to split one of the denominator strings,
making it look like a pair of unsummed currents, we use
(\slashsqr)  and the Weyl equation to write
$$
{
{ [p+\kappa(s{+}1,n)+Q]^2 }
\over
{ \bra{p}s{+}1,\ldots,n \ket{Q} }
} = {
{ u^{\alpha}(q)[p+\kappa(s{+}1,n)+Q]_{\alpha\dot\alpha}
[\bar p + \bar\kappa(s{+}1,n)]^{\dot\alpha\beta}u_{\beta}(Q) }
\over
{ \bra{p}s{+}1,\ldots,n \ket{Q} \thinspace\braket{q}{Q} }
}.
\eq
$$
Our goal is to break up $\bra{p}s{+}1,\ldots,n \ket{Q}$.  To
this end, we note that the proper endpoint momentum assignments
are ``$p$'' for ``$s$'' and ``$Q$'' for ``$n{+}1$''.  Hence
$$
{
{ [p+\kappa(s{+}1,n)+Q]^2 }
\over
{ \bra{p}s{+}1,\ldots,n \ket{Q} }
} ={
{ u^{\alpha}(q)[p+\kappa(s{+}1,n)+Q]_{\alpha\dot\alpha} }
\over
{ \bra{p}s{+}1,\ldots,n \ket{Q}  }
}
\sum_{v=s}^n k_v^{\dot\alpha\beta} u_{\beta}(Q)
{ {1}\over{\braket{q}{Q}} }.
\eq
$$
Supplying a factor of $\braket{v}{q}$ to numerator and denominator
and rearranging the factors in a suggestive manner produces
$$
{
{ [p+\kappa(s{+}1,n)+Q]^2 }
\over
{ \bra{p}s{+}1,\ldots,n \ket{Q} }
} = {
{u^{\alpha}(q)[p+\kappa(s{+}1,n)+Q]_{\alpha\dot\alpha}}
\over
{ \bra{p}s{+}1,\ldots,n \ket{Q} }
}
\sum_{v=s}^n
\bar u^{\dot\alpha}(k_v) \braket{v}{q}
\link{v}{q}{Q}.
\eq
$$
We may convert $\braket{v}{q} { \bra{v}{q}\ket{Q} }^{-1}$
into the factor required to
split the denominator by using (\linkidsummed) in reverse:
$$
{
{ [p+\kappa(s{+}1,n)+Q]^2 }
\over
{ \bra{p}s{+}1,\ldots,n \ket{Q} }
} = {
{u^{\alpha}(q)[p+\kappa(s{+}1,n)+Q]_{\alpha\dot\alpha}}
\over
{ \bra{p}s{+}1,\ldots,n \ket{Q} }
}
\sum_{v=s}^n \sum_{t=v}^{n}
k_v^{\dot\alpha\beta}u_{\beta}(q)
\link{t}{q}{t{+}1}.
\eq
$$
Interchanging the order of the sums, we see that
$$
{
{ [p+\kappa(s{+}1,n)+Q]^2 }
\over
{ \bra{p}s{+}1,\ldots,n \ket{Q} }
} = \sum_{t=s}^n
{
{ u^{\alpha}(q)[\kappa(t{+}1,n)+Q]_{\alpha\dot\alpha}
[\bar p + \bar\kappa(s{+}1,t)]^{\dot\alpha\beta}u_{\beta}(q) }
\over
{ \bra{p}s{+}1,\ldots,t\ket{q}
\bra{q}t{+}1,\ldots,n \ket{Q} }
},
\eqlabel\expandsquare
$$
where we have used (\slashsqr) to shorten the momentum sum
in the first factor.  Hence, we may write
$$
\eqalign{
\Lambda =
\permsum{2}{n} \sum_{s=1}^n  \sum_{t=s}^n
{
{ u^{\alpha}(q)[\kappa(t{+}1,n)+Q]_{\alpha\dot\alpha}
[\bar p + \bar\kappa(s{+}1,t)]^{\dot\alpha\beta}u_{\beta}(q) }
\over
{ \bra{P}2,\ldots,s\ket{q}
\bra{p}s{+}1,\ldots,t\ket{q}
\bra{q}t{+}1,\ldots,n \ket{Q} }
}.
}
\eq
$$
Since any set of labels for the dummy momenta appearing inside
the permutation sum is equivalent to any other, we choose to
relabel the momenta as follows:
$$
\eqalign{
\{t{+}1,\ldots,n\} &\longrightarrow \{s{+}1,\ldots,s{+}n{-}t\},
\cr
\{s{+}1,\ldots,t\} &\longrightarrow  \{s{+}n{-}t{+}1,\ldots,n\},
}
\eqlabel\relabelONE
$$
producing
$$
\eqalign{
\Lambda =
\permsum{2}{n} \sum_{s=1}^n  \sum_{t=s}^n
{
{ u^{\alpha}(q)[\kappa(s{+}1,s{+}n{-}t)+Q]_{\alpha\dot\alpha}
[\bar p + \bar\kappa(s{+}n{-}t{+}1,n)]^{\dot\alpha\beta}u_{\beta}(q) }
\over
{ \bra{P}2,\ldots,s\ket{q}
\bra{q}s{+}1,\ldots,s{+}n{-}t \ket{Q}
\bra{p}s{+}n{-}t{+}1,\ldots,t\ket{q} }
}.
}
\eq
$$
The replacement of $t$ by $r\equiv s+n-t$ immediately suggests
itself:
$$
\eqalign{
\Lambda =
\permsum{2}{n} \sum_{r=1}^n  \sum_{s=1}^r
{
{ u^{\alpha}(q)[\kappa(s{+}1,r)+Q]_{\alpha\dot\alpha}
[\bar p + \bar\kappa(r{+}1,n)]^{\dot\alpha\beta}u_{\beta}(q) }
\over
{ \bra{P}2,\ldots,s\ket{q}
\bra{q}s{+}1,\ldots,r \ket{Q}
\bra{p}r{+}1,\ldots,n\ket{q} }
}.
}
\eqlabel\waytogo
$$
At this stage, we use the Weyl to add $\bar q^{\dot\alpha\beta}$
to the second momentum factor in (\waytogo), causing it
to read
$$
[\bar p + \bar\kappa(r{+}1,n)+ \bar q]^{\dot\alpha\beta}
= - [\bar P + \bar\kappa(1,r)+ \bar Q]^{\dot\alpha\beta}.
\eq
$$
Next, we perform the sum on $s$ appearing in (\waytogo).
Extracting the relevant factors, we have
$$
\sigma_s \equiv \sum_{s=1}^r \sum_{w=s+1}^{r+1}
\link{s}{q}{s{+}1}
u^{\alpha}(q)(k_w)_{\alpha\dot\alpha},
\eq
$$
where ``$1$'' corresponds to ``$P$'' and ``$r{+}1$'' corresponds
to ``$Q$''.  The sum is easily performed using (\linkidsummed),
yielding
$$
\sigma_s =
{ {1}\over{\braket{P}{q}} }
u^{\alpha}(P)[P+\kappa(2,r)+Q]_{\alpha\dot\alpha}.
\eqlabel\shostakovich
$$
To get the final form, we add and subtract $(k_1)_{\alpha\dot\alpha}$
in (\shostakovich).  Inserting the result into (\waytogo) produces
$$
\eqalign{
\Lambda &=
-\permsum{2}{n} \sum_{r=1}^n
{
{ [P+\kappa(1,r)+Q]^2 }
\over
{ \bra{P}2,\ldots,r \ket{Q}
\bra{p}r{+}1,\ldots,n\ket{q} }
}
\cr & \quad +
\permsum{2}{n} \sum_{r=1}^n
{
{ \braket{P}{1} }
\over
{ \braket{P}{q} }
}
{
{ \bar u_{\dot\alpha}(k_1)
[\bar P + \bar\kappa(1,r)+\bar Q]^{\dot\alpha\beta}u_{\beta}(q) }
\over
{ \bra{P}2,\ldots,r \ket{Q}
\bra{p}r{+}1,\ldots,n\ket{q} }
},
}
\eq
$$
which is the desired identity.

It is somewhat more difficult to derive identities of this ilk
when the terms being targeted contain sums over $\pole$.
Simply put, the sum involving the various $\pole$'s gets
tangled up with the sum introduced for the purpose of splitting
the denominator.  To see how this works, consider
$$
\eqalign{
\curlyZ & \equiv
\permsum{2}{n} \sum_{s=1}^n
{
{ [P+\kappa(1,s)+q]^2 }
\over
{ \bra{P}2,\ldots,s\ket{q} \bra{p}s{+}1,\ldots,n \ket{Q}}
}
\cr & \quad\qquad\qquad\times
\sum_{j=2}^{s+1}
u^{\gamma}(k_1){{\pole}_{\gamma}}^{\delta}(P,1,2,\ldots,j)
u_{\delta}(k_1)
{\biggl\vert}_{j=s+1\equiv q}.
}
\eq
$$
We begin isolating the $j=s+1$ contribution to $\curlyZ$ and
setting it aside:
$$
\eqalign{
\curlyZ_1 & \equiv\negthinspace\negthinspace
\permsum{2}{n} \sum_{s=1}^n
{
{ [P+\kappa(1,s)+q]^2  \thinspace
u^{\gamma}(k_1){{\pole}_{\gamma}}^{\delta}(P,1,2,\ldots,s,q)
u_{\delta}(k_1)}
\over
{ \bra{P}2,\ldots,s\ket{q} \bra{p}s{+}1,\ldots,n \ket{Q}}
}.
}
\eqlabel\zeeone
$$
We apply momentum conservation to the remainder to obtain
$[p+\kappa(s{+}1,n)+Q]^2$ in the numerator and use
(\expandsquare) to write
$$
\eqalign{
\curlyZ_2 & =\negthinspace\negthinspace
\permsum{2}{n} \sum_{s=2}^n \sum_{t=s}^n
{
{ u^{\alpha}(q)[\kappa(t{+}1,n)+Q]_{\alpha\dot\alpha}
[\bar p + \bar\kappa(s{+}1,t)]^{\dot\alpha\beta}u_{\beta}(q) }
\over
{ \bra{P}2,\ldots,s\ket{q}\bra{p}s{+}1,\ldots,t\ket{q}
\bra{q}t{+}1,\ldots,n \ket{Q} }
}
\cr & \quad\qquad\qquad\times
\sum_{j=2}^s
u^{\gamma}(k_1){{\pole}_{\gamma}}^{\delta}(P,1,2,\ldots,j)
u_{\delta}(k_1).
}
\eqlabel\zeetwostart
$$
We now use (\relabelONE) to relabel the momenta appearing in
(\zeetwostart) and make the variable change, $r=s+n-t$:
$$
\eqalign{
\curlyZ_2 & =\negthinspace\negthinspace
\permsum{2}{n} \sum_{s=2}^n \sum_{r=s}^n
{
{ u^{\alpha}(q)[\kappa(s{+}1,r)+Q]_{\alpha\dot\alpha}
[\bar p + \bar\kappa(r{+}1,n)+\bar q]^{\dot\alpha\beta}u_{\beta}(q) }
\over
{ \bra{P}2,\ldots,s\ket{q}
\bra{q}s{+}1,\ldots,r \ket{Q} \bra{p}r{+}1,\ldots,n\ket{q}}
}
\cr & \quad\qquad\qquad\times
\sum_{j=2}^s
u^{\gamma}(k_1){{\pole}_{\gamma}}^{\delta}(P,1,2,\ldots,j)
u_{\delta}(k_1).
}
\eq
$$
Note that since for the range  of $j$ and $s$ considered,
$\{2,\ldots,j\} \subseteq \{2,\ldots,s\}$,
there has been  no change in the arguments of $\pole$.  In preparation
for doing the sum on $s$, we transpose the order of the matrix
multiplication and re-order the sums:
$$
\eqalign{
\curlyZ_2 & =-\negthinspace\negthinspace
\permsum{2}{n} \sum_{r=2}^n \sum_{j=2}^r \sum_{s=j}^r
{
{ u^{\beta}(q)[p + \kappa(r{+}1,n)+ q]_{\beta\dot\alpha}
[\bar \kappa(s{+}1,r)+\bar Q]^{\dot\alpha\alpha} u_{\alpha}(q) }
\over
{ \bra{P}2,\ldots,s\ket{q}
\bra{q}s{+}1,\ldots,r \ket{Q} \bra{p}r{+}1,\ldots,n\ket{q}}
}
\cr & \quad\qquad\qquad\times
u^{\gamma}(k_1){{\pole}_{\gamma}}^{\delta}(P,1,2,\ldots,j)
u_{\delta}(k_1).
}
\eqlabel\theproblemunmasked
$$
At this stage we see clearly that even though we tried to keep
$\pole$ out of this process, the appearance of $j$ as one of
the limits in the sum on $s$ will force it to become involved.
The sum required by (\theproblemunmasked)
is easily performed using (\linkidsummed):
$$
\sum_{s=j}^r \sum_{w=j+1}^{r+1}
\link{s}{q}{s{+}1}
\bar k_w^{\alpha\dot\alpha}u_{\alpha}(q) =
{ {-1}\over{\braket{q}{j}} }
[\bar\kappa(j{+}1,r)+\bar Q]_{\alpha\dot\alpha}u^{\alpha}(k_j).
\eqlabel\dmitri
$$
Combining (\dmitri) with (\theproblemunmasked) results in
$$
\eqalign{
\curlyZ_2 & =\negthinspace\negthinspace
\permsum{2}{n} \sum_{r=2}^n \sum_{j=2}^r
{
{ u^{\beta}(q)[p + \kappa(r{+}1,n)+ q]_{\beta\dot\alpha}
[\bar \kappa(j{+}1,r)+\bar Q]^{\dot\alpha\alpha}  }
\over
{ \bra{P}2,\ldots,r \ket{Q} \bra{p}r{+}1,\ldots,n\ket{q}}
}
\cr & \quad\qquad\qquad\times
{{\pole}_{\alpha}}^{\delta}(P,1,2,\ldots,j)
u_{\delta}(k_1)
{
{ \braket{1}{j} }
\over
{ \braket{q}{j} }
},
}
\eq
$$
where we have chosen to extract the implicit factor of $\braket{1}{j}$
from the $\pole$-structure and contract the momentum sum into
$\pole$ instead.

At this stage, we break $\curlyZ_2$ into two contributions by
writing
$$
\eqalign{
u^{\beta}(q)&[p + \kappa(r{+}1,n)+ q]_{\beta\dot\alpha}
[\bar \kappa(j{+}1,r)+\bar Q]^{\dot\alpha\alpha}  =
\cr & =
u^{\beta}(q)[p + \kappa(r{+}1,n)+ q]_{\beta\dot\alpha}
[\bar P + \bar\kappa(1,r)+\bar Q-\bar P
     -\bar\kappa(1,j)]^{\dot\alpha\alpha}
\cr & =
-[P+\kappa(1,r)+Q]^2 \thinspace u^{\alpha}(q)
\cr & \quad\enspace
+ u^{\beta}(q)[P+\kappa(1,r)+Q]_{\beta\dot\alpha}
[\bar P +\bar\kappa(1,j)]^{\dot\alpha\alpha},
}
\eqlabel\bReAkUp
$$
where we have used momentum conservation and (\slashsqr).
The first term in (\bReAkUp) produces
$$
\eqalign{
\curlyZ_{2 {\rm A}} & \equiv-\negthinspace\negthinspace
\permsum{2}{n} \sum_{r=2}^n
{
{ [P + \kappa(1,r) + Q ]^2  }
\over
{ \bra{P}2,\ldots,r \ket{Q} \bra{p}r{+}1,\ldots,n\ket{q}}
}
\cr & \quad\qquad\qquad\times
\sum_{j=2}^r
u^{\alpha}(k_1)
{{\pole}_{\alpha}}^{\delta}(P,1,2,\ldots,j)
u_{\delta}(k_1),
}
\eqlabel\zeetwoa
$$
which we set aside for the moment.  The rest of (\bReAkUp) gives
$$
\eqalign{
\curlyZ_{2 {\rm B}} & \equiv\negthinspace\negthinspace
\permsum{2}{n} \sum_{r=2}^n \sum_{j=2}^r
{
{ \braket{1}{j} }
\over
{ \braket{q}{j} }
}
{
{ u^{\beta}(q)[P + \kappa(1,r) + Q]_{\beta\dot\alpha}  }
\over
{ \bra{P}2,\ldots,r \ket{Q} \bra{p}r{+}1,\ldots,n\ket{q}}
}
\cr & \quad\qquad\qquad\times
[\bar P +\bar\kappa(1,j)]^{\dot\alpha\alpha}
{{\pole}_{\alpha}}^{\delta}(P,1,2,\ldots,j)
u_{\delta}(k_1).
}
\eqlabel\zeetwobstart
$$

We apply (\splitid) to (\zeetwobstart).  We cannot do the sum
over $j$ in this case because of the
factor $\braket{1}{j}/\braket{q}{j}$
appearing in (\zeetwobstart).  Instead, we must write
$$
\eqalign{
\curlyZ_{2 {\rm B}} & =\negthinspace\negthinspace
\permsum{2}{n} \sum_{r=2}^n \sum_{j=2}^r
{
{ \braket{1}{j} }
\over
{ \braket{q}{j} }
}
{
{ u^{\beta}(q)[P + \kappa(1,r) + Q]_{\beta\dot\alpha}  }
\over
{ \bra{P}2,\ldots,r \ket{Q} \bra{p}r{+}1,\ldots,n\ket{q}}
}
\cr & \quad\qquad\qquad\times
\Biggl[
{
{ [\bar P+\bar\kappa(1,j{-}1)]^{\dot\alpha\delta} u_{\delta}(k_1) }
\over
{ [P + \kappa(1,j{-}1)]^2 }
}
-
{
{ [\bar P+\bar\kappa(1,j)]^{\dot\alpha\delta} u_{\delta}(k_1) }
\over
{ [P + \kappa(1,j)]^2 }
}
\Biggr].
}
\eqlabel\zeetwobsplit
$$
Let us shift the sum over $j$ by one unit in the first of the two
terms in (\zeetwobsplit):
$$
\eqalign{
\curlyZ_{2 {\rm B}} & =\negthinspace\negthinspace
\permsum{2}{n} \sum_{r=2}^n \sum_{j=1}^{r-1}
{
{ \braket{1}{j{+}1} }
\over
{ \braket{q}{j{+}1} }
}
{
{ u^{\beta}(q)[P + \kappa(1,r) + Q]_{\beta\dot\alpha}
[\bar P+\bar\kappa(1,j)]^{\dot\alpha\delta} u_{\delta}(k_1)  }
\over
{ \bra{P}2,\ldots,r \ket{Q} \bra{p}r{+}1,\ldots,n\ket{q}
[P + \kappa(1,j)]^2}
}
\cr & \quad
-\permsum{2}{n} \sum_{r=2}^n \sum_{j=2}^{r}
{
{ \braket{1}{j} }
\over
{ \braket{q}{j} }
}
{
{ u^{\beta}(q)[P + \kappa(1,r) + Q]_{\beta\dot\alpha}
[\bar P+\bar\kappa(1,j)]^{\dot\alpha\delta} u_{\delta}(k_1)  }
\over
{ \bra{P}2,\ldots,r \ket{Q} \bra{p}r{+}1,\ldots,n\ket{q}
[P + \kappa(1,j)]^2}
}.
}
\eqlabel\zeetwobmismatch
$$
It is easy to show from (\fierz) that
$$
{
{ \braket{1}{j{+}1} }
\over
{ \braket{q}{j{+}1} }
} - {
{ \braket{1}{j} }
\over
{ \braket{q}{j} }
} =
\braket{1}{q} \link{j}{q}{j{+}1} .
\eq
$$
This combination will appear in (\zeetwobmismatch) if we make
the ranges of the summations be identical.  In addition to
extending the range of $j$ in both terms, we also add $r=1$:
$$
\eqalign{
\curlyZ_{2 {\rm B}} & =\negthinspace\negthinspace\negthinspace
\permsum{2}{n} \sum_{r=1}^n \sum_{j=1}^{r} \negthinspace
{
{ \braket{1}{q} u^{\beta}(q)[P + \kappa(1,r) + Q]_{\beta\dot\alpha}
[\bar P+\bar\kappa(1,j)]^{\dot\alpha\delta} u_{\delta}(k_1)  }
\over
{ \bra{P}2,\ldots,j\ket{q}
\bra{q}j{+}1,\ldots,r \ket{Q} \bra{p}r{+}1,\ldots,n\ket{q}
[P + \kappa(1,j)]^2}
}
\cr & \quad
-\permsum{2}{n} \sum_{r=1}^n
{
{ \braket{1}{Q} }
\over
{ \braket{q}{Q} }
}
{
{ u^{\beta}(q)[P + \kappa(1,r) + Q]_{\beta\dot\alpha}
[\bar P+\bar\kappa(1,r)]^{\dot\alpha\delta} u_{\delta}(k_1)  }
\over
{ \bra{P}2,\ldots,r \ket{Q} \bra{p}r{+}1,\ldots,n\ket{q}
[P + \kappa(1,r)]^2}
}
\cr & \quad
+\permsum{2}{n} \sum_{r=1}^n
{
{ \braket{1}{P} }
\over
{ \braket{q}{P} }
}
{
{ u^{\beta}(q)[P + \kappa(1,r) + Q]_{\beta\dot\alpha}
[\bar P+\bar k_1]^{\dot\alpha\delta} u_{\delta}(k_1)  }
\over
{ \bra{P}2,\ldots,r \ket{Q} \bra{p}r{+}1,\ldots,n\ket{q}
[P + k_1]^2}
}.
}
\eqlabel\noidea
$$
Notice that we have chosen the values of the endpoint terms
to be ``$P$'' for $j=1$ and ``$Q$'' for $j=r{+}1$.  This allowed
us to write
$$
{
{1}
\over
{\bra{P}2,\ldots,r\ket{Q}}
}
\link{j}{q}{j{+}1}
=
{
{1}
\over
{\bra{P}2,\ldots,j\ket{q} \bra{q}j{+}1,\ldots,r\ket{Q} }
}
\eq
$$
throughout the entire summation range in the first term of (\noidea).

We now concentrate on the first term of (\noidea), which we will
refer to as $\curlyZ_{2 {\rm B}1}$.  Momentum conservation plus
the Weyl equation
permits us to write
$$
u^{\beta}(q)[P + \kappa(1,r) + Q]_{\beta\dot\alpha} =
-u^{\beta}(q)[p + \kappa(r{+}1,n) ]_{\beta\dot\alpha}
\eq
$$
in the numerator.  At this stage,
we relabel the momenta once more:
$$
\eqalign{
\{r{+}1,\ldots,n\} &\longrightarrow \{j{+}1,\ldots,j{+}n{-}r\},
\cr
\{j{+}1,\ldots,r\} &\longrightarrow  \{j{+}n{-}r{+}1,\ldots,n\}.
}
\eqlabel\relabelUNDO
$$
This has the effect of restoring the original
ordering of the denominator
strings, as is readily apparent after eliminating $r$ in favor
of $i\equiv j+n-r$:
$$
\eqalign{
\curlyZ_{2 {\rm B}1} &\negthinspace =\negthinspace\negthinspace
\negthinspace
\permsum{2}{n} \sum_{j=1}^n \sum_{i=j}^{n}
\negthinspace
{
{ -\braket{1}{q} u^{\beta}(q)[p+\kappa(j{+}1,i)]_{\beta\dot\alpha}
[\bar P+\bar\kappa(1,j)]^{\dot\alpha\delta} u_{\delta}(k_1)  }
\over
{ \bra{P}2,\ldots,j\ket{q}\bra{p}j{+}1,\ldots,i\ket{q}
\bra{q}i{+}1,\ldots,n \ket{Q}
[P + \kappa(1,j)]^2}
}.
}
\eqlabel\zeetwobonestart
$$
Letting $i=j$ stand for $p$ and $i=n+1$ stand for $Q$ gives
us the following for the sum over $i$:
$$
\eqalign{
\sum_{i=j}^n \sum_{\ell=j}^i &
\link{i}{q}{i{+}1}
u^{\beta}(q)(k_{\ell})_{\beta\dot\alpha}
[\bar P+\bar\kappa(1,j)]^{\dot\alpha\delta} u_{\delta}(k_1)
\cr & =
\sum_{\ell=j}^n
\link{\ell}{q}{Q}
\braket{q}{\ell}
\bar u_{\dot\alpha}(k_\ell)
[\bar P+\bar\kappa(1,j)]^{\dot\alpha\delta} u_{\delta}(k_1)
\cr & =
{ {1}\over{\braket{q}{Q}} }
u^{\beta}(Q)[p+\kappa(j{+}1,n)]_{\beta\dot\alpha}
[\bar P+\bar\kappa(1,j)]^{\dot\alpha\delta} u_{\delta}(k_1)
\cr & =
{- {1}\over{\braket{q}{Q}} }
u^{\beta}(Q)[P+\kappa(1,j)+q]_{\beta\dot\alpha}
[\bar P+\bar\kappa(1,j)]^{\dot\alpha\delta} u_{\delta}(k_1)
\cr & =
-{ {\braket{Q}{1}} \over {\braket{q}{Q}} }
[P+\kappa(1,j)]^2
-
{ {1}\over{\braket{q}{Q}} }
u^{\beta}(Q)q_{\beta\dot\alpha}
[\bar P+\bar\kappa(1,j)]^{\dot\alpha\delta} u_{\delta}(k_1)
}
\eqlabel\SuMonI
$$
where we have made free use of momentum conservation, the
Weyl equation, and (\slashsqr).  Applying (\SuMonI) to
(\zeetwobonestart) gives us
$$
\eqalign{
\curlyZ_{2 {\rm B}1} & =\negthinspace
\negthinspace
\permsum{2}{n} \sum_{j=1}^n
\invlink{q}{1}{Q}
{
{ 1  }
\over
{ \bra{P}2,\ldots,j\ket{q}\bra{p}j{+}1,\ldots,n \ket{Q} }
}
\cr &  \quad -\negthinspace\negthinspace
\permsum{2}{n} \sum_{j=1}^n
{
{  u^{\beta}(k_1)q_{\beta\dot\alpha}
[\bar P+\bar\kappa(1,j)]^{\dot\alpha\delta} u_{\delta}(k_1)  }
\over
{ \bra{P}2,\ldots,j\ket{q}\bra{p}j{+}1,\ldots,n \ket{Q}
[P + \kappa(1,j)]^2}
}.
}
\eqlabel\zeetwobonealmostdone
$$
Supplying a factor of $[P+\kappa(1,j)+q]^2$ to the numerator
and denominator of  the second term in (\zeetwobonealmostdone)
allows us to recognize the presence of a factor of $\pole$:
$$
\eqalign{
\curlyZ_{2 {\rm B}1} & =\negthinspace
\negthinspace
\permsum{2}{n} \sum_{j=1}^n
\invlink{q}{1}{Q}
{
{ 1  }
\over
{ \bra{P}2,\ldots,j\ket{q}\bra{p}j{+}1,\ldots,n \ket{Q} }
}
\cr &  \quad -\negthinspace\negthinspace
\permsum{2}{n} \sum_{j=1}^n
{
{ [P+\kappa(1,j)+q]^2 \thinspace
u^{\gamma}(k_1){{\pole}_{\gamma}}^{\delta}(P,1,2,\ldots,j,q)
u_{\delta}(k_1) }
\over
{ \bra{P}2,\ldots,j\ket{q} \bra{p}j{+}1,\ldots,n \ket{Q}}
}.
}
\eqlabel\zeetwobone
$$
Thus, the point of the manipulations on $\curlyZ_{2 {\rm B}1}$
becomes apparent, for the second term of (\zeetwobone)
exactly cancels the contribution from $\curlyZ_1$,
as a comparison with (\zeeone) readily attests.

We now return to the two remaining terms of (\noidea):
$$
\eqalign{
\curlyZ_{2 {\rm B}2} & \equiv
-\negthinspace\negthinspace
\permsum{2}{n} \sum_{r=1}^n
{
{ \braket{1}{Q} }
\over
{ \braket{q}{Q} }
}
{
{ u^{\beta}(q)[P + \kappa(1,r) + Q]_{\beta\dot\alpha}
[\bar P+\bar\kappa(1,r)]^{\dot\alpha\delta} u_{\delta}(k_1)  }
\over
{ \bra{P}2,\ldots,r \ket{Q} \bra{p}r{+}1,\ldots,n\ket{q}
[P + \kappa(1,r)]^2}
}
\cr & \quad
+\negthinspace\negthinspace
\permsum{2}{n} \sum_{r=1}^n
{
{ \braket{1}{P} }
\over
{ \braket{q}{P} }
}
{
{ u^{\beta}(q)[P + \kappa(1,r) + Q]_{\beta\dot\alpha}
\bar P^{\dot\alpha\delta} u_{\delta}(k_1)  }
\over
{ \bra{P}2,\ldots,r \ket{Q} \bra{p}r{+}1,\ldots,n\ket{q}
[P + k_1]^2}
}.
}
\eqlabel\ABC
$$
Application of (\slashsqr) to the first term of (\ABC) yields
$$
\eqalign{
\curlyZ_{2 {\rm B}2} & =
-\negthinspace\negthinspace
\permsum{2}{n} \sum_{r=1}^n
\invlink{q}{1}{Q}
{
{ 1  }
\over
{ \bra{P}2,\ldots,r \ket{Q} \bra{p}r{+}1,\ldots,n\ket{q} }
}
\cr & \quad
-\negthinspace\negthinspace
\permsum{2}{n} \sum_{r=1}^n
{
{ \braket{1}{Q} }
\over
{ \braket{q}{Q} }
}
{
{ u^{\beta}(q)Q_{\beta\dot\alpha}
[\bar P+\bar\kappa(1,r)]^{\dot\alpha\delta} u_{\delta}(k_1)  }
\over
{ \bra{P}2,\ldots,r \ket{Q} \bra{p}r{+}1,\ldots,n\ket{q}
[P + \kappa(1,r)]^2}
}
\cr & \quad
+\negthinspace\negthinspace
\permsum{2}{n} \sum_{r=1}^n
{
{ \braket{1}{P} }
\over
{ \braket{q}{P} }
}
{
{ \bar u_{\dot\alpha}(P)
[\bar P + \bar\kappa(1,r) +\bar  Q]^{\dot\alpha\beta}
u_{\beta}(q)  }
\over
{ \bra{P}2,\ldots,r \ket{Q} \bra{p}r{+}1,\ldots,n\ket{q}}
}
{
{1}
\over
{ { \braket{P}{1} }^{*} }
}.
}
\eqlabel\DEF
$$
A factor of $[P+\kappa(1,r)+Q]^2$ supplied to the numerator and
denominator of the second term of (\DEF) converts this to
$$
\eqalign{
\curlyZ_{2 {\rm B}2} & =
-\negthinspace\negthinspace
\permsum{2}{n} \sum_{r=1}^n
\invlink{q}{1}{Q}
{
{ 1  }
\over
{ \bra{P}2,\ldots,r \ket{Q} \bra{p}r{+}1,\ldots,n\ket{q} }
}
\cr & \quad
-\negthinspace\negthinspace
\permsum{2}{n} \sum_{r=1}^n
{
{ [P+\kappa(1,r)+Q]^2 \thinspace
u^{\gamma}(k_1){{\pole}_{\gamma}}^{\delta}(P,1,2,\ldots,j,Q)
u_{\delta}(k_1)  }
\over
{ \bra{P}2,\ldots,r \ket{Q} \bra{p}r{+}1,\ldots,n\ket{q}}
}
\cr & \quad
+\negthinspace\negthinspace
\permsum{2}{n} \sum_{r=1}^n
{
{ \braket{1}{P} }
\over
{ \braket{q}{P} }
}
{
{ \bar u_{\dot\alpha}(P)
[\bar P + \bar\kappa(1,r) +\bar Q]^{\dot\alpha\beta}
u_{\beta}(q)  }
\over
{ \bra{P}2,\ldots,r \ket{Q} \bra{p}r{+}1,\ldots,n\ket{q}}
}
{
{1}
\over
{ { \braket{P}{1} }^{*} }
}.
}
\eqlabel\zeetwobtwo
$$
Comparing  to (\zeetwoa), we
see that  the second term of (\zeetwobtwo) is precisely
the contribution required to add a $j=r+1\equiv Q$ term
to $\curlyZ_{2 {\rm A}}$.

We now combine (\zeeone), (\zeetwoa),
(\zeetwobone) and (\zeetwobtwo) to obtain the final result:
$$
\eqalign{
\curlyZ & =
-\negthinspace\negthinspace
\permsum{2}{n} \sum_{r=1}^n
{
{ [P + \kappa(1,r) + Q ]^2  }
\over
{ \bra{P}2,\ldots,r \ket{Q} \bra{p}r{+}1,\ldots,n\ket{q}}
}
\cr & \quad\qquad\qquad\times
\sum_{j=2}^{r+1}
u^{\alpha}(k_1)
{{\pole}_{\alpha}}^{\delta}(P,1,2,\ldots,j)
u_{\delta}(k_1)
{\biggl\vert}_{j=r+1\equiv Q}
\cr & \quad
+ \negthinspace\negthinspace
\permsum{2}{n} \sum_{r=1}^n
\invlink{q}{1}{Q}
\Biggl[
{
{ 1  }
\over
{ \bra{P}2,\ldots,r \ket{q} \bra{p}r{+}1,\ldots,n\ket{Q} }
}
\cr & \qquad\qquad\qquad\qquad\qquad
-
{
{ 1  }
\over
{ \bra{P}2,\ldots,r \ket{Q} \bra{p}r{+}1,\ldots,n\ket{q} }
}
\Biggr]
\cr & \quad
+\negthinspace\negthinspace
\permsum{2}{n} \sum_{r=1}^n
{
{ \braket{P}{1} }
\over
{ \braket{P}{q} }
}
{
{ \bar u_{\dot\alpha}(P)
[\bar P + \bar \kappa(1,r) +\bar  Q]^{\dot\alpha\beta}
u_{\beta}(q)  }
\over
{ \bra{P}2,\ldots,r \ket{Q} \bra{p}r{+}1,\ldots,n\ket{q}}
}
{
{1}
\over
{ { \braket{P}{1} }^{*} }
}.
}
\eqlabel\zeedone
$$
Although (\zeedone) looks unwieldy,  large reductions
result when it is applied to actual amplitudes.   Typically, the
first term will cancel a corresponding contribution from
the uncrossed graphs. The remaining ``fragments'' may be
neatly reduced when all of the  pieces of the amplitude are combined.
Unfortunately, it is not always possible to produce a suitable
identity when $\pole$ is present:  it is very important to
have the ``right'' spinors
contracted into $\pole$.

\vfill\eject

\global \chap =1
\vfill \eject \vglue .2in
\centerline {\headingfont REFERENCES}
\vglue .5in\baselineskip =\single
\parindent =16pt \parskip =\single
\item {1.}C. Dunn and T.--M. Yan, Nucl.
Phys. {\fam \bffam \twelvbf B352},
402 (1991). \hfill \par
\item {2.}G. Mahlon and T.--M. Yan, Cornell preprint
CLNS 91/1119 (1992). \hfill \par
\item {3.}F. A. Berends and W. T. Giele, Nucl. Phys.
{\fam \bffam \twelvbf B306}, 759 (1988). \hfill \par
\item {4.}S. L. Glashow, Nucl.
Phys. {\fam \bffam \twelvbf 22}, 579 (1961);
S. Weinberg, Phys. Rev. Lett. {\fam \bffam \twelvbf 19}, 1264 (1967);
A. Salam in {\fam \itfam \twelvit Proc. 8th Nobel Symposium,
Aspen{\accent "7F a}sgarden,} edited by N. Svartholm,
(Almqvist and Wiksell, Stockholm, 1968), p. 367. \hfill \par
\item {5.}J. M. Cornwall, D. N. Levin, and G. Tiktopoulous, Phys. Rev.
{\fam \bffam \twelvbf D10}, 1145 (1974); B. W. Lee, C. Quigg, and
H. Thacker, Phys. Rev. {\fam \bffam \twelvbf D16}, 1519 (1977);
M. S. Chanowitz and M. K. Gaillard, Nucl. Phys.
{\fam \bffam \twelvbf B261}, 379 (1985); G. J. Gounaris, R. Kogerler,
and H. Neufeld, Phys. Rev.
{\fam \bffam \twelvbf D34}, 3257 (1986). \hfill \par
\item {6.}J. Schwinger, {\fam \itfam \twelvit Particles, Sources
and Fields,} (Addison-Wesley, Redwood City, 1970), Vol. I; Ann. Phys.
{\fam \bffam \twelvbf 119}, 192 (1979). \hfill \par
\item {7.} The spinor technique was first introduced by the CALCUL
collaboration, in the context of massless Abelian gauge theory:
P. De Causmaecker, R. Gastmans, W. Troost, and T.T. Wu, Phys. Lett.
{\fam \bffam \twelvbf 105B}, 215 (1981);
P. De Causmaecker, R. Gastmans,
W. Troost, and T.T. Wu, Nucl. Phys.
{\fam \bffam \twelvbf B206}, 53 (1982);
F. A. Berends, R. Kleiss, P. De Causmaecker, R. Gastmans, W. Troost,
and T.T. Wu, Nucl. Phys. {\fam \bffam \twelvbf B206}, 61 (1982);
F.A. Berends, P. De Causmaecker, R. Gastmans, R. Kleiss, W. Troost,
and T.T. Wu, Nucl. Phys. {\fam \bffam \twelvbf B239}, 382 (1984);
{\fam \bffam \twelvbf B239}, 395 (1984); {\fam \bffam \twelvbf B264},
243 (1986); {\fam \bffam \twelvbf B264}, 265 (1986).
\vskip \smallskipamount \item { } By now, many papers have been
published on the subject. A partial list of references follows.
\vskip \smallskipamount \item { } P. De Causmaecker, thesis, Leuven
University, 1983; R. Farrar and F. Neri, Phys. Lett.
{\fam \bffam \twelvbf 130B}, 109 (1983); R. Kleiss, Nucl. Phys.
{\fam \bffam \twelvbf B241}, 61 (1984); Z. Xu, D.H. Zhang and Z. Chang,
Tsingua University
preprint TUTP-84/3, 84/4, and 84/5a (1984); Nucl. Phys.
{\fam \bffam \twelvbf B291}, 392 (1987); J.F. Gunion and Z. Kunszt,
Phys. Lett. {\fam \bffam \twelvbf 161B}, 333 (1985); F.A. Berends,
P.H. Davereldt, and R. Kleiss, Nucl. Phys. {\fam \bffam \twelvbf B253},
441 (1985); R. Kleiss and W.J. Stirling, Nucl. Phys.
{\fam \bffam \twelvbf B262} 235 (1985); J.F. Gunion and Z. Kunszt,
Phys. Lett. {\fam \bffam \twelvbf 159B}, 167 (1985);
{\fam \bffam \twelvbf 161B}, 333 (1985); S.J. Parke and T.R. Taylor,
Phys. Rev. Lett. {\fam \bffam \twelvbf 56}, 2459 (1986); Z. Kunszt,
Nucl. Phys. {\fam \bffam \twelvbf B271}, 333 (1986);
J.F. Gunion and J. Kalinowski, Phys. Rev. {\fam \bffam \twelvbf D34},
2119 (1986); R. Kleiss and W.J. Stirling, Phys. Lett.
{\fam \bffam \twelvbf 179B}, 159 (1986); M. Mangano and S.J. Parke,
Nucl. Phys. {\fam \bffam \twelvbf B299}, 673 (1988); M. Mangano,
S.J. Parke, and Z. Xu, Nucl. Phys.
{\fam \bffam \twelvbf B298}, 653 (1988);
D.A. Kosower, B.--H. Lee, and V.P. Nair, Phys. Lett.
{\fam \bffam \twelvbf 201B}, 85 (1988); M. Mangano and S.J. Parke,
Nucl. Phys. {\fam \bffam \twelvbf B299}, 673 (1988); F.A. Berends
and W.T. Giele, Nucl. Phys. {\fam \bffam \twelvbf B313}, 595 (1989);
M. Mangano, Nucl. Phys. {\fam \bffam \twelvbf B315}, 391 (1989);
D.A. Kosower, Nucl. Phys. {\fam \bffam \twelvbf B335}, 23 (1990);
Phys. Lett. {\fam \bffam \twelvbf B254}, 439 (1991); Z. Bern and
D.A. Kosower, Nucl. Phys. {\fam \bffam \twelvbf B379}, 451 (1992);
C.S. Lam, McGill preprint McGill/92-32, 1992. \hfill \par
\item {8.}Many of the results for processes containing six or fewer
particles are collected in R. Gastmans and T.T. Wu,
{\fam \itfam \twelvit The Ubiquitous Photon: Helicity
Method for QED and QCD}
(Oxford University Press, New York, 1990). \hfill \par
\item {9.} The excellent review by Mangano and Parke provides a guide
to the various approaches to and extensive literature on the subject:
M. Mangano and S.J. Parke, Phys. Reports
{\fam \bffam \twelvbf 200}, 301 (1991). \hfill \par
\item {10.}G. Mahlon, T.--M. Yan and C. Dunn, Cornell
preprint CLNS 91/1120, 1992. \hfill \par
\item {11.}For a brief introduction to properties of two-component
Weyl-van der Waerden spinors, see, for example, M. F. Sohnius, Phys.
Reports {\fam \bffam \twelvbf 128}, 39 (1985). \hfill \par
\item {12.}M. Mangano, Nucl. Phys. {\fam \bffam \twelvbf B309}, 461
(1988). \hfill \par

\vfill\eject\bye

%
%
%
%

\font\twelvrm=cmr12
\font\ninerm=cmr9
\font\twelvi=cmmi12
\font\ninei=cmmi9
\font\twelvex=cmex10 scaled\magstep1
\font\twelvbf=cmbx12
\font\ninebf=cmbx9
\font\twelvit=cmti12
\font\twelvsy=cmsy10 scaled\magstep1
\font\ninesy=cmsy9
\font\twelvtt=cmtt12

\font\twelvsl=cmsl12

\def\twelvepoint{\def\rm{\fam0\twelvrm}
   \textfont0=\twelvrm \scriptfont0=\ninerm \scriptscriptfont0=\sevenrm
   \textfont1=\twelvi \scriptfont1=\ninei \scriptscriptfont1=\seveni
   \textfont2=\twelvsy \scriptfont2=\ninesy \scriptscriptfont2=\sevensy
   \textfont3=\twelvex \scriptfont3=\tenex \scriptscriptfont3=\tenex
        \textfont\itfam=\twelvit \def\it{\fam\itfam\twelvit}
        \textfont\slfam=\twelvsl \def\sl{\fam\slfam\twelvsl}
        \textfont\ttfam=\twelvtt \def\tt{\fam\ttfam\twelvtt}
        \textfont\bffam=\twelvbf \def\bf{\fam\bffam\twelvbf}
        \scriptfont\bffam=\ninebf  \scriptscriptfont\bffam=\sevenbf
}

\newdimen\normalwidth
\newdimen\double
\newdimen\single
\newdimen\indentlength          \indentlength=.5in

\newif\ifdrafton

\def\galley{
 \draftonfalse
 \twelvepoint
 \rm
 \font\chapterfont=cmbx10 scaled\magstep1
 \font\sectionfont=cmbx12
 \font\subsectionfont=cmbx12
 \font\headingfont=cmr10 scaled\magstep2
 \font\titlefont=cmbx10 scaled\magstep2
 \normalwidth=8.7in
 \double=.34in
 \single=.17in
 \hsize=\normalwidth
 \vsize=5.7in
 \hoffset=.1158in
 \voffset=.75in
 \hfuzz=0.5pt
 \baselineskip=\double plus 2pt minus 2pt }

\parindent=\indentlength
\clubpenalty=10000
\widowpenalty=10000
\displaywidowpenalty=500
\overfullrule=2pt
\tolerance=100

\newcount\chapterno     \chapterno=0
\newcount\sectionno     \sectionno=0
\newcount\appno         \appno=0
\newcount\subsectionno  \subsectionno=0
\newcount\eqnum \eqnum=0
\newcount\refno \refno=0
\newcount\chap
\newcount\figno \figno=0
\newcount\tableno \tableno=0
\newcount\lettno \lettno=0

\def\bodypaging{
 \headline={}
 \footline={}}

\newif\iftable \tablefalse
\newwrite\tablelist
\def\starttablelist{\iftable
 \immediate\openout\tablelist=tablelist
 \immediate\write\tablelist{\noexpand\centerline{\headingfont LIST
     OF TABLES}}
 \immediate\write\tablelist{\vskip\double\vskip\single\baselineskip=\single}
 \immediate\write\tablelist{\parskip=0pt \parindent=0pt}\else\relax\fi}
\def\inserttable#1#2{\global\advance\tableno by 1 \vfill\vskip\double
 \centerline{Table \the\tableno:\ #2}\nobreak\vskip\double\nobreak
 \begingroup #1\endgroup \vskip\double
 \iftable\immediate\write\tablelist{\hskip1cm
 \the\tableno. #2 \hfill\folio\vskip0cm}\else\relax\fi}
\def\andinserttable#1#2{\vfill\vskip\double
 \centerline{Table \the\tableno\ (continued):\ #2}
 \nobreak\vskip\double\nobreak
 \begingroup #1\endgroup \vskip\double}

\def\noadvancetable#1#2{\vfill\vskip\double
 \centerline{Table \the\tableno:\ #2}\nobreak\vskip\double\nobreak
 \begingroup #1\endgroup \vskip\double
 \iftable\immediate\write\tablelist{\hskip1cm
 \the\tableno. #2 \hfill\folio\vskip0cm}\else\relax\fi}
\def\continuetable#1#2{\vfill\vskip\double
 \centerline{Table \the\tableno{ }(continued):\ #2}\nobreak
 \vskip\double\nobreak
 \begingroup #1\endgroup \vskip\double}
\def\Tlabel#1{\global\advance\tableno by 1
  \xdef#1{\the\tableno} \the\tableno}
%

\input tables

\galley

%
\def \positronon{\bar\psi}
\def \electronon{\psi}

\def \Wnorm{W}
\def \Wms{{\cal{W}}}
\def \Wbar{{\overline{\Wms}}}

\def \PHI{{\mit\Phi}}

\def \pole{{\mit\Pi}}

\def \amp{{\cal M}}

%
%
\def \permsum#1#2{\sum_{{\cal{P}}(#1\ldots #2)}}

\def \braket#1#2{ \langle #1 \thinspace\thinspace #2 \rangle }
\def \bra#1{ \langle #1 | }
\def \ket#1{ | #1 \rangle }

\def \eps{\epsilon}
\def \vareps{\varepsilon}
\def \down{{}_\downarrow}
\def \up{{}_\uparrow}

\def \Poff{{\cal{P}}}
\def \Qoff{{\cal{Q}}}

\def \link#1#2#3{ {{\braket{#1}{#3}}\over{\bra{#1}#2\ket{#3}}} }
\def \invlink#1#2#3{ {{\bra{#1}#2\ket{#3}}\over{\braket{#1}{#3}}} }

\def\centeronto#1#2{{\setbox0=\hbox{#1}\setbox1=\hbox{#2}\ifdim
\wd1>\wd0\kern.5\wd1\kern-.5\wd0\fi
\copy0\kern-.5\wd0\kern-.5\wd1\copy1\ifdim\wd0>\wd1
\kern.5\wd0\kern-.5\wd1\fi}}
\def\slash#1{\centeronto{$#1$}{$/$}}

\def \varsp{\thinspace}


\pageno=1 \bodypaging 	 

%

\Tlabel\dummynumber1
\Tlabel\dummynumber2
\vfill\eject
\noadvancetable{
{\begintable
$P$  \vb $Q$  \vb $p$  \vb $q$  \|
${{F_1}^{\alpha}}_{\beta}(P,Q,p,q,1)$  \vb
${{F_2}^{\alpha}}_{\beta}(P,Q,p,q,1)$  \crthick
$W^{+}_{\up}$ \vb $W^{-}_{\up}$ \vb $\bar e_{\up}$ \vb $e_{\down}$ \|
$ 0 $ \vb $ 0 $
\cr
$W^{+}_{\up}$ \vb $W^{-}_{\up}$ \vb $W_L^{+}$ \vb $W_L^{-}$ \|
$ 0 $ \vb $ 0 $
\cr
$\bar e_{\up}$ \vb $e_{\down}$ \vb $\bar\mu_{\up}$ \vb $\mu_{\down}$ \|
${1\over2}\sec^2\theta_W {\braket{Q}{q}}^2  \thinspace
u^{\alpha}(k_1) \thinspace u_{\beta}(k_1)$     \vb
$-{1\over2}\sec^2\theta_W {\braket{1}{Q}}^2  \thinspace
u^{\alpha}(q) \thinspace u_{\beta}(q)$     \cr
$\bar e_{\down}$ \vb $e_{\up}$ \vb $\bar\mu_{\down}$ \vb $\mu_{\up}$ \|
$2\tan^2\theta_W {\braket{P}{1}}^2 \thinspace
u^{\alpha}(p) \thinspace u_{\beta}(p)$   \vb
$-2\tan^2\theta_W {\braket{P}{p}}^2 \thinspace
u^{\alpha}(k_1) \thinspace u_{\beta}(k_1)$   \cr
$\bar e_{\up}$ \vb $e_{\down}$ \vb $\bar\mu_{\down}$ \vb $\mu_{\up}$ \|
$-\tan^2\theta_W {\braket{p}{Q}}^2 \thinspace
u^{\alpha}(k_1) \thinspace u_{\beta}(k_1)$  \vb
$\tan^2\theta_W {\braket{1}{Q}}^2 \thinspace
u^{\alpha}(p) \thinspace u_{\beta}(p)$  \cr
$W_L^{+}$ \vb $W_L^{-}$ \vb $\bar e_{\up}$ \vb $ e_{\down}$ \|
${1\over2}\sec^2\theta_W \braket{P}{1}\braket{q}{Q} \thinspace
u^{\alpha}(q) \thinspace  u_{\beta}(k_1)$  \vb
$-{1\over2}\sec^2\theta_W \braket{1}{Q}\braket{P}{q} \thinspace
u^{\alpha}(q) \thinspace  u_{\beta}(k_1)$  \cr
$W_L^{+}$ \vb $W_L^{-}$ \vb $\bar e_{\down}$ \vb $ e_{\up}$ \|
$-\tan^2\theta_W \braket{P}{1} \braket{p}{Q} \thinspace
u^{\alpha}(p) \thinspace u_{\beta}(k_1)$   \vb
$\tan^2\theta_W \braket{P}{p} \braket{1}{Q} \thinspace
u^{\alpha}(p) \thinspace u_{\beta}(k_1)$   \cr
$W_L^{+}$ \vb $W_L^{-}$ \vb $W_L^{+}$ \vb $W_L^{-}$ \|
$-{1\over4}\sec^2\theta_W \braket{P}{1}
\biggl[  \braket{p}{Q} \thinspace u^{\alpha}(q)
       + \braket{q}{Q} \thinspace u^{\alpha}(p) \biggr]
\thinspace u_{\beta}(k_1)  \thinspace\thinspace $\vb
${1\over4}\sec^2\theta_W \braket{Q}{1}
\biggl[  \braket{p}{P} \thinspace u^{\alpha}(q)
       + \braket{q}{P} \thinspace u^{\alpha}(p) \biggr]
\thinspace u_{\beta}(k_1)  \thinspace\thinspace $\cr
$W_L^{+}$ \vb $W_L^{-}$ \vb $W_L^{+}$ \vb $W_L^{-}$ \|
$ 4\lambda g^{-2} \braket{P}{1} \braket{p}{q}\thinspace
u^{\alpha}(Q) \thinspace u_{\beta}(k_1)$  \vb
$- 4\lambda g^{-2} \braket{1}{Q} \braket{p}{q}\thinspace
u^{\alpha}(P) \thinspace u_{\beta}(k_1)$
\endtable}
}
{Group two quadruple current amplitude helicity functions}
\continuetable{
{\begintable
$P$  \vb $Q$  \vb $p$  \vb $q$  \|
${{F_3}^{\alpha}}_{\beta}(P,Q,p,q,1)$  \vb
${{F_4}^{\alpha}}_{\beta}(P,Q,p,q,1)$  \crthick
$W^{+}_{\up}$ \vb $W^{-}_{\up}$ \vb $\bar e_{\up}$ \vb $e_{\down}$ \|
$ 0 $ \vb $ 0 $
\cr
$W^{+}_{\up}$ \vb $W^{-}_{\up}$ \vb $W_L^{+}$ \vb $W_L^{-}$ \|
$ 0 $ \vb $ 0 $
\cr
$\bar e_{\up}$ \vb $e_{\down}$ \vb $\bar\mu_{\up}$ \vb $\mu_{\down}$ \|
${1\over2}\sec^2\theta_W {\braket{q}{Q}}^2  \thinspace
u^{\alpha}(k_1) \thinspace u_{\beta}(k_1)$     \vb
$-{1\over2}\sec^2\theta_W {\braket{1}{q}}^2  \thinspace
u^{\alpha}(Q) \thinspace u_{\beta}(Q)$     \cr
$\bar e_{\down}$ \vb $e_{\up}$ \vb $\bar\mu_{\down}$ \vb $\mu_{\up}$ \|
$2\tan^2\theta_W {\braket{p}{1}}^2 \thinspace
u^{\alpha}(P) \thinspace u_{\beta}(P)$   \vb
$-2\tan^2\theta_W {\braket{p}{P}}^2 \thinspace
u^{\alpha}(k_1) \thinspace u_{\beta}(k_1)$   \cr
$\bar e_{\up}$ \vb $e_{\down}$ \vb $\bar\mu_{\down}$ \vb $\mu_{\up}$ \|
$-\tan^2\theta_W {\braket{p}{1}}^2 \thinspace
u^{\alpha}(Q) \thinspace u_{\beta}(Q)$  \vb
$\tan^2\theta_W {\braket{p}{Q}}^2 \thinspace
u^{\alpha}(k_1) \thinspace u_{\beta}(k_1)$  \cr
$W_L^{+}$ \vb $W_L^{-}$ \vb $\bar e_{\up}$ \vb $ e_{\down}$ \|
${1\over2}\sec^2\theta_W \bra{P}q\ket{Q} \thinspace
u^{\alpha}(k_1) \thinspace  u_{\beta}(k_1)$  \vb
$-{1\over2}\sec^2\theta_W \bra{1}q\ket{1} \thinspace
u^{\alpha}(P) \thinspace  u_{\beta}(Q)$  \cr
$W_L^{+}$ \vb $W_L^{-}$ \vb $\bar e_{\down}$ \vb $ e_{\up}$ \|
$-\tan^2\theta_W \bra{1}p\ket{1} \thinspace
u^{\alpha}(Q) \thinspace u_{\beta}(P)$   \vb
$\tan^2\theta_W \bra{P}p \ket{Q} \thinspace
u^{\alpha}(k_1) \thinspace u_{\beta}(k_1)$   \cr
$W_L^{+}$ \vb $W_L^{-}$ \vb $W_L^{+}$ \vb $W_L^{-}$ \|
$-{1\over4}\sec^2\theta_W \braket{p}{1}
\biggl[  \braket{P}{q} \thinspace u^{\alpha}(Q)
       + \braket{Q}{q} \thinspace u^{\alpha}(P) \biggr]
\thinspace u_{\beta}(k_1)  \thinspace\thinspace $\vb
${1\over4}\sec^2\theta_W \braket{q}{1}
\biggl[  \braket{P}{p} \thinspace u^{\alpha}(Q)
       + \braket{Q}{p} \thinspace u^{\alpha}(P) \biggr]
\thinspace u_{\beta}(k_1)  \thinspace\thinspace $\cr
$W_L^{+}$ \vb $W_L^{-}$ \vb $W_L^{+}$ \vb $W_L^{-}$ \|
$ 4\lambda g^{-2} \braket{p}{1} \braket{P}{Q}\thinspace
u^{\alpha}(q) \thinspace u_{\beta}(k_1)$ \vb
$- 4\lambda g^{-2} \braket{1}{q} \braket{P}{Q}\thinspace
u^{\alpha}(p) \thinspace u_{\beta}(k_1)$
\endtable}
}
{Group two quadruple current amplitude helicity functions}
\continuetable{
{\begintable
$P$  \vb $Q$  \vb $p$  \vb $q$  \|
$f_5(P,Q,p,q,1)$ \vb $f_6(P,Q,p,q,1)$ \vb $f_7(P,Q,p,q,1)$ \crthick
$W^{+}_{\up}$ \vb $W^{-}_{\up}$ \vb $\bar e_{\up}$ \vb $e_{\down}$ \|
$ 0 $ \vb $ 0 $ \vb $ 0 $
\cr
$W^{+}_{\up}$ \vb $W^{-}_{\up}$ \vb $W_L^{+}$ \vb $W_L^{-}$ \|
$ 0 $ \vb $ 0 $ \vb $ 0 $
\cr
$\bar e_{\up}$ \vb $e_{\down}$ \vb $\bar\mu_{\up}$ \vb $\mu_{\down}$ \|
$-{1\over2}\sec^2\theta_W {\braket{Q}{q}}^2 $ \vb
$ {1\over2}\sec^2\theta_W \braket{Q}{q}\braket{1}{Q}\braket{1}{q} $\vb
$-{1\over2}\sec^2\theta_W \braket{Q}{q}\braket{1}{Q}\braket{1}{q} $
\cr
$\bar e_{\down}$ \vb $e_{\up}$ \vb $\bar\mu_{\down}$ \vb $\mu_{\up}$ \|
$ 2\tan^2\theta_W {\braket{P}{p}}^2 $ \vb
$-2\tan^2\theta_W \braket{P}{p} \braket{P}{1} \braket{p}{1} $ \vb
$ 2\tan^2\theta_W \braket{P}{p} \braket{P}{1} \braket{p}{1} $
\cr
$\bar e_{\up}$ \vb $e_{\down}$ \vb $\bar\mu_{\down}$ \vb $\mu_{\up}$ \|
$\tan^2\theta_W {\braket{p}{Q}}^2 $ \vb
$-\tan^2\theta_W \braket{p}{Q} \bra{p}1\ket{Q} $ \vb
$-\tan^2\theta_W \braket{p}{Q} \bra{p}1\ket{Q} $
\cr
$W_L^{+}$ \vb $W_L^{-}$ \vb $\bar e_{\up}$ \vb $ e_{\down}$ \|
$-{1\over2}\sec^2\theta_W \bra{P}q\ket{Q} $ \vb
$ {1\over2}\sec^2\theta_W \braket{P}{Q} {\braket{1}{q}}^2 $ \vb
$-{1\over2}\sec^2\theta_W \bra{P}q,1\ket{Q} $
\cr
$W_L^{+}$ \vb $W_L^{-}$ \vb $\bar e_{\down}$ \vb $ e_{\up}$ \|
$ \tan^2\theta_W \bra{P}p\ket{Q} $ \vb
$-\tan^2\theta_W \braket{P}{Q} {\braket{p}{1}}^2 $ \vb
$-\tan^2\theta_W \bra{P}p,1\ket{Q} $
\cr
$W_L^{+}$ \vb $W_L^{-}$ \vb $W_L^{+}$ \vb $W_L^{-}$ \|
${1\over4}\sec^2\theta_W
\biggl[  \braket{P}{p} \braket{q}{Q}
       + \braket{P}{q} \braket{p}{Q} \biggr] $ \vb
${1\over2}\sec^2\theta_W \braket{P}{Q} \bra{p}1\ket{q} $ \vb
${1\over2}\sec^2\theta_W \braket{p}{q} \bra{P}1\ket{Q} $
\cr
$W_L^{+}$ \vb $W_L^{-}$ \vb $W_L^{+}$ \vb $W_L^{-}$ \|
$- 4\lambda g^{-2} \braket{p}{q} \braket{P}{Q} $ \vb $0$ \vb $0$
\endtable}
}
{Group two quadruple current amplitude helicity functions}
\vfill\eject

\bye
%
%

\input pictex

%
%
\font\twelvrm=cmr12
\font\ninerm=cmr9
\font\twelvi=cmmi12
\font\ninei=cmmi9
\font\twelvex=cmex10 scaled\magstep1
\font\twelvbf=cmbx12
\font\ninebf=cmbx9
\font\twelvit=cmti12
\font\twelvsy=cmsy10 scaled\magstep1
\font\ninesy=cmsy9
\font\twelvtt=cmtt12

\font\twelvsl=cmsl12

\font\abstractfont=cmr10
\font\abstractitalfont=cmti10

\def\twelvepoint{\def\rm{\fam0\twelvrm}
   \textfont0=\twelvrm \scriptfont0=\ninerm \scriptscriptfont0=\sevenrm
   \textfont1=\twelvi \scriptfont1=\ninei \scriptscriptfont1=\seveni
   \textfont2=\twelvsy \scriptfont2=\ninesy \scriptscriptfont2=\sevensy
   \textfont3=\twelvex \scriptfont3=\tenex \scriptscriptfont3=\tenex
        \textfont\itfam=\twelvit \def\it{\fam\itfam\twelvit}
        \textfont\slfam=\twelvsl \def\sl{\fam\slfam\twelvsl}
        \textfont\ttfam=\twelvtt \def\tt{\fam\ttfam\twelvtt}
        \textfont\bffam=\twelvbf \def\bf{\fam\bffam\twelvbf}
        \scriptfont\bffam=\ninebf  \scriptscriptfont\bffam=\sevenbf
        \skewchar\ninei='177
        \skewchar\twelvi='177
        \skewchar\seveni='177
}

\newdimen\normalwidth
\newdimen\double
\newdimen\single
\newdimen\indentlength          \indentlength=.5in

\newif\ifdrafton

\def\galley{
 \draftonfalse
 \twelvepoint
 \rm
 \font\chapterfont=cmbx10 scaled\magstep1
 \font\sectionfont=cmbx12
 \font\subsectionfont=cmbx12
 \font\headingfont=cmr10 scaled\magstep2
 \font\titlefont=cmbx10 scaled\magstep2
 \normalwidth=5.7in
 \double=.34in
 \single=.17in
 \hsize=\normalwidth
 \vsize=8.7in
 \hoffset=0.48in
 \voffset=0.1in
 \hfuzz=0.5pt
 \baselineskip=\double plus 2pt minus 2pt }

\parindent=\indentlength
\clubpenalty=10000
\widowpenalty=10000
\displaywidowpenalty=500
\overfullrule=2pt
\tolerance=100

\newcount\chapterno     \chapterno=0
\newcount\sectionno     \sectionno=0
\newcount\appno         \appno=0
\newcount\subsectionno  \subsectionno=0
\newcount\eqnum \eqnum=0
\newcount\refno \refno=0
\newcount\chap
\newcount\figno \figno=0
\newcount\tableno \tableno=0
\newcount\lettno \lettno=0

\def\bodypaging{
 \headline={\ifodd\chap \hfil \else \tenrm\hfil\twelvrm\folio \fi}
 \footline={\rm \ifodd\chap \global\chap=0 \tenrm\hfil\twelvrm\folio\hfil
 \else \hfil \fi}}

\galley             

\pageno=74 \bodypaging 	 

$$
\beginpicture

\setcoordinatesystem units <1in,1in> point at 3.5 9.0


\ellipticalarc axes ratio 2:1 360 degrees from 2.50 8.40 center at 2.15 8.40

\ellipticalarc axes ratio 2:1 360 degrees from 2.50 6.60 center at 2.15 6.60

\ellipticalarc axes ratio 2:1 360 degrees from 5.20 8.40 center at 4.85 8.40

\ellipticalarc axes ratio 2:1 360 degrees from 5.20 6.60 center at 4.85 6.60


\setlinear

\plot 1.00 6.60 1.80 6.60 /
\plot 2.50 6.60 4.50 6.60 /
\plot 5.20 6.60 6.00 6.60 /


\plot 1.00 8.40  1.05 8.45  1.15 8.35
                 1.25 8.45  1.35 8.35
                 1.45 8.45  1.55 8.35
                 1.65 8.45  1.75 8.35  1.80 8.40 /

\plot 2.50 8.40  2.55 8.45  2.65 8.35
                 2.75 8.45  2.85 8.35
                 2.95 8.45  3.05 8.35
                 3.15 8.45  3.25 8.35
                 3.35 8.45  3.45 8.35
                 3.55 8.45  3.65 8.35
                 3.75 8.45  3.85 8.35
                 3.95 8.45  4.05 8.35
                 4.15 8.45  4.25 8.35
                 4.35 8.45  4.45 8.35  4.50 8.40 /

\plot 5.20 8.40  5.25 8.45  5.35 8.35
                 5.45 8.45  5.55 8.35
                 5.65 8.45  5.75 8.35
                 5.85 8.45  5.95 8.35  6.00 8.40 /


\setquadratic

\plot 3.50 8.40  3.55 8.35  3.50 8.30  3.45 8.25
      3.50 8.20  3.55 8.15  3.50 8.10  3.45 8.05
      3.50 8.00  3.55 7.95  3.50 7.90  3.45 7.85
      3.50 7.80  3.55 7.75  3.50 7.70  3.45 7.65
      3.50 7.60  3.55 7.55  3.50 7.50  3.45 7.45
      3.50 7.40  3.55 7.35  3.50 7.30  3.45 7.25
      3.50 7.20  3.55 7.15  3.50 7.10  3.45 7.05
      3.50 7.00  3.55 6.95  3.50 6.90  3.45 6.85
      3.50 6.80  3.55 6.75  3.50 6.70  3.45 6.65
      3.50 6.60 /


\setshadesymbol <z,z,z,z> ({\fiverm .})
\setshadegrid span <.2pt>

\setlinear

\vshade 1.35 6.60 6.60
        1.40 6.55 6.65 /

\vshade 2.95 6.60 6.60
        3.00 6.55 6.65 /

\vshade 4.00 6.60 6.60
        4.05 6.55 6.65 /

\vshade 5.60 6.60 6.60
        5.65 6.55 6.65 /


\setshadesymbol <z,z,z,z> ({\fiverm .})
\setshadegrid span <1.75pt>

\setquadratic

\vshade 1.8000 8.4000000 8.4000000
        1.8875 8.2842484 8.5157516
        1.9750 8.2484456 8.5515544
        2.0625 8.2305570 8.5694430
        2.1500 8.2250000 8.5750000
        2.2375 8.2305570 8.5694430
        2.3250 8.2484456 8.5515544
        2.4125 8.2842484 8.5157516
        2.5000 8.4000000 8.4000000 /

\vshade 4.5000 8.4000000 8.4000000
        4.5875 8.2842484 8.5157516
        4.6750 8.2484456 8.5515544
        4.7625 8.2305570 8.5694430
        4.8500 8.2250000 8.5750000
        4.9375 8.2305570 8.5694430
        5.0250 8.2484456 8.5515544
        5.1125 8.2842484 8.5157516
        5.2000 8.4000000 8.4000000 /

\vshade 1.8000 6.6000000 6.6000000
        1.8875 6.4842480 6.7157516
        1.9750 6.4484456 6.7515544
        2.0625 6.4305570 6.7694430
        2.1500 6.4250000 6.7750000
        2.2375 6.4305570 6.7694430
        2.3250 6.4484456 6.7515544
        2.4125 6.4842480 6.7157516
        2.5000 6.6000000 6.6000000 /

\vshade 4.5000 6.6000000 6.6000000
        4.5875 6.4842480 6.7157516
        4.6750 6.4484456 6.7515544
        4.7625 6.4305570 6.7694430
        4.8500 6.4250000 6.7750000
        4.9375 6.4305570 6.7694430
        5.0250 6.4484456 6.7515544
        5.1125 6.4842480 6.7157516
        5.2000 6.6000000 6.6000000 /

 \startrotation by  0.000000   1.000000 about      2.150000       8.575000
 \setquadratic
 \plot
     2.150000       8.575000
     2.200000       8.625000
     2.250000       8.575000
     2.300000       8.525000
     2.350000       8.575000
     2.400000       8.625000
     2.450000       8.575000
     2.500000       8.525000
     2.550000       8.575000
     2.600000       8.625000
     2.650000       8.575000
 /
 \stoprotation

 \startrotation by  0.000000  -1.000000 about      2.150000       8.225000
 \setquadratic
 \plot
     2.150000       8.225000
     2.200000       8.275001
     2.250000       8.225000
     2.300000       8.175000
     2.350000       8.225000
     2.400000       8.275001
     2.450000       8.225000
     2.500000       8.175000
     2.550000       8.225000
     2.600000       8.275001
     2.650000       8.225000
 /
 \stoprotation

 \startrotation by -0.422618  -0.906308 about      1.975000       8.248446
 \setquadratic
 \plot
     1.975000       8.248446
     2.025000       8.298446
     2.075000       8.248446
     2.125000       8.198445
     2.175000       8.248446
     2.225000       8.298446
     2.275000       8.248446
     2.325000       8.198445
     2.375000       8.248446
     2.425000       8.298446
     2.475000       8.248446
 /
 \stoprotation

 \startrotation by  0.422618  -0.906308 about      2.325000       8.248446
 \setquadratic
 \plot
     2.325000       8.248446
     2.375000       8.298446
     2.425000       8.248446
     2.475000       8.198445
     2.525000       8.248446
     2.575000       8.298446
     2.625000       8.248446
     2.675000       8.198445
     2.725000       8.248446
     2.775000       8.298446
     2.825000       8.248446
 /
 \stoprotation

 \startrotation by -0.422618   0.906308 about      1.975000       8.551555
 \setquadratic
 \plot
     1.975000       8.551555
     2.025000       8.601555
     2.075000       8.551555
     2.125000       8.501554
     2.175000       8.551555
     2.225000       8.601555
     2.275000       8.551555
     2.325000       8.501554
     2.375000       8.551555
     2.425000       8.601555
     2.475000       8.551555
 /
 \stoprotation

 \startrotation by  0.422618   0.906308 about      2.325000       8.551555
 \setquadratic
 \plot
     2.325000       8.551555
     2.375000       8.601555
     2.425000       8.551555
     2.475000       8.501554
     2.525000       8.551555
     2.575000       8.601555
     2.625000       8.551555
     2.675000       8.501554
     2.725000       8.551555
     2.775000       8.601555
     2.825000       8.551555
 /
 \stoprotation


 \startrotation by  0.000000  -1.000000 about      4.850000       6.425000
 \setquadratic
 \plot
     4.850000       6.425000
     4.900000       6.475000
     4.950000       6.425000
     5.000000       6.375000
     5.050000       6.425000
     5.100000       6.475000
     5.150000       6.425000
     5.200000       6.375000
     5.250000       6.425000
     5.300000       6.475000
     5.350000       6.425000
 /
 \stoprotation

 \startrotation by -0.422618  -0.906308 about      4.675000       6.448446
 \setquadratic
 \plot
     4.675000       6.448446
     4.725000       6.498446
     4.775000       6.448446
     4.825000       6.398446
     4.875000       6.448446
     4.925000       6.498446
     4.975000       6.448446
     5.025000       6.398446
     5.075000       6.448446
     5.125000       6.498446
     5.175000       6.448446
 /
 \stoprotation

 \startrotation by  0.422618  -0.906308 about      5.025000       6.448446
 \setquadratic
 \plot
     5.025000       6.448446
     5.075000       6.498446
     5.125000       6.448446
     5.175000       6.398446
     5.225000       6.448446
     5.275000       6.498446
     5.325000       6.448446
     5.375000       6.398446
     5.425000       6.448446
     5.475000       6.498446
     5.525000       6.448446
 /
 \stoprotation

 \startrotation by -0.216440   0.976296 about      4.762500       6.769443
 \setquadratic
 \plot
     4.762500       6.769443
     4.812500       6.819443
     4.862500       6.769443
     4.912500       6.719443
     4.962500       6.769443
     5.012500       6.819443
     5.062500       6.769443
     5.112500       6.719443
     5.162500       6.769443
     5.212500       6.819443
     5.262500       6.769443
 /
 \stoprotation

 \startrotation by  0.199368   0.979925 about      4.937500       6.769443
 \setquadratic
 \plot
     4.937500       6.769443
     4.987500       6.819443
     5.037500       6.769443
     5.087500       6.719443
     5.137500       6.769443
     5.187500       6.819443
     5.237500       6.769443
     5.287500       6.719443
     5.337500       6.769443
     5.387500       6.819443
     5.437500       6.769443
 /
 \stoprotation

 \startrotation by  0.766044   0.642788 about      5.112500       6.715752
 \setquadratic
 \plot
     5.112500       6.715752
     5.162500       6.765752
     5.212500       6.715752
     5.262500       6.665751
     5.312500       6.715752
     5.362500       6.765752
     5.412500       6.715752
     5.462500       6.665751
     5.512500       6.715752
     5.562500       6.765752
     5.612500       6.715752
 /
 \stoprotation

 \startrotation by -0.766045   0.642787 about      4.587500       6.715752
 \setquadratic
 \plot
     4.587500       6.715752
     4.637500       6.765752
     4.687500       6.715752
     4.737500       6.665751
     4.787500       6.715752
     4.837500       6.765752
     4.887500       6.715752
     4.937500       6.665751
     4.987500       6.715752
     5.037500       6.765752
     5.087500       6.715752
 /
 \stoprotation


 \startrotation by -0.216440   0.976296 about      4.762500       8.569443
 \setquadratic
 \plot
     4.762500       8.569443
     4.812500       8.619443
     4.862500       8.569443
     4.912500       8.519443
     4.962500       8.569443
     5.012500       8.619443
     5.062500       8.569443
     5.112500       8.519443
     5.162500       8.569443
     5.212500       8.619443
     5.262500       8.569443
 /
 \stoprotation

 \startrotation by  0.199368   0.979925 about      4.937500       8.569443
 \setquadratic
 \plot
     4.937500       8.569443
     4.987500       8.619443
     5.037500       8.569443
     5.087500       8.519443
     5.137500       8.569443
     5.187500       8.619443
     5.237500       8.569443
     5.287500       8.519443
     5.337500       8.569443
     5.387500       8.619443
     5.437500       8.569443
 /
 \stoprotation

 \startrotation by  0.000000  -1.000000 about      4.850000       8.225000
 \setquadratic
 \plot
     4.850000       8.225000
     4.900000       8.275001
     4.950000       8.225000
     5.000000       8.175000
     5.050000       8.225000
     5.100000       8.275001
     5.150000       8.225000
     5.200000       8.175000
     5.250000       8.225000
     5.300000       8.275001
     5.350000       8.225000
 /
 \stoprotation

 \startrotation by -0.422618  -0.906308 about      4.675000       8.248446
 \setquadratic
 \plot
     4.675000       8.248446
     4.725000       8.298446
     4.775000       8.248446
     4.825000       8.198445
     4.875000       8.248446
     4.925000       8.298446
     4.975000       8.248446
     5.025000       8.198445
     5.075000       8.248446
     5.125000       8.298446
     5.175000       8.248446
 /
 \stoprotation

 \startrotation by  0.422618  -0.906308 about      5.025000       8.248446
 \setquadratic
 \plot
     5.025000       8.248446
     5.075000       8.298446
     5.125000       8.248446
     5.175000       8.198445
     5.225000       8.248446
     5.275000       8.298446
     5.325000       8.248446
     5.375000       8.198445
     5.425000       8.248446
     5.475000       8.298446
     5.525000       8.248446
 /
 \stoprotation


 \startrotation by  0.000000  -1.000000 about      2.150000       6.425000
 \setquadratic
 \plot
     2.150000       6.425000
     2.200000       6.475000
     2.250000       6.425000
     2.300000       6.375000
     2.350000       6.425000
     2.400000       6.475000
     2.450000       6.425000
     2.500000       6.375000
     2.550000       6.425000
     2.600000       6.475000
     2.650000       6.425000
 /
 \stoprotation

 \startrotation by -0.216440   0.976296 about      2.062500       6.769443
 \setquadratic
 \plot
     2.062500       6.769443
     2.112500       6.819443
     2.162500       6.769443
     2.212500       6.719443
     2.262500       6.769443
     2.312500       6.819443
     2.362500       6.769443
     2.412500       6.719443
     2.462500       6.769443
     2.512500       6.819443
     2.562500       6.769443
 /
 \stoprotation

 \startrotation by  0.199368   0.979925 about      2.237500       6.769445
 \setquadratic
 \plot
     2.237500       6.769445
     2.287500       6.819445
     2.337500       6.769445
     2.387500       6.719445
     2.437500       6.769445
     2.487500       6.819445
     2.537500       6.769445
     2.587500       6.719445
     2.637500       6.769445
     2.687500       6.819445
     2.737500       6.769445
 /
 \stoprotation

 \startrotation by -0.422618  -0.906308 about      1.975000       6.448446
 \setquadratic
 \plot
     1.975000       6.448446
     2.025000       6.498446
     2.075000       6.448446
     2.125000       6.398446
     2.175000       6.448446
     2.225000       6.498446
     2.275000       6.448446
     2.325000       6.398446
     2.375000       6.448446
     2.425000       6.498446
     2.475000       6.448446
 /
 \stoprotation

 \startrotation by  0.422618  -0.906308 about      2.325000       6.448446
 \setquadratic
 \plot
     2.325000       6.448446
     2.375000       6.498446
     2.425000       6.448446
     2.475000       6.398446
     2.525000       6.448446
     2.575000       6.498446
     2.625000       6.448446
     2.675000       6.398446
     2.725000       6.448446
     2.775000       6.498446
     2.825000       6.448446
 /
 \stoprotation


\put {$\gamma$ or $Z$} at 3.9 7.5

\put {$W^{+}$} at 0.84 8.4

\put {$W^{-}$} at 6.16 8.4

\put {$e^{+}$} at 0.84 6.6

\put {$e^{-}$} at 6.16 6.6

\put {$\longrightarrow$} at 1.25 8.6

\put {$P$} at 1.25 8.75

\put {$\longleftarrow$} at 5.75 8.6

\put {$Q$} at 5.75 8.75

\put {$\longrightarrow$} at 1.25 6.4

\put {$p$} at 1.25 6.25

\put {$\longleftarrow$} at 5.75 6.4

\put {$q$} at 5.75 6.25

\put {$1,\ldots,r$} at 2.15  9.25

\put {$r{+}1,\ldots,s$} at 4.85 9.25

\put {$s{+}1,\ldots,t$} at 2.15 5.75

\put {$t{+}1,\ldots,n$} at 4.85 5.75



\setlinear

\plot 3.50 2.60  3.50 4.40 /


\ellipticalarc axes ratio 2:1 360 degrees from 2.50 4.40 center at 2.15 4.40

\ellipticalarc axes ratio 2:1 360 degrees from 2.50 2.60 center at 2.15 2.60

\ellipticalarc axes ratio 2:1 360 degrees from 5.20 4.40 center at 4.85 4.40

\ellipticalarc axes ratio 2:1 360 degrees from 5.20 2.60 center at 4.85 2.60


\plot 3.50 2.60 4.50 2.60 /
\plot 5.20 2.60 6.00 2.60 /

\plot 3.50 4.40 4.50 4.40 /
\plot 5.20 4.40 6.00 4.40 /


\plot 1.00 4.40  1.05 4.45  1.15 4.35
                 1.25 4.45  1.35 4.35
                 1.45 4.45  1.55 4.35
                 1.65 4.45  1.75 4.35  1.80 4.40 /

\plot 2.50 4.40  2.55 4.45  2.65 4.35
                 2.75 4.45  2.85 4.35
                 2.95 4.45  3.05 4.35
                 3.15 4.45  3.25 4.35
                 3.35 4.45  3.45 4.35  3.50 4.40 /

\plot 1.00 2.60  1.05 2.65  1.15 2.55
                 1.25 2.65  1.35 2.55
                 1.45 2.65  1.55 2.55
                 1.65 2.65  1.75 2.55  1.80 2.60 /

\plot 2.50 2.60  2.55 2.65  2.65 2.55
                 2.75 2.65  2.85 2.55
                 2.95 2.65  3.05 2.55
                 3.15 2.65  3.25 2.55
                 3.35 2.65  3.45 2.55  3.50 2.60 /


\setshadesymbol <z,z,z,z> ({\fiverm .})
\setshadegrid span <.2pt>

\setlinear

\vshade 4.00 2.55 2.65
        4.05 2.60 2.60 /

\vshade 5.60 2.55 2.65
        5.65 2.60 2.60 /

\vshade 4.00 4.40 4.40
        4.05 4.35 4.45 /

\vshade 5.60 4.40 4.40
        5.65 4.35 4.45 /

\hshade 3.45 3.50 3.50
        3.55 3.45 3.55 /


\setshadesymbol <z,z,z,z> ({\fiverm .})
\setshadegrid span <1.75pt>

\setquadratic

\vshade 1.8000 4.4000000 4.4000000
        1.8875 4.2842484 4.5157516
        1.9750 4.2484456 4.5515544
        2.0625 4.2305570 4.5694430
        2.1500 4.2250000 4.5750000
        2.2375 4.2305570 4.5694430
        2.3250 4.2484456 4.5515544
        2.4125 4.2842484 4.5157516
        2.5000 4.4000000 4.4000000 /

\vshade 4.5000 4.4000000 4.4000000
        4.5875 4.2842484 4.5157516
        4.6750 4.2484456 4.5515544
        4.7625 4.2305570 4.5694430
        4.8500 4.2250000 4.5750000
        4.9375 4.2305570 4.5694430
        5.0250 4.2484456 4.5515544
        5.1125 4.2842484 4.5157516
        5.2000 4.4000000 4.4000000 /

\vshade 1.8000 2.6000000 2.6000000
        1.8875 2.4842480 2.7157516
        1.9750 2.4484456 2.7515544
        2.0625 2.4305570 2.7694430
        2.1500 2.4250000 2.7750000
        2.2375 2.4305570 2.7694430
        2.3250 2.4484456 2.7515544
        2.4125 2.4842480 2.7157516
        2.5000 2.6000000 2.6000000 /

\vshade 4.5000 2.6000000 2.6000000
        4.5875 2.4842480 2.7157516
        4.6750 2.4484456 2.7515544
        4.7625 2.4305570 2.7694430
        4.8500 2.4250000 2.7750000
        4.9375 2.4305570 2.7694430
        5.0250 2.4484456 2.7515544
        5.1125 2.4842480 2.7157516
        5.2000 2.6000000 2.6000000 /


 \startrotation by -0.216440   0.976296 about      2.062500       4.569443
 \setquadratic
 \plot
     2.062500       4.569443
     2.112500       4.619443
     2.162500       4.569443
     2.212500       4.519443
     2.262500       4.569443
     2.312500       4.619443
     2.362500       4.569443
     2.412500       4.519443
     2.462500       4.569443
     2.512500       4.619443
     2.562500       4.569443
 /
 \stoprotation

 \startrotation by  0.199368   0.979925 about      2.237500       4.569443
 \setquadratic
 \plot
     2.237500       4.569443
     2.287500       4.619443
     2.337500       4.569443
     2.387500       4.519443
     2.437500       4.569443
     2.487500       4.619443
     2.537500       4.569443
     2.587500       4.519443
     2.637500       4.569443
     2.687500       4.619443
     2.737500       4.569443
 /
 \stoprotation

 \startrotation by -0.766044  -0.642788 about      1.887500       4.284248
 \setquadratic
 \plot
     1.887500       4.284248
     1.937500       4.334249
     1.987500       4.284248
     2.037500       4.234248
     2.087500       4.284248
     2.137500       4.334249
     2.187500       4.284248
     2.237500       4.234248
     2.287500       4.284248
     2.337500       4.334249
     2.387500       4.284248
 /
 \stoprotation

 \startrotation by -0.216439  -0.976296 about      2.062500       4.230557
 \setquadratic
 \plot
     2.062500       4.230557
     2.112500       4.280557
     2.162500       4.230557
     2.212500       4.180557
     2.262500       4.230557
     2.312500       4.280557
     2.362500       4.230557
     2.412500       4.180557
     2.462500       4.230557
     2.512500       4.280557
     2.562500       4.230557
 /
 \stoprotation

 \startrotation by  0.216440  -0.976296 about      2.237500       4.230557
 \setquadratic
 \plot
     2.237500       4.230557
     2.287500       4.280557
     2.337500       4.230557
     2.387500       4.180557
     2.437500       4.230557
     2.487500       4.280557
     2.537500       4.230557
     2.587500       4.180557
     2.637500       4.230557
     2.687500       4.280557
     2.737500       4.230557
 /
 \stoprotation

 \startrotation by  0.766045  -0.642787 about      2.412500       4.284248
 \setquadratic
 \plot
     2.412500       4.284248
     2.462500       4.334249
     2.512500       4.284248
     2.562500       4.234248
     2.612500       4.284248
     2.662500       4.334249
     2.712500       4.284248
     2.762500       4.234248
     2.812500       4.284248
     2.862500       4.334249
     2.912500       4.284248
 /
 \stoprotation


 \startrotation by -0.422618   0.906308 about      4.675000       4.551554
 \setquadratic
 \plot
     4.675000       4.551554
     4.725000       4.601554
     4.775000       4.551554
     4.825000       4.501554
     4.875000       4.551554
     4.925000       4.601554
     4.975000       4.551554
     5.025000       4.501554
     5.075000       4.551554
     5.125000       4.601554
     5.175000       4.551554
 /
 \stoprotation

 \startrotation by  0.000000   1.000000 about      4.850000       4.575000
 \setquadratic
 \plot
     4.850000       4.575000
     4.900000       4.625000
     4.950000       4.575000
     5.000000       4.525000
     5.050000       4.575000
     5.100000       4.625000
     5.150000       4.575000
     5.200000       4.525000
     5.250000       4.575000
     5.300000       4.625000
     5.350000       4.575000
 /
 \stoprotation

 \startrotation by  0.422618   0.906308 about      5.025000       4.551554
 \setquadratic
 \plot
     5.025000       4.551554
     5.075000       4.601554
     5.125000       4.551554
     5.175000       4.501554
     5.225000       4.551554
     5.275000       4.601554
     5.325000       4.551554
     5.375000       4.501554
     5.425000       4.551554
     5.475000       4.601554
     5.525000       4.551554
 /
 \stoprotation

 \startrotation by -0.422618  -0.906308 about      4.675000       4.248446
 \setquadratic
 \plot
     4.675000       4.248446
     4.725000       4.298446
     4.775000       4.248446
     4.825000       4.198445
     4.875000       4.248446
     4.925000       4.298446
     4.975000       4.248446
     5.025000       4.198445
     5.075000       4.248446
     5.125000       4.298446
     5.175000       4.248446
 /
 \stoprotation

 \startrotation by  0.000000  -1.000000 about      4.850000       4.225000
 \setquadratic
 \plot
     4.850000       4.225000
     4.900000       4.275000
     4.950000       4.225000
     5.000000       4.175000
     5.050000       4.225000
     5.100000       4.275000
     5.150000       4.225000
     5.200000       4.175000
     5.250000       4.225000
     5.300000       4.275000
     5.350000       4.225000
 /
 \stoprotation

 \startrotation by  0.422618  -0.906308 about      5.025000       4.248446
 \setquadratic
 \plot
     5.025000       4.248446
     5.075000       4.298446
     5.125000       4.248446
     5.175000       4.198445
     5.225000       4.248446
     5.275000       4.298446
     5.325000       4.248446
     5.375000       4.198445
     5.425000       4.248446
     5.475000       4.298446
     5.525000       4.248446
 /
 \stoprotation


 \startrotation by -0.216440   0.976296 about      2.062500       2.769443
 \setquadratic
 \plot
     2.062500       2.769443
     2.112500       2.819443
     2.162500       2.769443
     2.212500       2.719443
     2.262500       2.769443
     2.312500       2.819443
     2.362500       2.769443
     2.412500       2.719443
     2.462500       2.769443
     2.512500       2.819443
     2.562500       2.769443
 /
 \stoprotation

 \startrotation by  0.216440   0.976296 about      2.237500       2.769443
 \setquadratic
 \plot
     2.237500       2.769443
     2.287500       2.819443
     2.337500       2.769443
     2.387500       2.719443
     2.437500       2.769443
     2.487500       2.819443
     2.537500       2.769443
     2.587500       2.719443
     2.637500       2.769443
     2.687500       2.819443
     2.737500       2.769443
 /
 \stoprotation

 \startrotation by -0.216439  -0.976296 about      2.062500       2.430557
 \setquadratic
 \plot
     2.062500       2.430557
     2.112500       2.480557
     2.162500       2.430557
     2.212500       2.380557
     2.262500       2.430557
     2.312500       2.480557
     2.362500       2.430557
     2.412500       2.380557
     2.462500       2.430557
     2.512500       2.480557
     2.562500       2.430557
 /
 \stoprotation

 \startrotation by  0.216440  -0.976296 about      2.237500       2.430557
 \setquadratic
 \plot
     2.237500       2.430557
     2.287500       2.480557
     2.337500       2.430557
     2.387500       2.380557
     2.437500       2.430557
     2.487500       2.480557
     2.537500       2.430557
     2.587500       2.380557
     2.637500       2.430557
     2.687500       2.480557
     2.737500       2.430557
 /
 \stoprotation


 \startrotation by -0.216440   0.976296 about      4.762500       2.769443
 \setquadratic
 \plot
     4.762500       2.769443
     4.812500       2.819443
     4.862500       2.769443
     4.912500       2.719443
     4.962500       2.769443
     5.012500       2.819443
     5.062500       2.769443
     5.112500       2.719443
     5.162500       2.769443
     5.212500       2.819443
     5.262500       2.769443
 /
 \stoprotation

 \startrotation by  0.216440   0.976296 about      4.937500       2.769443
 \setquadratic
 \plot
     4.937500       2.769443
     4.987500       2.819443
     5.037500       2.769443
     5.087500       2.719443
     5.137500       2.769443
     5.187500       2.819443
     5.237500       2.769443
     5.287500       2.719443
     5.337500       2.769443
     5.387500       2.819443
     5.437500       2.769443
 /
 \stoprotation

 \startrotation by -0.422618  -0.906308 about      4.675000       2.448446
 \setquadratic
 \plot
     4.675000       2.448446
     4.725000       2.498446
     4.775000       2.448446
     4.825000       2.398446
     4.875000       2.448446
     4.925000       2.498446
     4.975000       2.448446
     5.025000       2.398446
     5.075000       2.448446
     5.125000       2.498446
     5.175000       2.448446
 /
 \stoprotation

 \startrotation by  0.000000  -1.000000 about      4.850000       2.425000
 \setquadratic
 \plot
     4.850000       2.425000
     4.900000       2.475000
     4.950000       2.425000
     5.000000       2.375000
     5.050000       2.425000
     5.100000       2.475000
     5.150000       2.425000
     5.200000       2.375000
     5.250000       2.425000
     5.300000       2.475000
     5.350000       2.425000
 /
 \stoprotation

 \startrotation by  0.422618  -0.906308 about      5.025000       2.448446
 \setquadratic
 \plot
     5.025000       2.448446
     5.075000       2.498446
     5.125000       2.448446
     5.175000       2.398446
     5.225000       2.448446
     5.275000       2.498446
     5.325000       2.448446
     5.375000       2.398446
     5.425000       2.448446
     5.475000       2.498446
     5.525000       2.448446
 /
 \stoprotation


\put {$\nu$} at 3.7 3.5

\put {$W^{+}$} at 0.84 4.4

\put {$e^{-}$} at 6.16 4.4

\put {$W^{-}$} at 0.84 2.6

\put {$e^{+}$} at 6.16 2.6

\put {$\longrightarrow$} at 1.25 4.6

\put {$P$} at 1.25 4.75

\put {$\longleftarrow$} at 5.75 4.6

\put {$q$} at 5.75 4.75

\put {$\longrightarrow$} at 1.25 2.4

\put {$Q$} at 1.25 2.25

\put {$\longleftarrow$} at 5.75 2.4

\put {$p$}  at 5.75 2.25

\put {$1,\ldots,r$} at 2.15 5.25

\put {$r{+}1,\ldots,s$} at 4.85 5.25

\put {$s{+}1,\ldots,t$} at 2.15 1.75

\put {$t{+}1,\ldots,n$} at 4.85 1.75

\put {Figure 1.  Contributions to
$e^{+} \thinspace e^{-} \longrightarrow W^{+} \thinspace W^{-}
\thinspace \gamma \thinspace \gamma \cdots \gamma$.} at 3.5 0.9
\endpicture
$$
\vfil\eject\bye
%
%

\input pictex

%
%
\font\twelvrm=cmr12
\font\ninerm=cmr9
\font\twelvi=cmmi12
\font\ninei=cmmi9
\font\twelvex=cmex10 scaled\magstep1
\font\twelvbf=cmbx12
\font\ninebf=cmbx9
\font\twelvit=cmti12
\font\twelvsy=cmsy10 scaled\magstep1
\font\ninesy=cmsy9
\font\twelvtt=cmtt12

\font\twelvsl=cmsl12

\font\abstractfont=cmr10
\font\abstractitalfont=cmti10

\def\twelvepoint{\def\rm{\fam0\twelvrm}
   \textfont0=\twelvrm \scriptfont0=\ninerm \scriptscriptfont0=\sevenrm
   \textfont1=\twelvi \scriptfont1=\ninei \scriptscriptfont1=\seveni
   \textfont2=\twelvsy \scriptfont2=\ninesy \scriptscriptfont2=\sevensy
   \textfont3=\twelvex \scriptfont3=\tenex \scriptscriptfont3=\tenex
        \textfont\itfam=\twelvit \def\it{\fam\itfam\twelvit}
        \textfont\slfam=\twelvsl \def\sl{\fam\slfam\twelvsl}
        \textfont\ttfam=\twelvtt \def\tt{\fam\ttfam\twelvtt}
        \textfont\bffam=\twelvbf \def\bf{\fam\bffam\twelvbf}
        \scriptfont\bffam=\ninebf  \scriptscriptfont\bffam=\sevenbf
        \skewchar\ninei='177
        \skewchar\twelvi='177
        \skewchar\seveni='177
}

\newdimen\normalwidth
\newdimen\double
\newdimen\single
\newdimen\indentlength          \indentlength=.5in

\newif\ifdrafton

\def\galley{
 \draftonfalse
 \twelvepoint
 \rm
 \font\chapterfont=cmbx10 scaled\magstep1
 \font\sectionfont=cmbx12
 \font\subsectionfont=cmbx12
 \font\headingfont=cmr10 scaled\magstep2
 \font\titlefont=cmbx10 scaled\magstep2
 \normalwidth=5.7in
 \double=.34in
 \single=.17in
 \hsize=\normalwidth
 \vsize=8.7in
 \hoffset=0.48in
 \voffset=0.1in
 \hfuzz=0.5pt
 \baselineskip=\double plus 2pt minus 2pt }

\parindent=\indentlength
\clubpenalty=10000
\widowpenalty=10000
\displaywidowpenalty=500
\overfullrule=2pt
\tolerance=100

\newcount\chapterno     \chapterno=0
\newcount\sectionno     \sectionno=0
\newcount\appno         \appno=0
\newcount\subsectionno  \subsectionno=0
\newcount\eqnum \eqnum=0
\newcount\refno \refno=0
\newcount\chap
\newcount\figno \figno=0
\newcount\tableno \tableno=0
\newcount\lettno \lettno=0

\def\bodypaging{
 \headline={\ifodd\chap \hfil \else \tenrm\hfil\twelvrm\folio \fi}
 \footline={\rm \ifodd\chap \global\chap=0 \tenrm\hfil\twelvrm\folio\hfil
 \else \hfil \fi}}

\galley             

\pageno=75 \bodypaging 	 

\vfil
$$
\beginpicture

\setcoordinatesystem units <1in,1in> point at 3.5 9.0


\ellipticalarc axes ratio 1:2 360 degrees from 3.675 5.5 center at 3.5 5.5

\ellipticalarc axes ratio 2:1 360 degrees from 2.5 7.25 center at 2.15 7.25

\ellipticalarc axes ratio 2:1 360 degrees from 5.20 3.75 center at 4.85 3.75


\setlinear

\plot 1.00 7.25  1.80 7.25 /
\plot 2.50 7.25  6.00 7.25 /


\setdashes <5pt>

\plot 1.00 3.75  4.50 3.75 /
\plot 5.20 3.75  6.00 3.75 /

\setsolid


\plot 3.50 7.25  3.55 7.20  3.45 7.10
                 3.55 7.00  3.45 6.90
                 3.55 6.80  3.45 6.70
                 3.55 6.60  3.45 6.50
                 3.55 6.40  3.45 6.30
                 3.55 6.20  3.45 6.10
                 3.55 6.00  3.45 5.90  3.50 5.85 /

\plot 3.50 5.15  3.55 5.10  3.45 5.00
                 3.55 4.90  3.45 4.80
                 3.55 4.70  3.45 4.60
                 3.55 4.50  3.45 4.40
                 3.55 4.30  3.45 4.20
                 3.55 4.10  3.45 4.00
                 3.55 3.90  3.45 3.80  3.50 3.75 /


\setshadesymbol <z,z,z,z> ({\fiverm .})

\setshadegrid span <.2pt>

\setlinear

\vshade 1.35 7.25 7.25
        1.40 7.20 7.30 /

\vshade 2.95 7.25 7.25
        3.00 7.20 7.30 /

\vshade 4.70 7.25 7.25
        4.75 7.20 7.30 /


\setshadesymbol <z,z,z,z> ({\fiverm .})

\setshadegrid span <1.75pt>

\setquadratic

\vshade 1.8000 7.2500000 7.2500000
        1.8875 7.1342484 7.3657516
        1.9750 7.0984456 7.4015544
        2.0625 7.0805570 7.4194430
        2.1500 7.0750000 7.4250000
        2.2375 7.0805570 7.4194430
        2.3250 7.0984456 7.4015544
        2.4125 7.1342484 7.3657516
        2.5000 7.2500000 7.2500000 /

\vshade 4.5000 3.7500000 3.7500000
        4.5875 3.6342484 3.8657516
        4.6750 3.5984456 3.9015544
        4.7625 3.5805570 3.9194430
        4.8500 3.5750000 3.9250000
        4.9375 3.5805570 3.9194430
        5.0250 3.5984456 3.9015544
        5.1125 3.6342484 3.8657516
        5.2000 3.7500000 3.7500000 /

\hshade 5.1500 3.5000000 3.5000000
        5.2375 3.3842484 3.6157516
        5.3250 3.3484456 3.6515544
        5.4125 3.3305570 3.6694430
        5.5000 3.3250000 3.6750000
        5.5875 3.3305570 3.6694430
        5.6750 3.3484456 3.6515544
        5.7625 3.3842484 3.6157516
        5.8500 3.5000000 3.5000000 /


 \startrotation by  0.965926   0.258819 about      3.669443       5.587500
 \setquadratic
 \plot
     3.669443       5.587500
     3.719443       5.637500
     3.769443       5.587500
     3.819443       5.537500
     3.869443       5.587500
     3.919443       5.637500
     3.969443       5.587500
     4.019443       5.537500
     4.069443       5.587500
     4.119443       5.637500
     4.169443       5.587500
 /
 \stoprotation

 \startrotation by  0.965926  -0.258819 about      3.669443       5.412500
 \setquadratic
 \plot
     3.669443       5.412500
     3.719443       5.462500
     3.769443       5.412500
     3.819443       5.362500
     3.869443       5.412500
     3.919443       5.462500
     3.969443       5.412500
     4.019443       5.362500
     4.069443       5.412500
     4.119443       5.462500
     4.169443       5.412500
 /
 \stoprotation

 \startrotation by -1.000000   0.000000 about      3.325000       5.500000
 \setquadratic
 \plot
     3.325000       5.500000
     3.375000       5.550000
     3.425000       5.500000
     3.475000       5.450000
     3.525000       5.500000
     3.575000       5.550000
     3.625000       5.500000
     3.675000       5.450000
     3.725000       5.500000
     3.775000       5.550000
     3.825000       5.500000
 /
 \stoprotation

 \startrotation by -0.906308  -0.422618 about      3.348446       5.325000
 \setquadratic
 \plot
     3.348446       5.325000
     3.398446       5.375000
     3.448446       5.325000
     3.498446       5.275000
     3.548446       5.325000
     3.598446       5.375000
     3.648446       5.325000
     3.698446       5.275000
     3.748446       5.325000
     3.798446       5.375000
     3.848446       5.325000
 /
 \stoprotation

 \startrotation by -0.906308   0.422618 about      3.348446       5.675000
 \setquadratic
 \plot
     3.348446       5.675000
     3.398446       5.725000
     3.448446       5.675000
     3.498446       5.625000
     3.548446       5.675000
     3.598446       5.725000
     3.648446       5.675000
     3.698446       5.625000
     3.748446       5.675000
     3.798446       5.725000
     3.848446       5.675000
 /
 \stoprotation


 \startrotation by  0.000000  -1.000000 about      2.150000       7.075000
 \setquadratic
 \plot
     2.150000       7.075000
     2.200000       7.125000
     2.250000       7.075000
     2.300000       7.025000
     2.350000       7.075000
     2.400000       7.125000
     2.450000       7.075000
     2.500000       7.025000
     2.550000       7.075000
     2.600000       7.125000
     2.650000       7.075000
 /
 \stoprotation

 \startrotation by -0.422618  -0.906308 about      1.975000       7.094844
 \setquadratic
 \plot
     1.975000       7.094844
     2.025000       7.144845
     2.075000       7.094844
     2.125000       7.044844
     2.175000       7.094844
     2.225000       7.144845
     2.275000       7.094844
     2.325000       7.044844
     2.375000       7.094844
     2.425000       7.144845
     2.475000       7.094844
 /
 \stoprotation

 \startrotation by  0.422618  -0.906308 about      2.325000       7.094844
 \setquadratic
 \plot
     2.325000       7.094844
     2.375000       7.144845
     2.425000       7.094844
     2.475000       7.044844
     2.525000       7.094844
     2.575000       7.144845
     2.625000       7.094844
     2.675000       7.044844
     2.725000       7.094844
     2.775000       7.144845
     2.825000       7.094844
 /
 \stoprotation

 \startrotation by  0.000000   1.000000 about      2.150000       7.425000
 \setquadratic
 \plot
     2.150000       7.425000
     2.200000       7.475000
     2.250000       7.425000
     2.300000       7.375000
     2.350000       7.425000
     2.400000       7.475000
     2.450000       7.425000
     2.500000       7.375000
     2.550000       7.425000
     2.600000       7.475000
     2.650000       7.425000
 /
 \stoprotation

 \startrotation by -0.422618   0.906308 about      1.975000       7.401555
 \setquadratic
 \plot
     1.975000       7.401555
     2.025000       7.451555
     2.075000       7.401555
     2.125000       7.351554
     2.175000       7.401555
     2.225000       7.451555
     2.275000       7.401555
     2.325000       7.351554
     2.375000       7.401555
     2.425000       7.451555
     2.475000       7.401555
 /
 \stoprotation

 \startrotation by  0.422618   0.906308 about      2.325000       7.401555
 \setquadratic
 \plot
     2.325000       7.401555
     2.375000       7.451555
     2.425000       7.401555
     2.475000       7.351554
     2.525000       7.401555
     2.575000       7.451555
     2.625000       7.401555
     2.675000       7.351554
     2.725000       7.401555
     2.775000       7.451555
     2.825000       7.401555
 /
 \stoprotation


 \startrotation by -0.766044  -0.642788 about      4.587500       3.634248
 \setquadratic
 \plot
     4.587500       3.634248
     4.637500       3.684248
     4.687500       3.634248
     4.737500       3.584249
     4.787500       3.634248
     4.837500       3.684248
     4.887500       3.634248
     4.937500       3.584249
     4.987500       3.634248
     5.037500       3.684248
     5.087500       3.634248
 /
 \stoprotation

 \startrotation by -0.224951  -0.974370 about      4.762500       3.580557
 \setquadratic
 \plot
     4.762500       3.580557
     4.812500       3.630557
     4.862500       3.580557
     4.912500       3.530557
     4.962500       3.580557
     5.012500       3.630557
     5.062500       3.580557
     5.112500       3.530557
     5.162500       3.580557
     5.212500       3.630557
     5.262500       3.580557
 /
 \stoprotation

 \startrotation by  0.224951  -0.974370 about      4.937500       3.580557
 \setquadratic
 \plot
     4.937500       3.580557
     4.987500       3.630557
     5.037500       3.580557
     5.087500       3.530557
     5.137500       3.580557
     5.187500       3.630557
     5.237500       3.580557
     5.287500       3.530557
     5.337500       3.580557
     5.387500       3.630557
     5.437500       3.580557
 /
 \stoprotation

 \startrotation by  0.766045  -0.642787 about      5.112500       3.634248
 \setquadratic
 \plot
     5.112500       3.634248
     5.162500       3.684248
     5.212500       3.634248
     5.262500       3.584249
     5.312500       3.634248
     5.362500       3.684248
     5.412500       3.634248
     5.462500       3.584249
     5.512500       3.634248
     5.562500       3.684248
     5.612500       3.634248
 /
 \stoprotation


\put {$e^{+}$} at 0.84 7.25

\put {$H$} at 0.84 3.75

\put {$\nu$} at 6.16 7.25

\put {$W^{-}_L$} at 6.16 3.75

\put {$\longrightarrow$} at 1.25 7.45

\put {$p$} at 1.25 7.6

\put {$\longleftarrow$} at 5.75 7.45

\put {$\nu$} at 5.75 7.6

\put {$\longrightarrow$} at 1.25 3.55

\put {$H$} at 1.25 3.4

\put {$\longleftarrow$} at 5.75 3.55

\put {$q$} at 5.75 3.4

\put {$1,\ldots,s$} at 2.15 8.10

\put {$t{+}1,\ldots,n$} at 4.85 2.90

\put {$s{+}1,\ldots,t$} at 4.65 5.5

\put {$\downarrow\enspace
{\cal{P}}=p+\kappa(1,s)+\nu$} at 4.4671473 6.50

\put {$\uparrow\enspace
{\cal{Q}}=H+\kappa(t{+}1,n)+q$} at 4.6 4.5

\endpicture
$$
\bigskip\noindent
Figure 2.  Contributions to
$e^{+} \thinspace \nu \longrightarrow W_L^{+} \thinspace H
\thinspace \gamma \thinspace \gamma \cdots \gamma$.
The seagull graph with a photon attached at the $\phi^{+}
W^{-}H$ vertex is not shown.

\vfil\eject\bye